\documentclass[traditabstract]{aa}

\usepackage{comment}
\usepackage{lscape}

\usepackage{natbib}
\usepackage{amssymb}
\usepackage{url}
\usepackage{deluxetable}
\usepackage{graphicx}
 \usepackage{longtable}
\newcommand{\plotone}[1]{\includegraphics[width=0.9\textwidth]{#1}}

\newcommand{\zrangecol}{1}
\newcommand{\volcol}{2}
\newcommand{\lirdcol}{3}
\newcommand{\sfrdcol}{4}
\newcommand{\sfrdcontrcol}{5}
\newcommand{\mstardcol}{6}
\newcommand{\mstardcontrcol}{7}
\newcommand{\qavcol}{8}
\newcommand{\mdustdcol}{9}

\newcommand{\sfruvcol}{4}
\newcommand{\sfrircol}{5}
\newcommand{\ssfrcol}{7}
\newcommand{\mstarcol}{8}
\newcommand{\mburstcol}{9}
\newcommand{\mlkcol}{10}
\newcommand{\tdcol}{13}
\newcommand{\avcol}{14}
\newcommand{\agncol}{16}
\newcommand{\submm}{submillimeter}

\begin{document}

\title{Cosmic evolution  of submillimeter galaxies and their contribution to stellar mass assembly}
 
\titlerunning{Evolution of submillimeter galaxies}
\authorrunning{Micha{\l}owski et al.}

\author{Micha{\l}~J.~Micha{\l}owski\inst{1,}\inst{2}
\and
Jens Hjorth\inst{1}
\and 
Darach Watson\inst{1}
	}

\institute{Dark Cosmology Centre, Niels Bohr Institute, University of Copenhagen, Juliane Maries Vej 30, DK-2100 Copenhagen \O, Denmark, {\tt michal@dark-cosmology.dk}
\and
{Scottish Universities Physics Alliance}, Institute for Astronomy, University of Edinburgh, Royal Observatory, Edinburgh, EH9 3HJ, UK
}

\abstract{
The nature of galaxies selected at {\submm} wavelengths (SMGs, $S_{850}\gtrsim3$ mJy), some of the bolometrically most luminous objects at high redshifts, is still elusive. In particular their star formation histories and source of emission are not accurately constrained. In this paper we { introduce a new approach to analyse the SMG data. Namely, we present the first  self-consistent UV-to-radio} spectral energy distribution fits of 76 SMGs with spectroscopic redshifts using all photometric datapoints from ultraviolet to radio simultaneously. We find that they are highly star-forming (median star formation rate $713\,M_\odot\mbox{ yr}^{-1}$ { for SMGs at $z>0.5$}), { moderately dust-obscured (median $ A_V\sim2$ mag)}, hosting significant stellar populations (median stellar mass $3.7\times10^{11}\,M_\odot$) of which only a minor part has been formed in the ongoing starburst episode. This implies that in the past, SMGs experienced either another starburst episode or merger with several galaxies. The properties of SMGs suggest that they are progenitors of present-day elliptical galaxies. { We find that these bright SMGs  contribute significantly to the cosmic star formation rate density { ($ \sim20$\%)}  and stellar mass density { ($ \sim30$--$ 50$\%)} at redshifts $2$--$4$. }
Using  number counts at low fluxes we find that as much as { $ 80$\%} of the cosmic star formation at these redshifts took place in SMGs brighter than $0.1$ mJy. 
{  We find evidence that  a linear infrared-radio correlation holds for SMGs in an unchanged form up to redshift of $3.6$, though its normalization is offset from the local relation  by a factor of $\sim2.1$ towards higher radio luminosities.}
{ We present a compilation of photometry data of SMGs and determinations of cosmic SFR and stellar mass densities.}

}

\keywords{galaxies: active -- galaxies: evolution  -- galaxies: high-redshift -- galaxies: ISM -- galaxies: starburst -- submillimeter}

\maketitle

\section{Introduction}
\label{sec:intro}

Submillimeter galaxies \citep[SMGs; see][]{blain02} were discovered at $850\,\mu$m ($S_{850}\gtrsim3$ mJy) by the Submillimetre Common-User Bolometer Array \citep[SCUBA;][]{hollandscuba} mounted on the James Clerk Maxwell Telescope (JCMT). Due to the coarse resolution of SCUBA, localizations derived from high-resolution radio maps had to be used to measure their spectroscopic redshifts \citep{chapman05}. Lots of studies have addressed the issue of characterizing the nature of SMGs 
 \citep[][{ some of these works were based on surveys with sensitivity worse than $3$ mJy quoted above}]{egami04,greve04,greve05,smail04,swinbank04,swinbank06,swinbank08,takagi04,%
alexander05,borys05,%
kovacs06,laurent06,pope06,tacconi06,tacconi08,takata06,%
younger07,younger08b,younger09b,%
clements08,coppin08,dye08,dye09,hainline08phd,hainline09,perera08,scott08,%
austermann09,devlin09,eales09,murphy09,murphy09b,tamura09,weiss09b,aravena10}.
 However they were usually based on limited samples ($\lesssim\mbox{}$20 sources), limited wavelength coverage or photometric redshifts. These limitations have made it difficult to solve several issues, including the characterization of the star formation histories of SMGs and their dominant source of emission.

An important open question concerns the contribution of SMGs to cosmic stellar mass assembly.
This is important, because in order to understand galaxy evolution, the build-up of stellar mass must be mapped out to high redshifts. It is usually parametrized by the total star formation rate (SFR) density per unit comoving volume, \citep[$\rho_{\rm SFR}$; see e.g.][]{hopkins04,hopkins06}. 
At high redshifts it is difficult to disentangle the contribution to $\rho_{\rm SFR}$ from galaxy populations of different masses due to incompleteness at low luminosities. 

Another approach to study stellar mass assembly is to consider directly the stellar mass density per unit comoving volume, $\rho_*$, which is equivalent to the integrated $\rho_{\rm SFR}$ over the age of the Universe. It is established that $\rho_*$ grows with cosmic time  \citep[stellar mass is accumulating;][]{drory05,fontana06,elsner08,perezgonzalez08,marchesini09}, but the contribution from different galaxy populations is not well-determined. { {\it Spitzer} observations of SMGs \citep{egami04,frayer04,ivison04,borys05,ashby06,laurent06,pope06,dye08,hainline08phd,hainline09} have enabled studies of the rest-frame near-infrared (near-IR) part of the spectrum, where old stellar populations are dominant -- an important step forward in getting full spectral energy distributions  and accurate estimates of stellar masses of SMGs. The results indicate that SMGs are among the most massive galaxies in the Universe.}

The dominant source of emission from SMGs is dust reprocessed emission  either from young stars or active galactic nuclei (AGNs). One way to test it is to compare the infrared (IR) and radio luminosities of SMGs, because, at least locally,
star-forming galaxies follow a remarkably tight correlation between IR and radio luminosities \citep{helou85,condon}. The correlation is believed to result from the fact that both IR and radio emissions are related to short-lived massive stars: the former originates from dust heated by  ultraviolet (UV) light from blue, massive stars and the latter from synchrotron emission of electrons produced in supernova remnants. Therefore, a relation consistent with the local one is an indication of star formation dominating both the IR and radio emissions. There is growing evidence that the correlation holds at redshifts $z\lesssim1$ \citep{garrett02,gruppioni03,appleton04,boyle07,marleau07,vlahakis07,yang07}. At higher redshifts sample sizes are small making it difficult to draw  robust conclusions \citep{appleton04,kovacs06,beswick08,ibar08,sajina08,garn09,murphy09,murphy09b,rieke09,seymour09,younger09}.
The only sign of evolution was reported by \citet{ivison09} based on stacking analysis of the $24\,\mu$m-selected galaxies, though possibly interpreted as a selection effect.

The objective of this paper { is to model for the first time the entire UV-to-radio spectral energy distributions of a statistically significant sample of SMGs in a self-consistent way. Using these models we}
{\it i}) consistently  derive the properties of  SMGs using all available data to characterize their nature and determine the dominant emission mechanism; {\it ii}) estimate the contribution of SMGs to the cosmic SFR and stellar mass densities; {\it iii}) investigate whether the local IR-radio correlation holds at high redshifts in an unchanged form. In Section~\ref{sec:sample} our SMG sample is presented. Our methodology is outlined in Section~\ref{sec:method}. We derive the properties of SMGs in Section~\ref{sec:results} and discuss the implications in Section~\ref{sec:discussion}. Section~\ref{sec:conclusion} closes with our conclusions.
We use a cosmological model with $H_0=70$ km s$^{-1}$ Mpc$^{-1}$,  $\Omega_\Lambda=0.7$ and $\Omega_m=0.3$.

\section{Sample}
\label{sec:sample}

We base our analysis on 76 SMGs ($S_{850}\gtrsim3$ mJy) from the sample of \citet{chapman05}, all with spectroscopically measured redshifts spanning a range of $0.080$--$3.623$. 

The way the sample is selected involves complex biases, which are difficult to fully quantify and account for. The parent sample of \citet{chapman05} consists of 150 SMGs out of which 104 have radio identifications. The sample discussed here (76 galaxies) consists of the SMGs for which redshifts have been measured (spectroscopic completeness $\sim75$\%).  All this implies that the sample is biased against: {\it i}) faint {\submm} emitters (low dust content and/or hot dust, influence mostly the low-$z$ portion of the sample); {\it ii}) faint radio emitters \citep[high-$z$ and cold dust, see Figure~3 of][]{chapman05}; {\it iii}) faint optical emitters (difficult to obtain spectra); {\it iv}) $z\sim1.2$--$1.8$ (``redshift desert'' where no emission lines enter the observable wavelengths). At low redshifts ($z<1$) the sample may also be incomplete due to a limited sky area (and therefore -- volume) coverage making it difficult to detect rare strong {\submm} emitters (for details on the SMG selection effects see also Figure~2 of \citet{blainT} and discussion in Section~4.4 of \citet{michalowski08}). 

It is important to estimate what  the influence of these selection effects on our results is. In total we analyse $\sim50$\% (76/150) of the parent sample. Additionally, 25 radio-detected SMGs without spectroscopic redshifts have similar long-wavelength properties compared to the redshift sample  \citep[see Figure~1 of][]{chapman05}, so their absence from the sample probably does not significantly bias our results. The same is true for the SMGs in the ``redshift desert'', since they are missed not due to their inherent properties. The remaining  46 radio-nondetected SMGs ($\sim30$\%)  could in principle have very different properties than our sample resulting in a potential limitation in our analysis.

Even if most of the SMGs without spectroscopic redshifts are similar to those in our sample,  the incompleteness at $z<1.8$ implies that the estimates of SMG densities (Sections~\ref{sec:sfrd}, \ref{sec:mstard} and \ref{sec:dust}) in the three low-redshift bins (see Section~\ref{sec:bins}) are strict lower limits.

{
Due to the negative K-correction at {\submm} wavelengths, SMGs at $z\gtrsim0.5$ form a sample with homogenous IR luminosity \citep{blain96,blain97}. However, SCUBA sources at $z\lesssim0.5$ belong to a different population of objects and are intrinsically fainter.  
The limited volume coverage at these low redshifts makes the sample of these objects small and incomplete. This prevents a separate study of their properties.  
We did not take into account these sources when we computed median values of the properties of SMGs.
}

The photometric datapoints (Tables~\ref{tab:phot} and \ref{tab:lim} in appendix%
\footnote{ For convenience we make the compilation available in electronic form. We suggest that the original data source be consulted and referred to appropriately.}%
) were collected from the literature: \citet[][$IK$, radio]{ivison02}, \citet[][$R$, $1.2$ mm]{ivison05}, \citet[][$VI$]{chapman03b}, \citet[][$BR$, $850\,\mu$m, radio]{chapman05}, \citet[][$UBVRIzHK$]{capak04}, \citet[][$UBVIK$]{clements04}, \citet[][$24\,\mu$m]{egami04}, \citet[][$1.2$ mm]{greve04}, \citet[][$IJK$]{smail04}, \citet[][$Rz$]{fomalont06}, \citet[][$350\,\mu$m, $1.2$ mm]{kovacs06}, \citet[][$350\,\mu$m, $1.1$ mm]{laurent06}, \citet[][$1.3$ mm]{tacconi06}, \citet[][$R$, $24\,\mu$m]{pope06}, \citet[][$160\,\mu$m]{huynh07}, \citet[][$3.6, 4.8, 5.6, 8.0, 24, 70\,\mu$m]{hainline08phd}. 
{ We have not used the existing mid-IR spectra \citep{valiante07,pope08,menendezdelmestre07,menendezdelmestre09}, but for completeness  we have indicated in Table~\ref{tab:phot} those SMGs for which {\it Spitzer}/IRS spectra exist.}

\begin{figure*}
\begin{center}
\plotone{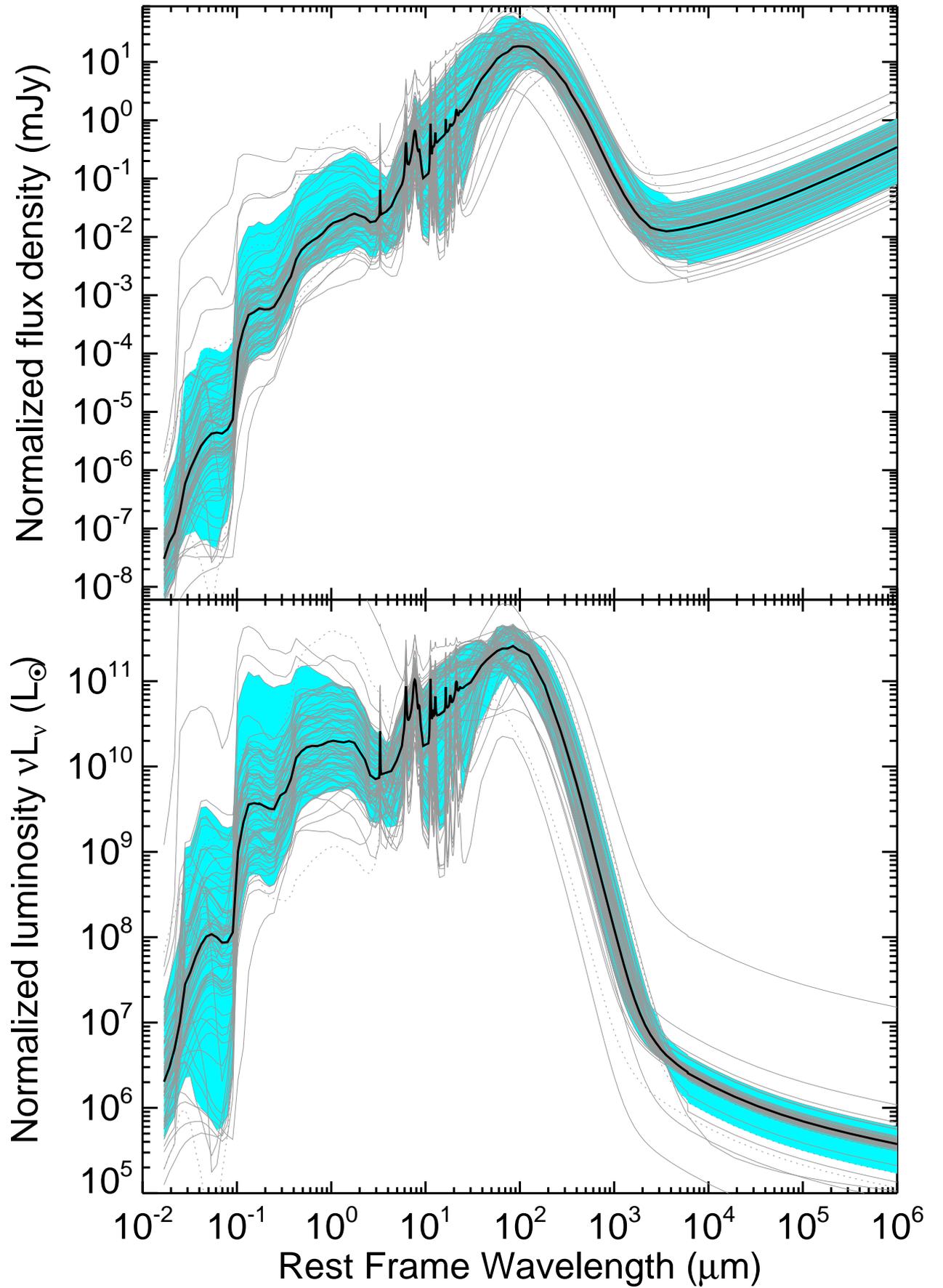}
\end{center}
\caption{Median spectral energy distribution (SED) of SMGs ({\it thick lines}) and SEDs of individual SMGs ({\it thin lines}). { {\it Dotted lines} indicate $z<0.5$ objects.} {\it Shaded areas} enclose $90$\% of the SEDs. {\it Top}: all SEDs were divided by the corresponding $850\,\mu$m datapoint and scaled, so that the median SED has a flux of $5$ mJy at the rest-frame $283\,\mu$m (observed $850\,\mu$m at $z=2$). {\it Bottom}: SEDs were normalized to an infrared star formation rate of $100\,M_\odot$ yr$^{-1}$.
}
\label{fig:avsed}
\end{figure*}

\begin{figure*}
\begin{center}
\includegraphics[width=0.85\textwidth]{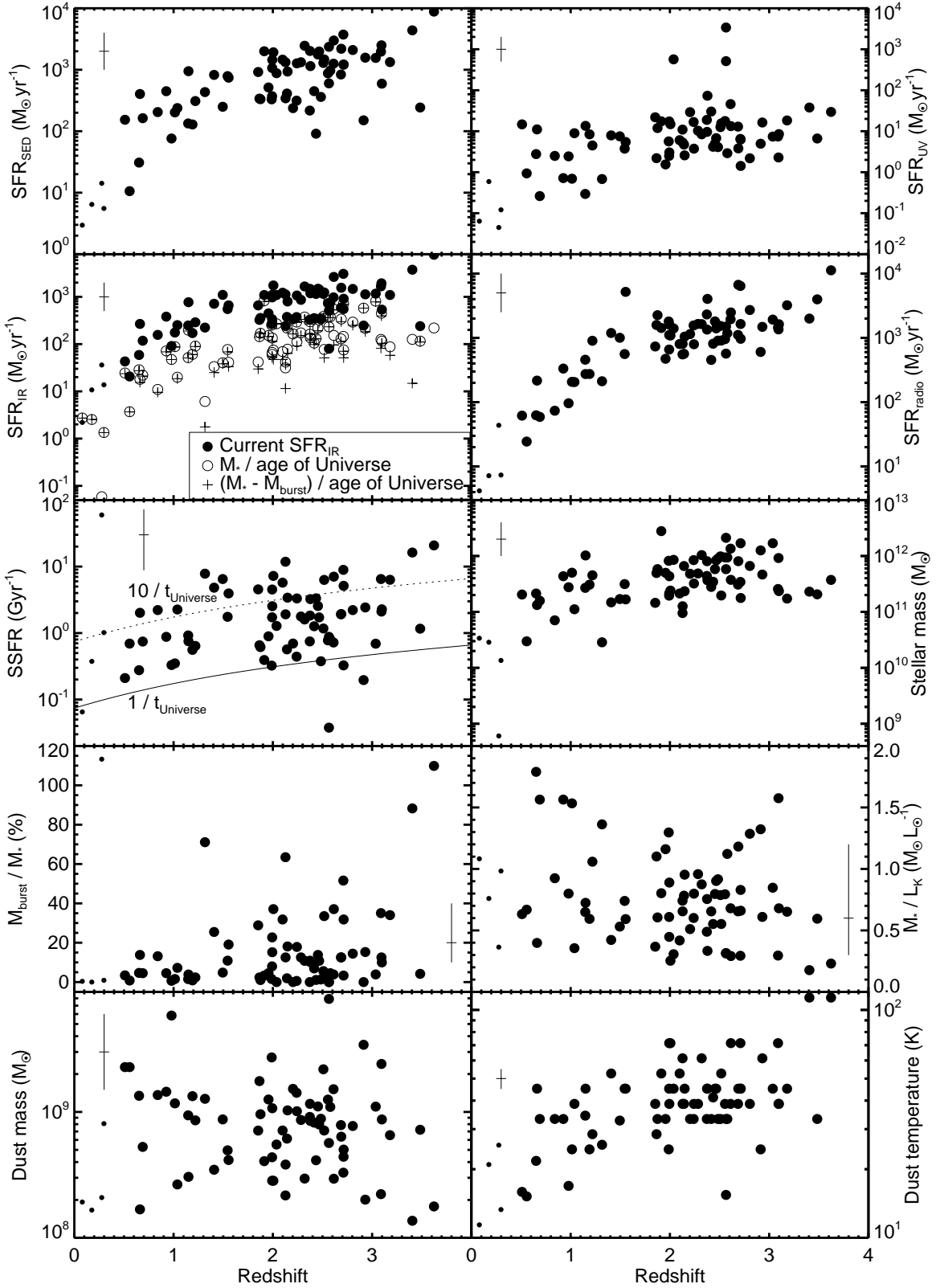}
\end{center}
\caption{Redshift evolution of the properties ({\it full circles}, see Table~\ref{tab:grasilres} in appendix) of the sample of 76 SMGs with spectroscopic redshifts \citep{chapman05}. { {\it Small symbols} indicate $z<0.5$ objects.}  Typical errors (Section~\ref{sec:results}) are shown as {\it crosses}. From top-left to bottom-right: star formation rate (SFR) derived from spectral energy distribution modeling, ultraviolet, infrared and radio emission,  SFR per unit stellar mass ($\equiv\mbox{SFR}_{\rm IR}/M_*$), stellar mass, fraction of stellar population formed during the ongoing starburst, stellar mass-to-light ratio, dust mass and temperature.  In the SFR$_{\rm IR}$ panel, we also show the minimum average SFRs (see Section~\ref{sec:sfr}) required to build up the total stellar mass within the age of the Universe at a given redshift ({\it empty circles}) and to build up the fraction of stellar population that was not formed during the ongoing starburst ({\it plus signs}). The location of plus signs indicates that SMGs must have been highly star-forming even before the onset of the ongoing starburst. When empty circles and plus signs overlap, the contribution of the ongoing starburst to the total stellar mass of a galaxy is  negligible (i.e.~$M_{\rm burst}/M_*\sim0$).
}
\label{fig:zevol}
\end{figure*}

\section{Methodology}
\label{sec:method}

\subsection{SED modeling}
\label{sec:sed}

{ In order to model the spectral energy distributions (SEDs) of SMGs, we use all the photometric datapoints simultaneously. This has the advantage that all the galaxy properties are derived consistently regardless of the wavelength regime in which those properties shape the SEDs (for example, recent star formation governs the UV and far-IR parts of a spectrum of a galaxy, whereas accumulated stellar mass is responsible for near-IR emission). Moreover in the full SED modeling no single datapoint drives the fit alone}. 

We utilized the set of 35\,000 models from \citet{iglesias07} developed in GRASIL \citep{silva98}\footnote{\url{http://adlibitum.oat.ts.astro.it/silva/default.html}} based on numerical calculations of radiative transfer within a galaxy. They cover a broad range of galaxy properties from quiescent to starburst. Their star formation histories are assumed to be a smooth Schmidt-type law \citep[SFR proportional to the gas mass to some power, see][for details]{silva98} with a starburst (if any) on top of that starting $50$ Myr { before the time of the evolution of a galaxy at which the SED is computed}. Additionally we fitted templates based on nearby galaxies \citep{silva98} and gamma-ray burst host galaxies \citep{michalowski08}. We simultaneously  used all the photometric datapoints from UV to radio (Tables~\ref{tab:phot} and \ref{tab:lim}). In cases where the data given by different authors were contradictory, we disregarded the obvious outliers. We scaled the SEDs to match the data and chose the one with the lowest $\chi^2$.

Based on the best fits we derived the properties of the galaxies as explained in \citet{michalowski08,michalowski09}. In particular,  SFRs, stellar ($M_*$) and starburst ($M_{\rm burst}$) masses  were given as output from GRASIL, { rest-frame UV and $K$ ($L_K$) monochromatic luminosities were interpolated from the best-fitting SEDs, whereas IR luminosities ($L_{\rm IR}$) were integrated in a range $8$--$1000\,\mu$m, UV and IR SFRs (SFR$_{\rm IR}$ was adopted for all subsequent calculations, because SFR$_{\rm UV}$ is on average two orders of magnitude lower) were calculated using \citet{kennicutt}, } dust masses ($M_d$) were calculated from the $850\,\mu$m detections using equation~(5) of \citet{michalowski09} and radio SFRs were calculated from the $20$~cm detections using the empirical formula of \citet{bell03} \citep[see Section~4.2 of][]{michalowski09}. Dust temperatures ($T_d$) were estimated by identifying the peak of the dust emission and assuming an emissivity index $\beta=1.3$. { The  average extinction in the rest-frame $V$-band was calculated from the unextinguished starlight given in GRASIL:  $ A_V=2.5\log$(unextinguished $V$-band starlight / observed  $V$-band starlight).} IR-radio correlation parameters were calculated according to the formula $q=\log(L_{\rm IR}[L_\odot] / 3.75\times10^{12}/L_{\nu\, 1.4\, {\rm GHz}} [L_\odot \mbox{Hz}^{-1}])$, where $L_{\nu\, 1.4\, {\rm GHz}}$ is a rest-frame $1.4$ GHz luminosity density computed from the observed $1.4$ GHz flux assuming a spectral slope of $-0.75$.

\subsection{Volume densities}
\label{sec:bins}

In order to calculate the SFR density, the stellar density and the dust mass densities per unit comoving volume, $\rho_{\rm SFR}$, $\rho_*$ and $\rho_{\rm dust}$, we used the following angular areas for the {\submm} surveys \citep[Table~1 of][]{chapman05}: CFRS-03: $60$ arcmin$^2$ and CFRS-14: $48$ arcmin$^2$  \citep{webb03}, Lockman Hole: $122$ arcmin$^2$ and ELAIS-N2: $102$ arcmin$^2$ \citep{scott}, HDF-N: $100$ arcmin$^2$ \citep{chapman01}, SSA-13 and SSA-22: $100$  arcmin$^2$ each \citep{chapman03c}, totaling $632$ arcmin$^2$.

{ We divided our sample into four high-redshift bins (Table~\ref{tab:zbins}) with approximately the same number of SMGs plus an additional bin for $z<0.5$ sources (see Section \ref{sec:sample}).} 
The densities in each bin were calculated as a sum of SFR$_{\rm IR}$ (or $M_*$, or $M_{d}$) of all SMGs in this bin divided by its comoving volume { \citep[a similar approach to calculate the SFR and number volume densities of SMGs was taken by][]{coppin09,daddi09,younger09b,wang09}. The volume densities (Column~{\volcol})} were found using the total area from the previous paragraph. 

{
We removed the contribution of ten SMGs\footnote{\object{SMMJ123553.26+621337.7}, \object{SMMJ123555.14+620901.7}, \object{SMMJ123600.10+620253.5}, \object{SMMJ123600.15+621047.2}, \object{SMMJ123606.85+621021.4}, \object{SMMJ123716.01+620323.3}, \object{SMMJ163706.51+405313.8}, \object{SMMJ221804.42+002154.4}, \object{SMMJ221806.77+001245.7}
},
which were observed by SCUBA in the photometry mode (as opposed to the blank-field mapping mode) targeting optically-faint radio galaxies \citep{chapman05}. These objects fall outside the fields discussed here.
}

{
The method is therefore to analyse the fraction of the sky observed by SCUBA and estimate the number  of SMGs and their volume densities. However, the true number of SMGs in our fields could be higher. On the other hand, regardless of the selection effects, the true number of SMGs in our fields
cannot be lower than the number of SMGs in our sample. In turn, the true values of SFR and
$M_*$ densities cannot be lower than the values we derive. Therefore our results on volume densities should be regarded as robust lower limits.
}

{
Having this in mind we note that the parent sample of \citet{chapman05} includes only $29$\% of all the SMGs detected in the used survey fields \citep[compare with][]{scott,webb03,webb03b}. Therefore { even if we analysed the full parent sample} the estimated densities  { would be} conservative lower limits. We attempt to correct for this incompleteness by assuming that the parent sample of \citet{chapman05} is a fair representation of the total population. In this case our numbers should be multiplied by $3.5$ { ($\sim1/29$\%). This correction should in principle be derived separately for each redshift bin, but the missing redshift information for the majority of  the SMGs in the used survey fields makes such calculation impossible.  We note that this correction does not remove the bias against SMGs that are faint at radio and optical wavelengths, as discussed in Section~\ref{sec:sample}}. 
}

{
We have not applied a volume density correction for the AGN contribution, because it is at most minor. Even though a fraction
of SMGs host AGNs and a few individual SMGs have been shown to exhibit a significant AGN contribution to their emission, it is established that on average AGN activity is responsible for at most $\sim10$--$20$\% of the bolometric infrared emission of SMGs \citep{alexander05,alexander08,menendezdelmestre07,menendezdelmestre09,valiante07,pope08,hainline09,murphy09, watabe09}.
Therefore a potential error associated with the AGN contribution in our analysis
of a statistically significant sample is smaller than the systematic
uncertainty \citep[e.g.~30\% error of luminosity-SFR conversion;][]{kennicutt}.}

The percentage contribution of SMGs to the SFR and $M_*$ densities (Columns~{\sfrdcontrcol} and {\mstardcontrcol} of Table~\ref{tab:zbins}) was calculated as $\rho_{\rm SMG}/(\rho_{\rm SMG}+\rho_{\rm other})$, where $\rho_{\rm SMG}$ is the density of SMGs at each redshift bin (Columns~{\sfrdcol} and {\mstardcol}) and $\rho_{\rm other}$ is the density of other galaxies assumed to be an average of determinations (excluding lower limits) reported by other authors (Figure~\ref{fig:sfrdz}; Tables~\ref{tab:sfrd} and \ref{tab:mstard}  in appendix), for which the redshift ranges overlap with our bins. This way of calculating the contribution is justified if SMGs do not enter the ``other'' samples of galaxies. This is usually the case because SMGs are faint in the optical. However, if this was not fulfilled, the real percentage contribution of SMGs would be even higher. 

\section{Results}
\label{sec:results}

 The best fits\footnote{The SED fits can be downloaded from\\ \protect\url{http://archive.dark-cosmology.dk}} 
are shown in Figure~\ref{fig:sed} and the median SEDs (in flux and luminosity domains) are shown in Figure~\ref{fig:avsed}. 

The resulting properties of the galaxies are listed in Table~\ref{tab:grasilres} and shown in Figure~\ref{fig:zevol} as a function of redshift. { We notice similar trends to \citet{hainline08phd} that lower-$z$ SMGs are less luminous and colder (see her Figures~4.7 and 4.9).}

In two cases we obtained much better fits using the templates of \citet{silva98} instead of those of \citet{iglesias07}, namely, an \object {HR 10} template for \object{SMMJ105151.69+572636.0} and a spiral Sc template for \object{SMMJ221733.12+001120.2}. In 9 cases\footnote{\object{SMMJ030226.17+000624.5}, \object{SMMJ030231.81+001031.3}, \object{SMMJ030236.15+000817.1}, \object{SMMJ030238.62+001106.3}, \object{SMMJ123636.75+621156.1}, \object{SMMJ123651.76+621221.3}, \object{SMMJ123721.87+621035.3}, \object{SMMJ163639.01+405635.9}, \object{SMMJ221724.69+001242.1}} where our fits strongly underpredict the $850\,\mu$m datapoint we adopted the $L_{\rm IR}$ and $T_d$ estimates of \citet{chapman05}.

The determination of the IR luminosity suffers from systematic uncertainties depending on the choice of the SED template. Our approach of using all the optical, {\submm} and radio data to constrain the shape of the SED results in a moderate systematic error in the IR luminosity \citep[less than a factor of $\sim2$;][]{bell07}.
The choice of a \citet{salpeter} IMF with cutoffs of $0.15$ and $120\,{\rm M}_\odot$ introduces a maximum systematic error of a factor of $\sim2$ in the determination of the stellar masses and SFRs \citep{erb06}. \citet{bell07} have also found that random errors in stellar mass are less than a factor of $\sim2$. 
Estimates of dust temperatures  have uncertainties of $\sim5$--$10$ K dominated by the unknown value of the emissivity index, $\beta$. The SFR determination  based on radio observations is accurate up to $30$\% since it agrees with the detailed spectrophotometric SED fitting \citep{michalowski06}. The uncertainties in $q$ (defined in Section~\ref{sec:sed}) are $\sim0.3$ \citep[see also][]{kovacs06}, dominated by the error in  $L_{\rm IR}$.

\begin{figure}
\begin{center}
\includegraphics[width=0.45\textwidth]{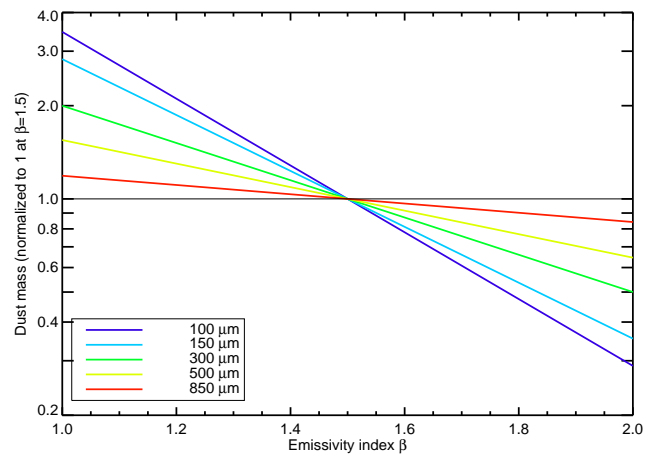}
\end{center}
\caption{{ Derived dust mass of a mock galaxy with dust temperature $T_d=40$~K  and a flux density of $5$ mJy at several infrared rest-wavelengths as a function of the assumed emissivity index $\beta$. For each wavelength the dust masses were normalized to 1 at $\beta=1.5$. The spread of the derived dust masses shows that} the uncertainty of the dust mass resulting from unknown $\beta$ is a factor of a few.
}
\label{fig:beta}
\end{figure}

{ 
In order to assess the influence of the choice of emissivity index $\beta=1.3$ on the dust mass estimates, we recalculated the dust temperatures and masses in a range of $\beta$ of $1$--$2$. The resulting error was less than a factor of $3.5$.

This is illustrated on Figure~\ref{fig:beta} where we present a more systematic analysis of this problem. We calculated the dust mass  of a mock galaxy with $T_d=40$~K (this choice does not influence the results) using $\beta$ in the range $1$--$2$ assuming a flux density of $5$~mJy at a variety of infrared rest-wavelengths { probed by observations}. Then we normalized dust masses to $1$ at $\beta=1.5$. We conclude that as long as the observations probe wavelengths longer than $\sim150\,\mu$m ($z\lesssim4.7$ for observed wavelength of $850\,\mu$m), then the error on the dust mass resulting from unknown $\beta$ is less than a factor of $\sim5$.
} 

None of these errors significantly affects our conclusions, because the inferred nature of SMGs would not be different even in the worst case scenario when all systematic errors work in one direction (increasing or decreasing the obtained values). Moreover, we analyse a statistically significant sample of 76 galaxies, so random errors of a factor of $2$ are reduced to $<20$\% when an error of a mean is considered. 

\begin{table*}
\caption{Mean values for SMGs in redshift bins \label{tab:zbins}	}
\centering
\begin{tabular}{ccccccccc}
\hline\hline
 & Volume & $\log\rho_{\rm IR}$  & $\rho_{\rm SFR}$ & & $\log\rho_{*}$ & & & $\log\rho_{\rm dust}$\\
$z$  & ($10^6$ Mpc$^3$) & $(L_\odot\,\mbox{Mpc}^{-3})$ & $(M_\odot\,\mbox{yr}^{-1}\,\mbox{Mpc}^{-3})$ & \multicolumn{1}{c}{\%} & $(M_\odot\,\mbox{Mpc}^{-3})$ & \% & $q$ & $(M_\odot\,\mbox{Mpc}^{-3})$\\
(1)  & (2)              &  (3) & (4) &  (5) & (6) & (7) & (8) & (9)\\
\hline
$0.080-0.500$ & $0.12$ & $7.03_{-0.14}^{+0.11}$ & $0.0018\pm0.0005$ & $
           5_{-           2}^{+           3}$ & $6.35_{-0.16}^{+0.21}$ & $
           1_{-           0.2}^{+           0.4}$ & $2.54\pm 0.12$ & $
4.06_{-0.06}^{+0.05}$ \\
$0.510-1.316$ & $1.03$ & $7.81_{-0.07}^{+0.06}$ & $0.0111\pm0.0016$ & $
           9_{-           2}^{+           2}$ & $7.16_{-0.08}^{+0.13}$ & $
           5_{-           1}^{+           2}$ & $2.52\pm 0.06$ & $
4.32_{-0.05}^{+0.04}$ \\
$1.408-2.142$ & $1.63$ & $8.12_{-0.05}^{+0.05}$ & $0.0228\pm0.0027$ & $          11_{-           3}^{+           3}$ & $7.18_{-0.11}^{+0.16}$ & $
          11_{-           3}^{+           6}$ & $2.29\pm 0.08$ & $3.84_{-0.04}^{+0.04}$ \\
$2.148-2.565$ & $0.89$ & $8.45_{-0.05}^{+0.05}$ & $0.0486\pm0.0054$ & $          18_{-           5}^{+           7}$ & $7.61_{-0.08}^{+0.12}$ & $
          51_{-          11}^{+          20}$ & $2.27\pm 0.09$ & $4.35_{-0.04}^{+0.04}$ \\
$2.578-3.623$ & $2.17$ & $8.30_{-0.05}^{+0.05}$ & $0.0341\pm0.0040$ & $          20_{-           4}^{+           5}$ & $7.28_{-0.07}^{+0.12}$ & $
          31_{-           7}^{+          14}$ & $2.25\pm 0.08$ & $3.86_{-0.03}^{+0.03}$ \\
%
\hline
\tablecomments{Column~(\zrangecol):~redshift range of the bins. Column~(\volcol):~comoving volume of these bins (calculated in Section~\ref{sec:bins}). Column~(\lirdcol):~IR luminosty density of SMGs. Column~(\sfrdcol):~Resulting IR SFR density of SMGs (Section~\ref{sec:sfrd}). Column~(\sfrdcontrcol):~contribution of SMGs to the cosmic SFR density (calculated in Section~\ref{sec:bins}). Column~(\mstardcol):~stellar mass density of SMGs (Section~\ref{sec:mstard}). Column~(\mstardcontrcol):~contribution of SMGs to the cosmic $M_*$ density (calculated in Section~\ref{sec:bins}). Column~(\qavcol):~mean (and error of the mean) FIR-radio correlation parameter for SMGs (Section~\ref{sec:firr}). Column~(\mdustdcol):~dust mass density of SMGs (Section~\ref{sec:dust}). Columns 3-7 and 9 have been corrected for incompleteness by a factor of $3.5$ (Section~\ref{sec:bins}).  }
\end{tabular}
\end{table*}

Table~\ref{tab:zbins} contains the volume densities and mean IR-radio correlation parameter divided into five redshift bins (see Section~\ref{sec:bins}). 
The uncertainties quoted on $\rho_{\rm SFR}$ and $\rho_{*}$ include the systematic $30$\%  uncertainty of the $L_{\rm IR}$ to SFR conversion \citep{kennicutt} and a factor of $\sim2$ systematic uncertainty in the stellar mass \citep{michalowski08}. { The systematic error resulting from our incompleteness correction (Section~\ref{sec:bins}) is likely a factor of a few.}

\section{Discussion}
\label{sec:discussion}

\subsection{Spectral energy distributions of SMGs}

{
We have presented the first successful attempt  to fit the entire UV-to-radio SEDs of SMGs in a self-consistent way taking into account
all the available data simultaneously.
Our study provides evidence that GRASIL models can reproduce the SMG data. Namely, we found good fits for all  SMGs in our sample with the best IR/{\submm} wavelength coverage%
\footnote{SMMJ105201.25+572445.7, SMMJ105230.73+572209.5,
SMMJ163650.43+405734.5, SMMJ163658.19+410523.8,
SMMJ163706.51+405313.8} except of SMMJ105238.30+572435.8. 

As is evident from Figure~\ref{fig:avsed}, regardless of whether SEDs were normalized to the same observed $850\,\mu$m datapoint or SFR$_{\rm IR}$, the scatter at optical and near-IR wavelengths is significant, showing that SMGs exhibit a wide range of stellar population properties \citep[as also noted by][]{ivison02}. This implies the need for an SED template library in SMG studies, as opposed to single-template fitting.

Having constrained the SEDs of SMGs we now turn to a discussion of what we can learn about these galaxies using the best-fitting models.
}

\subsection{Properties of SMGs}

\subsubsection{Star formation rates}
\label{sec:sfr}

The very high (current) SFRs of SMGs (median $713\,M_\odot$ yr$^{-1}$, Column~{\sfrircol} of Table~\ref{tab:grasilres} and Figure~\ref{fig:zevol}) place them among the most powerful starburst galaxies in the Universe. Such extreme SFRs likely result from major mergers \citep[e.g.][]{chapman04b,swinbank04,greve05,tacconi06,tacconi08,younger07,younger08b,bercianoalba10,narayanan09,narayanan10}  and cannot be sustained for a long period { \citep[after a few hundred  Myr { at most} the gas reservoir { should} be depleted; see][]{greve05,hainline06}. }

On the other hand, their { extinction-uncorrected} UV SFRs are two orders of magnitude lower (median $\sim7\,M_\odot$ yr$^{-1}$, Column~{\sfruvcol}). This implies that the majority of star formation in SMGs is hidden by dust. Therefore, optical observations alone are not sufficient to investigate their nature and contribution to cosmic star formation.

Using stellar masses of SMGs we placed lower limits on the time-averaged SFRs required to build their stellar masses within the age of the Universe ($\equiv M_*/\mbox{age of the Universe at given redshift}$), shown as empty circles on Figure~\ref{fig:zevol}. Their median value of $\sim130\,M_\odot\mbox{ yr}^{-1}$ indicates that SMGs had to be relatively highly star-forming throughout the age of the Universe to build up their stellar populations at a constant rate. { Even if our estimates of stellar masses were underestimated by a factor of a few due to systematic uncertainties (Section~\ref{sec:results}), the SMGs would have had to be luminous infrared galaxies (LIRGs with SFR$\mbox{}\gtrsim20\,M_\odot$ yr$^{-1}$) during their evolution.}

Having constrained the mass of stars formed during the ongoing starburst episode, $M_{\rm burst}$, we can further constrain the minimum average SFR of SMGs {\it before} the onset of this starburst, $\equiv(M_*-M_{\rm burst})/ \mbox{age of the Universe}$ (plus signs on Figure~\ref{fig:zevol}). The median is still high, $\sim 100\, M_\odot\mbox{ yr}^{-1}$, so SMGs must have been highly star-forming in the past too. At redshifts $2$--$3$ the age of the Universe is $\sim3$--$2$ Gyr and it is unlikely that a galaxy can sustain this high SFR over such a long period. {\em Therefore we conclude that either the stellar masses of SMGs have been formed in at least two strong} ($>100\, M_\odot\mbox{ yr}^{-1}$) {\em starburst episodes or continuously over the period of $2$-$3$ Gyr but in several smaller galaxies that eventually merged}. In order to build up the stellar mass of one SMG, five such galaxies would need to form stars continuously at a rate of $20\, M_\odot\mbox{ yr}^{-1}$, a value more likely to be sustainable over several Gyr. The latter scenario is consistent with the results of \citet{dye08} based on observed optical to mid-IR data of 51 SMGs with photometric redshifts. They found that approximately half the stellar mass in SMGs has been formed over a long ($\sim1$--$2$ Gyr) period of approximately constant star formation activity. { The possibility that a significant part of stellar mass in SMGs was formed before the ongoing starburst has also been suggested by \citet{hainline08phd}, who compared the build-up timescale of stellar mass  and the duration of the SMG phase.}

The median value of the SFR per unit stellar mass (SSFR$\mbox{}\equiv\mbox{}$SFR$_{\rm IR}/M_*$, Column~{\ssfrcol} of Table~\ref{tab:grasilres}) of $\sim1.8\,\mbox{Gyr}^{-1}$ is within the range for other high-$z$ star-forming samples \citep[compare with Figures~2 and 4 of][respectively]{castroceron06,castroceron09}. This indicates that SMGs are forming stars intensely. 

SSFRs are compared with (the inverse of) the age of the Universe in Figure~\ref{fig:zevol}. The SMGs  close to the solid line could have formed their stellar populations at the present rate within the age of the Universe. However, the SMGs close to, or above the dashed line could have formed their stars at the present rate within less than $10$\% of the age of the Universe, i.e., within $\lesssim300$ Myr at $z=2$. These galaxies are experiencing a powerful starburst episode.

At the extreme there are three high-$z$ SMGs\footnote{SMMJ131201.17+424208.1, SMMJ141802.87+523011.1, SMMJ221806.77+001245.7 plus a low-mass, low-$z$ case, SMMJ030238.62+001106.3} with very high SSFRs$\mbox{}>10\,\mbox{Gyr}^{-1}$ (Column~{\ssfrcol} of Table~\ref{tab:grasilres}). They are all hot ($T_d>60$ K, Column~{\tdcol}) and formed the majority of their stellar populations during the ongoing starburst ($M_{\rm burst}/M_*>60$\%, Column~{\mburstcol}). Therefore they are likely the most powerful cases of SMGs formed in major mergers of galaxies with huge gas reservoirs that were subsequently converted into stars.

Our median SSFR at $z>1.7$ ($1.83\,\mbox{Gyr}^{-1}$) is a factor of $\sim2$ lower than that of \citet[][$3$--$4.5\,\mbox{Gyr}^{-1}$; see their Figure~12b]{dunne09} for $10^{11}<M_*<10^{12}\,M_\odot$ galaxies at these redshifts. 
This difference can
be explained if the radio luminosities \citep[used by][to estimate SFRs]{dunne09} are boosted by AGN activity more than the IR luminosities used here. Indeed, if we use SFR$_{\rm radio}$ instead of SFR$_{\rm IR}$ to calculate SSFRs the median for the SMGs at $z>1.7$ increases to $3.20\,\mbox{Gyr}^{-1}$ (see Section~\ref{sec:agn} for discussion of AGN contamination in our sample).

In order to assess the accuracy of SFR estimates based on radio emission (independent of SED modeling) we compared the ratio of $\mbox{SFR}_{\rm radio} / \mbox{SFR}_{\rm IR}$. Its median value is equal to $\sim1.3$. Hence, assuming that IR emission is a good proxy for SFR, then radio estimates suffer from a $\sim30$\% systematic error. This is illustrated on Figure~\ref{fig:lirrad} where the dashed line denotes the relation between IR and radio luminosities required to make SFR$_{\rm IR}=\mbox{}$SFR$_{\rm radio}$. Indeed the radio luminosity gives systematically higher SFRs for SMGs (most of the points are above the line). 
This can be caused by a significant AGN contamination boosting radio flux (see Section~\ref{sec:agn}), or a strong bias favouring radio-bright galaxies, because those non-detected at radio do not enter our sample (Section~\ref{sec:sample}). Alternatively, it could be that for luminous galaxies either the IR conversion of \citet{kennicutt} should be scaled up by a factor of $1.3$, or the radio conversion of \citet{bell03} scaled down.

\subsubsection{Stellar masses}
\label{sec:mstar}

SMGs having stellar masses of $\sim10^{11}$--$10^{12}\,M_\odot$ (Column~{\mstarcol} of Table~\ref{tab:grasilres} and Figure~\ref{fig:zevol}) are among the most massive galaxies in the Universe, regardless of redshift \citep[compare with Figures~2 and 4 of][respectively]{castroceron06,castroceron09}. This property makes them natural candidates for the progenitors of the present-day ellipticals.

The relatively tight range of stellar masses is likely not a result of sensitivity limits at optical and near-IR. This is because {\it i}) galaxies with stellar mass as low as $\sim10^9\,M_\odot$ would have been detected in deep {\it Spitzer} imaging at redshifts $z\sim2$ \citep[e.g.][]{reddy06} {\it ii}) our sample accounts for 50\% { of the parent \citet{chapman05} sample} (and only 30\% of the parent sample may have different properties than our sample, see Section~\ref{sec:sample}), so it is unlikely that we miss only the low-mass objects. 
Therefore, high $M_*$ seems to be an intrinsic property of {\submm}-selected galaxies. Mergers of less massive galaxies could not result in a powerful starburst giving rise to detectable {\submm} emission \citep[see also][]{dave09}.

{\em Only a minor part} (median $\sim8$\%, Column~{\mburstcol} of Table~\ref{tab:grasilres} and Figure~\ref{fig:zevol}) {\em of the stellar populations present in SMGs has been formed during the ongoing starburst episodes}. Hence, even though SMGs probably evolve into ellipticals, the majority of the stellar mass in such ellipticals had been created before the {\submm}-bright phase. 

This could mean that the current SFRs and stellar masses of SMGs are only loosely connected and indeed this manifests itself in a very high spread (around two orders of magnitude) in SSFRs in our sample even though the stellar mass range is relatively tight: $\sim10^{11}$--$10^{12}\,M_\odot$ (Figure~\ref{fig:zevol}). This behaviour is unusual  compared to other galaxies \citep[see][]{castroceron06,castroceron09}.

However we note that the low stellar masses created in the ongoing starburst may partially be an effect of the assumed starburst ages of $50$ Myr. If a starburst duration of $100$--$200$ Myr were adopted { \citep{smail04,borys05,hainline08phd,tacconi08}} the resulting $M_{\rm burst}$ could be higher by a factor of $\sim2$--$4$.

The mass-to-light ratios, $M_*/L_K$, of SMGs (Column~{\mlkcol} of Table~\ref{tab:grasilres} and Figure~\ref{fig:zevol}) are typical for massive galaxies. Specifically, the median ($0.68\,M_\odot\,L_\odot^{-1}$) is similar to the values  for $M_*>10^{11}\,M_\odot$ galaxies  \citep[][their Table~1]{drory04}  and to simulated massive galaxies at $z\sim1$ \citep[][their Figure~4]{courty07}.

\subsubsection{Dust properties}
\label{sec:dust}

{
Our fits suggest that SMGs are moderately dust-obscured with a median $A_V\sim2$ mag (Column~{\avcol} of Table~\ref{tab:grasilres}). Our estimates are consistent within $1$--$2\sigma$ with the mean/median values obtained by \citet[][$1.70$--$2.44$]{smail04}, \citet[][$3.0 \pm 1.0$]{swinbank04}, \citet[][$1.7 \pm 0.2$]{borys05} and \citet[][$1.7\pm0.1$]{hainline08phd} based on near-IR data. For individual SMGs we obtained systematically larger extinction (median difference of $\sim0.3$ mag) than \citet{hainline08phd}. The difference may be accounted for  if there is significant extinction even in {\it Spitzer} IRAC data.
}

The dust density of SMGs at $z<0.5$ (Column~{\mdustdcol} of Table~\ref{tab:zbins}) is { approximately $3$\% of the total local ($0.013<z<0.18$) dust budget of $\log\rho_{\rm dust}=5.57_{-0.17}^{+0.12}\, M_\odot\,\mbox{Mpc}^{-3}$ given by \citet{driver07} based on an assumed dust-to-light ratio.
Therefore SMGs contribute  very little to the dust budget at low redshifts. 
}

 In our sample of SMGs $\rho_{\rm dust}$  does not change significantly from $z\sim3.6$ to $z\sim0.5$. 
We do not detect any evolution of  dust mass in SMGs across the entire redshift range (Figure~\ref{fig:zevol}).
{  A constant dust mass density across redshifts $0$--$3.5$ was also found by \citet{pascale09} based on a stacking analysis at {\submm} wavelengths of galaxies selected at $24\,\mu$m. }
 
The question is what happened to the dust produced in SMGs. If they evolve into dust-poor ellipticals, then the dust  is not simply stored in their end-products (as is probably the case for stellar masses). It is therefore plausible that dust is either blown away (by stellar and/or AGN winds) { or absorbed in star formation,} or destroyed during subsequent evolution after the SMG event.

\begin{figure*}
\begin{center}
\includegraphics[width=0.85\textwidth]{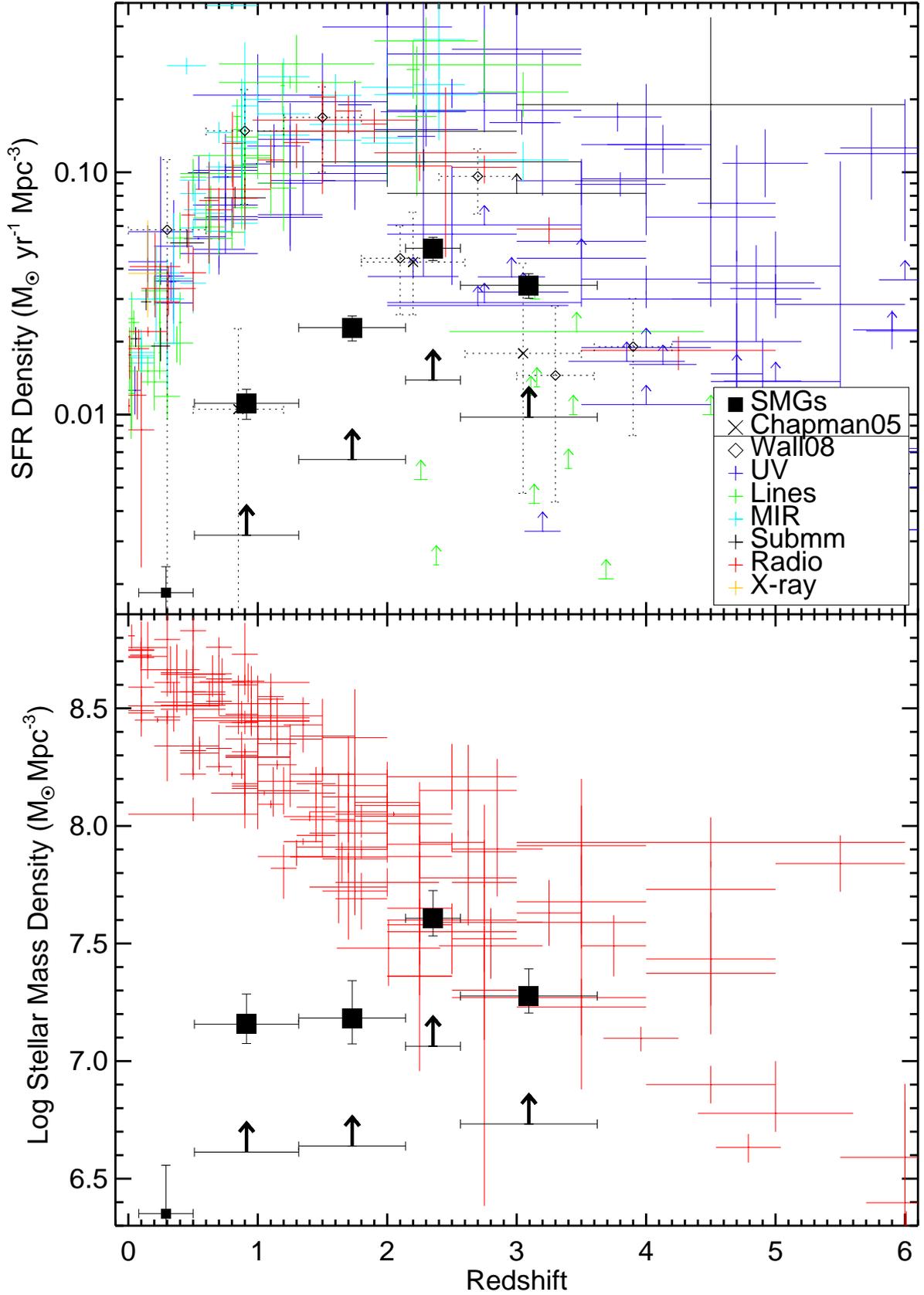}
\end{center}
\caption{{\it Top}: Cosmic star formation density. The SMGs' contribution rises with redshift from $ \sim9$\% to { $  \sim20$\%} (Section~\ref{sec:sfrd} and Table~\ref{tab:zbins}).
{\it Filled Squares}: data for SMGs at $z>0.5$  in four bins (Table~\ref{tab:zbins} and Section~\ref{sec:bins}). { {\it Small Squares}: data for SMGs at $z<0.5$}. { {\it Thick black arrows}: the SMG data without incompleteness correction (factor of $3.5$, Section~\ref{sec:bins}).} { {\it Black crosses} and {\it diamonds}: star formation density of SMGs determined by \citet{chapman05} and \citet{wall08}, respectively.} {\it Colored points with error bars}: determination of the cosmic value based on different estimates -- ultraviolet ({\it violet}), emission lines: [\ion{O}{2}], [\ion{O}{3}], H$\alpha$, H$\beta$ ({\it green}), mid-IR ({\it light blue}), {\submm} ({\it black}), radio ({\it red}), X-ray ({\it yellow}). Extinction correction and, in many cases, incompleteness correction have been applied by the authors. {\it Arrows}: lower limits. \newline
{\it Bottom:} Cosmic stellar mass density. The SMGs' contribution rises with redshift from $\sim 5$\% to { $ \sim 50$\%} (Section~\ref{sec:mstard} and Table~\ref{tab:zbins}). 
 {\it Red points with error bars}: determination  of the cosmic value from literature.
The data and the references are listed in Tables~\ref{tab:sfrd} and \ref{tab:mstard}  in appendix.
}
\label{fig:sfrdz}
\label{fig:mstardz}
\end{figure*}

\subsubsection{Comparison with GRB hosts 
}

In \citet{michalowski08} we presented a hypothesis that gamma-ray burst (GRB) host galaxies may  constitute a subsample of hotter/less luminous counterparts of SMGs.
Indeed, the UV-to-IR SEDs of three $z\sim2$--$3$ SMGs\footnote{%
\object{SMMJ141750.50+523101.0}, 
\object{SMMJ141802.87+523011.1}, 
\object{SMMJ163627.94+405811.2}%
}
are consistent with $z\sim1$ {\submm}/radio bright GRB hosts \citep[dashed lines on Figure~\ref{fig:sed} from][]{michalowski08}, but $1.2$--$3.9$ times more luminous. These three SMGs are similar to GRB hosts with respect to their hot dust temperatures ($\sim40$--$60$ K), high SSFRs ($\gtrsim2$ Gyr$^{-1}$, high fraction of stellar mass formed in the ongoing starburst ($>10$\%) and blue optical colors.

If larger samples of GRB hosts shows a similar tendency that their brightest members overlap with the hotter subsample of SMGs, then GRB events will provide an effective way of selecting hot SMGs, otherwise difficult to localize.

\subsection{Contribution to stellar mass assembly}

\subsubsection{Star formation rate volume density}
\label{sec:sfrd}

SFR densities of SMGs were calculated as described in Section~\ref{sec:bins}.
{
In order to assess the accuracy of our simplified method of dividing the sum of the SFRs of the detected SMGs by the total survey volume, we compare our estimates with those resulting from detailed calculation of the volume contribution of individual SMGs done by  \citet[][based on the same sample as we analyse]{chapman05}   and \citet[][based on 35 SMGs in GOODS-N field of which 17 have spectroscopic redshifts]{wall08}. The comparison is shown in Figure~\ref{fig:sfrdz}. Our results in two high-redshift bins ($z>2$) corrected for incompleteness (Section~\ref{sec:bins}) are consistent with that of  \citet{chapman05} and \citet{wall08}. At lower redshifts we find values similar  to \citet{chapman05}, but an order of magnitude lower than \citet{wall08}. Therefore we conclude that {\it i}) our method to calculate volumes is accurate, since it gives consistent results with other estimates; and {\it ii}) our sample is incomplete in the three low-redshift bins as anticipated in Section~\ref{sec:sample}.
}

From Figure~\ref{fig:sfrdz} (and Columns~{\sfrdcol} and {\sfrdcontrcol} of Table~\ref{tab:zbins}) it is apparent that a $\rho_{\rm SFR}$ of SMGs starts to decline (with cosmic time) earlier ({ about} $z\sim2$) than that of other galaxies ($z\sim1$). 
More quantitatively, SMGs harbour { $ \sim20$\%} of the cosmic $\rho_{\rm SFR}$ at $z\sim2$--$3.6$ (Column~{\sfrdcontrcol}), but their contribution drops to $ \sim9$\% at $0.5<z<1.4$. It is likely 
that at lower redshifts, due to the decreased rate of mergers \citep[e.g.][]{rawat08,deravel09}, there are fewer galaxies left that can still sustain high SFRs to be detected at {\submm} wavelengths. However, part of the decrease of SMG $\rho_{\rm SFR}$  can be explained by the ``redshift desert'', which makes it difficult to { measure redshifts of}  $z\sim1.2$--$1.8$ SMGs (see Section~\ref{sec:sample}).

{\em A high value of $\rho_{\rm SFR}$ of SMGs at $z\sim2$--$3$ and the subsequent decline are consistent with the hypothesis that the SMG population is a manifestation of powerful starburst episodes  evolving into the present-day ellipticals} (as discussed in Section~\ref{sec:mstar}). In this scenario galaxies detected in the {\submm} at high-$z$ do not enter the sample of SMGs at low-$z$ because they have already evolved into passive galaxies.  It has indeed been  found that ellipticals contain old stars formed at $z\sim1.5$--$4$ \citep{daddi00,dokkum01,ven03}.
The evolution of SMGs into ellipticals has also been claimed by several authors based on their luminosity function \citep{smail04}, huge luminosities \citep{eales99} and gas reservoirs \citep{smail02,greve05}, strong clustering \citep{ivison00,almaini03}, space density and morphology \citep{barger99,lilly99,trentham99,swinbank06} and evolutionary SED models \citep{takagi04}.

\citet{knudsen08} analysed number counts of SMGs fainter than the SCUBA confusion limit, using those behind clusters of galaxies magnified by lensing. They concluded that the integrated light produced by the SMGs brighter than $0.1$ mJy 
(i.e.~LIRGs and ULIRGs with roughly $L_{\rm IR}>8\times10^{10}\,L_\odot$ and SFR$\mbox{}>15\,M_\odot$ yr$^{-1}$) 
is comparable to the extragalactic background light (EBL) at $850\,\mu$m \citep[see also][]{blain99,cowie02}. This means that these galaxies host the majority of the cosmic {\em obscured} star formation. \citet{knudsen08} also found that sources brighter than $2.5$ mJy (roughly the limit of the survey considered here) contribute $\sim25$\% to the to EBL at $850\,\mu$m \citep[see also][]{hughes98,barger99,wang04,coppin06}. Together with our results this implies that as much as { $ \sim80$\% ($ 4\times20$\%)} of the cosmic star formation at $z\sim2$--$3.6$ reside in SMGs brighter than $0.1$ mJy. This is only true if the faint ($<2$ mJy) SMGs have similar dust temperatures to the brighter ones. If they are colder (hotter) their {\submm} fluxes corresponds to lower (higher) SFRs (because it is calibrated to total IR emission) and therefore the total SMG population contribute less (more) than { $ 80$\%} to the cosmic $\rho_{\rm SFR}$. {  This picture is however complicated, because based on stacking analysis it has been claimed that the distribution of the faint SMGs  peaks at lower redshifts \citep[$z<1.5$;][]{wang06,serjeant08}.
}

{\em Our overall conclusion is that the SMG population plays a significant role at redshifts $z\sim2$--$4$, namely sources brighter than} $\sim3$ ($0.1$) mJy at $850\,\mu$m {\em host} { $ 20$\% ($ 80$\%)} {\em of cosmic star formation}. Their contribution  can however be lower in reality if very small (but numerous) galaxies are missed in all high-$z$ flux-limited galaxy surveys. In such a case the total SFR density (color points on Figure~\ref{fig:sfrdz}) would be underestimated. To solve this issue much deeper surveys at high-$z$ are necessary, either blank-field or for well-selected dwarf galaxy samples (e.g., GRB hosts or Ly$\alpha$ emitters).

\citet{zheng07} estimated $\rho_{\rm SFR}$ at $z\sim0.9$ for massive galaxies ($M_*>10^{11}\,M_\odot$) down to $R<24$ mag (only $\sim40$\% of SMGs satisfy the latter criterion) equal to $0.0052_{-0.0021}^{+0.0020}\,M_\odot\mbox{ yr}^{-1}\mbox{ Mpc}^{-3}$. This value is only a factor of $2$ lower than our estimate for the SMGs at $0.5<z<1.4$ (Table~\ref{tab:zbins}). Therefore, although SMGs do not host a major fraction of the cosmic SFR at these redshifts, they contribute significantly ($ 0.0102/(0.0052\times0.6+0.0102)\sim66$\%) to the SFR budget of massive galaxies.

\subsubsection{Stellar mass volume density}
\label{sec:mstard}

Stellar mass densities of SMGs were calculated as described in Section~\ref{sec:bins}.
Figure~\ref{fig:mstardz} and Table~\ref{tab:zbins} (Columns~{\mstardcol} and {\mstardcontrcol}) show that {\em at $z\sim2$--$3.6$ a significant part { ($ \sim30$--$ 50$\%)} of the cosmic stellar mass had been formed in the progenitors of SMGs}. At lower redshifts $\rho_{*}$ of SMGs (and hence their contribution to the cosmic $\rho_{*}$) drops, likely because the majority of SMGs at higher redshifts had already evolved into passive galaxies at $z\sim1.5$, and so dropped out of our {\submm}-selected sample. Moreover the sample is incomplete at $z\sim1.2$--$1.8$ due to the ``redshift desert'' (see Section~\ref{sec:sample}). This brings down the densities of SMGs in the low-$z$ bins.

Since most of the stellar mass of SMGs has not been formed in the ongoing starburst (Section~\ref{sec:mstar}), their $\rho_*$ reflects the integrated contribution of SMGs to the cosmic $\rho_{\rm SFR}$. Therefore the relatively high contribution of SMGs  to the cosmic $\rho_*$ in the last redshift bin ($ \sim31$\%, Column~{\mstardcontrcol} of Table~\ref{tab:zbins}) means that SMGs play a non-negligible role in the cosmic stellar assembly even at $z>3.6$. This can be tested by analysis of a sample of $z\gtrsim4$ SMGs in a defined survey sky area \citep[e.g.][{ note that these results are likely affected by cosmic variance}]{michalowski10smg4,younger09b}.  It has been confirmed that such distant SMGs exist \citep{capak08,knudsen08b,knudsen09,schinnerer08,coppin09,daddi09,daddi09b}. 

\subsection{Source of emission}

\subsubsection{IR-radio correlation}
\label{sec:firr}

\begin{figure*}
\begin{center}
\plotone{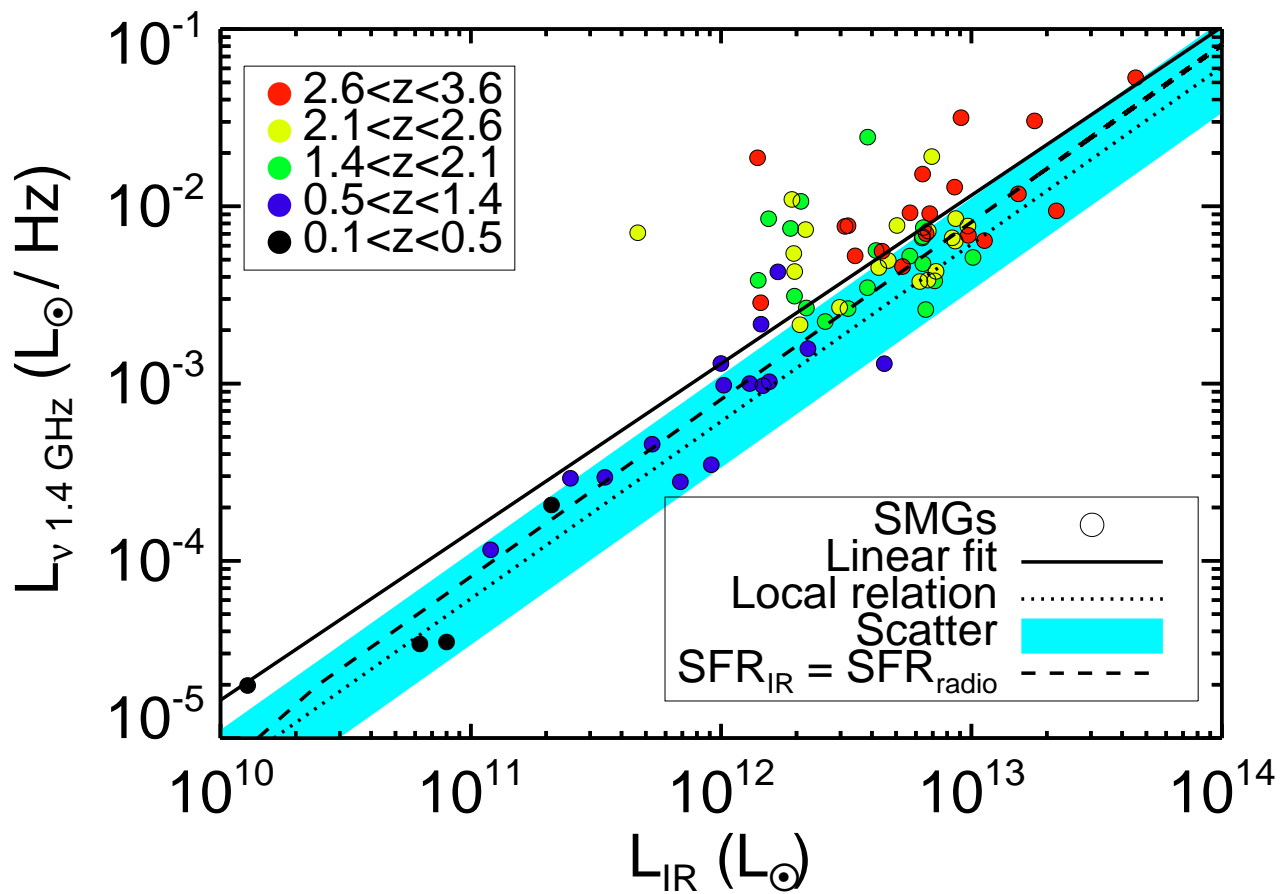}
\end{center}
\caption{Radio luminosity density as a function of infrared ($8$--$1000\,\mu$m) luminosity of SMGs showing 
{ a linear relation, though with a normalization offset from the local relation by a factor of $\sim2.1$ towards higher radio luminosities} (Section~\ref{sec:firr}). {\it Circles}: values for individual SMGs color-coded by redshift. {\it Solid line}: linear fit to the data (eq.~\ref{eq:lirrad}). {\it Dotted line}: the mean local relation \citep{bell03}. {\it Shaded area}: its scatter. {\it Dashed line}: the track where SFR$_{\rm IR}$ \citep{kennicutt} is equal to SFR$_{\rm radio}$ \citep{bell03}. The strong outliers (above the line) at high-luminosity end are probably caused by AGN activity increasing radio luminosities.
}
\label{fig:lirrad}
\end{figure*}

\begin{figure*}
\begin{center}
\plotone{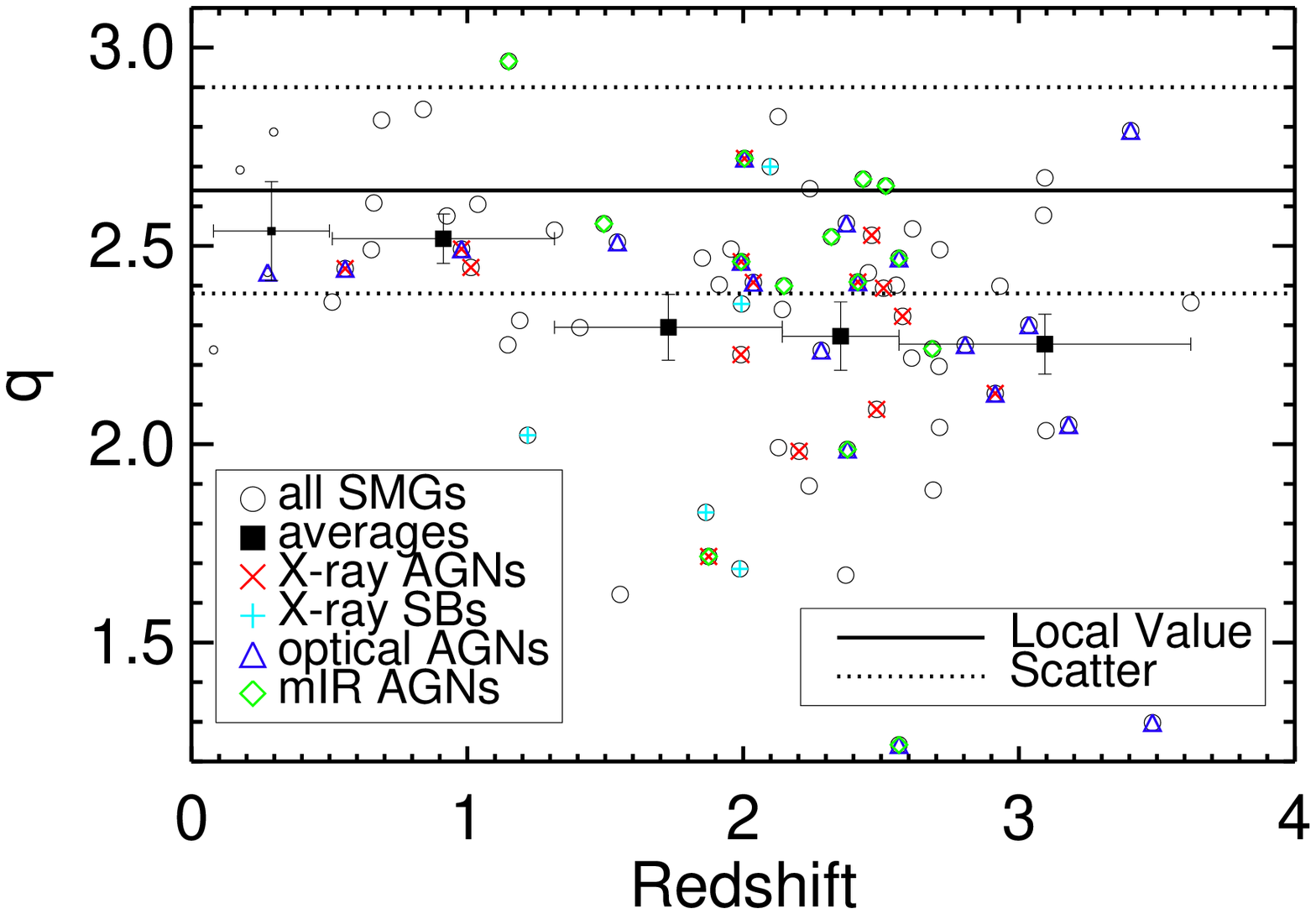}
\end{center}
\caption{The ratio of the infrared ($8$--$1000\,\mu$m) and radio luminosities $q$ (defined in Section~\ref{sec:sed}) as a function of redshift of SMGs. It provides evidence that 
{ a linear IR-radio correlation holds for SMGs up to $z\sim3.6$, though with a normalization offset from the local relation by a factor of $\sim2.1$ ($\Delta q\sim-0.32$) towards higher radio luminosities}
 (Section~\ref{sec:firr}). {\it Circles}: values for individual SMGs. {\it Squares}: the mean values (and errors on the mean) in five redshift bins containing equal number of galaxies (Table~\ref{tab:zbins} and Section~\ref{sec:bins}). { {\it Small symbols} indicate $z<0.5$ objects.} {\it Red crosses}: SMGs classified as AGNs based on X-ray emission \citep{alexander05}. {\it Light blue plus signs}: SMGs classified as starbursts based on X-ray emission \citep{alexander05}. {\it Violet triangles}: SMGs classified as AGNs based on optical spectra \citep{chapman05}. {\it Green diamonds}: SMGs classified as AGNs based on a mid-IR power-law (Section~\ref{sec:agn}).  The mean local $q=2.64$ \citep{bell03} is shown as a {\it solid line} with $0.26$ scatter  ({\it dotted lines}). The $q$ values for majority of AGN-classified SMGs 
 { do not differ from the rest of the SMG population}
 (see Section~\ref{sec:agn}). 
}
\label{fig:qz}
\end{figure*}

{ With our full SED modelling of 76 SMGs we confirm the results of \citet{hainline08phd} on the correlation between IR and radio luminosities.} Figure~\ref{fig:lirrad} shows that SMGs follow 
a linear IR-radio correlation. 
The two outliers (with $q\sim1.3$, see Section~\ref{sec:agn}) are probably caused by AGN activity contributing significantly to radio luminosities. 
A linear fit gives:
\begin{equation}
\begin{array}{rcl}
\log (L_{\nu\, 1.4\, {\rm GHz}} / L_\odot \mbox{Hz}^{-1}) &=& (0.95\pm0.07) \log (L_{\rm IR} / L_\odot) \\
							&&\mbox{}- (14.3\pm0.8).\\
\end{array}
\label{eq:lirrad}
\end{equation}
The slope is consistent (within errors) with unity, suggestive of the linear relation between $L_{\nu\, 1.4\, {\rm GHz}}$ and $L_{\rm IR}$ at the high-end ($L_{\rm IR}\gtrsim10^{11}\,L_\odot$) of the galaxy luminosity function {  \citep[a similar value of $1.064 \pm 0.025$ was found by][]{hainline08phd}. }

The IR-radio correlation is usually quantified by the ratio of IR and radio luminosities, $q$ (see Section~\ref{sec:sed}). The mean $q$ for SMGs ($2.32\pm0.04$, scatter: $0.34$) is 
{ significantly lower than that of local star-forming galaxies \citep[$2.64$ with a scatter of $0.26$;][]{bell03}. Similar offsets were reported by \citet{kovacs06}, \citet{murphy09} and \citet{murphy09b} based on smaller samples of SMGs. {\em We conclude that at $z>1.4$ SMGs have radio luminosities on average a factor of $\sim2.1$ larger  ($\Delta q\sim-0.32$) than what would result from the local relation.} The difference is significant at the level of $4$--$5\sigma$ and can be explained in three ways.

Radio-loud AGNs have on average low $q$ values \citep[see e.g.][]{miller01,yun01,yang07}.
If $\gtrsim50$\% of the radio emission of SMGs is powered by AGNs, then the radio luminosities of SMGs higher by a factor of $\sim2.1$ can be accounted for. However, there are indications that SMGs are starburst-dominated (see Section~\ref{sec:agn}), so we deem this explanation less likely.

Another explanation is that the radio excess is a result of the bias against radio-faint sources in our sample (see Section~\ref{sec:sample}). This can be tested when a sample of SMGs with localizations (and hence redshifts) independent of radio detections is available \citep[e.g.][]{daddi09,daddi09b,knudsen09,weiss09}.

The third possibility is that some properties influencing the IR or radio emission are intrinsically different for SMGs and local galaxies. The sample of \citet{bell03} includes local normal, star-forming spiral and irregular  galaxies, blue compact dwarfs, starburst galaxies and ULIRGs.
Therefore the difference in the properties between this sample and such extreme galaxies as SMGs is expected.
 Such explanation was offered by \citet{lacki09} and \citet{lacki09b}. Their numerical modelling showed that cosmic-ray electrons in ``puffy starbursts'' (vertically and radially extended galaxies with vertical scale heights $\sim1$ kpc) experience weaker 
bremsstrahlung and ionization
 losses 
 resulting in stronger radio emission. Indeed, there are indications that SMGs are extended on vertical scales of $\sim1$ kpc \citep{lacki09b,tacconi06,tacconi08,genzel08, younger08b,law09}, so we find this explanation probable.
}

{ The systematic uncertainties in the determination of $L_{\rm IR}$ (factor of $\lesssim2$, Section~\ref{sec:results}) may in principle also explain the offset. However, we find this unlikely because similar offsets were found by other authors using different fitting methods \citep{kovacs06,murphy09,murphy09b}. 
}

The $q$ values for SMGs are shown in Figure~\ref{fig:qz} as a function of redshift. We do not detect any significant evolution across the redshift range $1.4$--$3.6$. 
{ 
The only sign of evolution is that the mean $q$ in the low-redshift bin ($0.5<z<1.4$) is above the value found at higher redshifts ($\sim4\sigma$). 
This 
can be explained either by the contribution of reprocessed emission from low-mass stars \citep[cirrus emission, e.g.][and references therein]{yun01} to the IR, or by the fact that at low redshifts SMGs are more similar to other local galaxies and do not exhibit large vertical scale heights characteristic for ``puffy starbursts'' (see above). 
}

It is important to note that the derived linear IR-radio correlation for SMGs is not a consequence of the use of the SED templates (which were tuned to fulfill this correlation locally), because the radio luminosities used here were derived based on the observational data only, independent of the SED modeling.

\subsubsection{AGN activity}
\label{sec:agn}

As discussed in Section~\ref{sec:firr}, 
{
AGN activity could explain low $q$ values of SMGs.
This is at least true for
}
the two SMGs with lowest $q$\footnote{\object{SMMJ131215.27+423900.9}, \object{SMMJ141813.54+522923.4}}, spectroscopically classified as AGN \citep{chapman05}.

{ In the SEDs of SMGs} there are clear signs that some of them host AGNs (though, not necessarily a bolometrically dominant ones). Radio datapoints are higher than model predictions by more than $3\sigma$ in $36$\% (27/76) of SMGs, whereas they are lower than models only for  $8$\% (6/76). This may hint at an AGN contribution in these galaxies. However, 4 out of 5 X-ray identified starbursts (Column~{\agncol} of Table~\ref{tab:grasilres}) also exhibit radio excess, { so we find other explanations of radio excess presented in Section~\ref{sec:firr} more reliable}.

Another indication of an AGN contribution is that $18$\% (14/76) of SMGs show a mid-IR power-law AGN feature incompatible with our starburst models (see Figure~\ref{fig:sed} and Column~{\agncol} of Table~\ref{tab:grasilres}). However, rest-frame $2$--$5\,\mu$m excess was also interpreted as a tracer of  recent star formation \citep{mentuch09}.

Finally, three SMGs\footnote{SMMJ123716.01+620323.3, SMMJ131215.27+423900.9, SMMJ131222.35+423814.1} have exceptionally high SFR$_{\rm UV}$ ($>500\,M_\odot\mbox{ yr}^{-1}$, Column~{\sfruvcol} of Table~\ref{tab:grasilres}). Strikingly, all of them were fitted with non-starburst models ($M_{\rm burst}=0$, Column~{\mburstcol}), so modeling is consistent with these high SFRs being continuous (the same is true for three other non-starburst SMGs with high SFR$_{\rm IR}$). Such a scenario is unlikely, so this hints at an AGN contribution to the UV/IR emission.

{ However, the fact that we obtained reasonable SED fits for most of the SMGs using purely star-forming models (Figure~\ref{fig:sed}) hints at the conclusion that AGN activity is not dominant in our sample.}

We investigated the issue of AGN activity further by analysing the average $q$ values of the following subsamples (see also Figure~\ref{fig:qz}): X-ray identified \citep{alexander05} AGNs: $2.32\pm0.06$ and starbursts:  $2.12\pm0.18$; optically identified AGNs \citep{chapman05}: $2.27\pm0.09$; and mid-IR identified AGNs (see above): $2.36\pm0.12$. All subsamples are consistent with 
{ the value derived for the entire sample ($2.32$)}
{ Hence, we confirm the finding of \citet{hainline08phd} that even the AGN-classified SMGs follow a linear IR-radio correlation. }  This means that even if an AGN is present it does not contribute to the emission of an SMG significantly (with the exception of the two $q\sim1.3$ sources). 

This is in line with {\it i}) the X-ray studies of SMGs indicating that the contribution of AGN activity to their IR emission is only $\sim8$\% on average  \citep{alexander05}; {\it ii}) { { mid-IR colors of SMGs
indicating that AGNs dominate the emission at these wavelengths only in $13$--$19$\% cases \citep{hainline09};}
{\it iii}) mid-IR spectroscopy of SMGs revealing only weak AGN-like continua \citep{valiante07,pope08,menendezdelmestre07,menendezdelmestre09,murphy09,watabe09}; {\it iv}) near-IR spectroscopy revealing that starbursts dominate the emission of SMGs \citep{swinbank04}.  } Moreover, \citet{devries07} found that star formation processes (if present) account for at least $75$\% of the radio luminosities of optically-selected AGNs.

{\em Therefore we conclude that AGNs are present in a significant fraction of SMGs, but their contribution to the IR emission is at most minor.}

\subsection{Comparison of our results with the literature}

For the sample of SMGs discussed in this paper there are previous estimates of some of their properties. In this section we compare them with our results.

\citet{chapman05} derived $L_{\rm IR}$ and $T_d$ based only on the $850\,\mu$m and $1.4$ GHz data. There is no systematic difference between the determinations of $T_d$ (our median of $38.7$ K, theirs: $38.3$ K). The mean difference between individual datapoints is $4$ K ($\sim10$\%).
However, our values for $L_{\rm IR}$ are systematically lower than theirs (the median ratio of individual datapoints is
$1.7$). We find our values more reliable since they are based on data spanning a wider wavelength range. Overestimation of $L_{\rm IR}$ when using only  $850\,\mu$m and $1.4$ GHz was also noticed by \citet{kovacs06} and \citet{pope06}.

\citet{kovacs06} investigated a subsample observed at $350\,\mu$m. Their median dust mass ($9.04\,\log M_\odot$) and $q$ value ($2.20$) are consistent with our estimates ($9.01$ and $2.35$, respectively).  The median difference between individual datapoints  is $\sim30$\% for dust masses and $\sim13$\% for $q$. 

The median stellar mass for a subsample of 13 SMGs investigated by \citet[][$11.51\,\log M_\odot$]{borys05} is close to our value ($11.70$). However,  estimates of \citet[][median $10.82\,\log M_\odot$]{hainline08phd} for  64 SMGs are a factor of $\sim5.6$ smaller than our values ($11.57$). \citet{hainline08phd} postulated that the discrepancy between her results and those of \citet{borys05} arose from a combination of systematic differences between the applied SED models and a higher AGN contribution in the $K$-band \citep[used by][]{borys05} with respect to the $H$-band. Our estimates are based on all the available photometric data, and so we find the former explanation more likely. { In particular, the differences in the applied stellar population models and their ages may explain the discrepancy.}
 
\section{Conclusions}
\label{sec:conclusion}

We have investigated the UV-to-radio SEDs of 76 SMGs ($S_{850}\gtrsim3$ mJy)  with spectroscopic redshifts ($0.080$--$3.623$). For the first time the properties of such a significant sample has been derived consistently using all available data. The resulting SFRs (median $713\,M_\odot\mbox{ yr}^{-1}$) and stellar masses ($11.57\,\log M_\odot$) are among the highest in the Universe.

Such high stellar masses, already present at redshifts $\sim2$--$3$, require that SMGs experienced either at least two starburst episodes, or a merger of several smaller galaxies. Our modeling suggests that only a minor fraction ($8$\%) of their stellar populations was formed during the ongoing starburst episodes. This is supported by the fact that the SFRs and $M_*$ of SMGs are basically disconnected, i.e.~we observe two orders of magnitude spread in SSFRs whereas the range of $M_*$ is relatively narrow: $10^{11}$--$10^{12}\,M_\odot$. We concluded that dust is  blown away or destroyed during the evolution of SMGs, since it is not stored in the likely end-products of SMGs, elliptical galaxies.

Indeed, the high stellar masses and the evolution of the SFR and stellar mass densities of SMGs are consistent with a scenario in which SMGs are progenitors of present-day ellipticals.

We found that SMGs contribute significantly  to the cosmic SFR, $\rho_{\rm SFR}$ { ($ \sim20$\%)} and stellar mass, $\rho_*$ { ($ 30$--$ 50$\%)} densities at $z\sim2$--$4$. If we consider {\submm} sources down to $0.1$ mJy the contribution to $\rho_{\rm SFR}$ rises to { $ \sim80$\%}. 

Our analysis suggests that a linear IR-radio correlation holds for SMGs at least up to a redshift of  $3.6$, { but they are $\sim2.1$ times brighter at radio wavelengths than what would result from the local correlation}. 

\begin{acknowledgements}

We  thank Joanna Baradziej, Jos\'{e} Mar\'{i}a Castro Cer\'{o}n, Thomas Greve, Brian Lacki,  
Kim Nilsson, Jesper Sommer-Larsen, Sune Toft,
B\"arbel Tress and Gunther Tress for discussion and comments; our referee for help with improving this paper; Scott Chapman for information on the survey areas; Jorge Iglesias-P\'{a}ramo for kindly providing his SED templates; and Fabio Fontanot for kindly providing his model and data on star formation density. 
{ We acknowledge use of the extensive {\it Spitzer} database in the PhD thesis of Laura Hainline.}

M.~J.~M.~would like to acknowledge support from The Faculty of Science, University of Copenhagen. The Dark Cosmology Centre is funded by the Danish National Research Foundation. This research has made use of NASA's Astrophysics Data System Bibliographic Services.

\end{acknowledgements}



\appendix
\section{Long tables and figures}

\begin{table*}
\caption{Photometry detections of SMGs \label{tab:phot}	}
\centering
\begin{tabular}{lrrrrl}
\hline\hline
 & & $\lambda_{\rm obs}$ & flux & error & \\
		SMG & $z$ & ($\mu$m) & ($\mu$Jy) & ($\mu$Jy) & Reference \\
\hline
SMMJ030226.17+000624.5 & 0.080 & 0.365 & 33.419 & 0.306 & \citet{clements04} \\ 
SMMJ030226.17+000624.5 & 0.080 & 0.428 & 83.176 & 7.319 & \citet{chapman05} \\ 
SMMJ030226.17+000624.5 & 0.080 & 0.440 & 83.946 & 0.232 & \citet{clements04} \\ 
SMMJ030226.17+000624.5 & 0.080 & 0.550 & 181.970 & 0.335
 & \citet{clements04} \\ 
SMMJ030226.17+000624.5 & 0.080 & 0.656 & 275.423 & 24.234
 & \citet{chapman05} \\ 
\hline
\tablecomments{The second reference in the first entry for each SMG indicates that there exists a mid-IR Spitzer/IRS spectrum of this object \citep{valiante07,pope08,menendezdelmestre07,menendezdelmestre09}. This table is available in its entirety in a machine-readable form in the electronic edition of the Journal. A portion is shown here for guidance regarding its form and content.}
\tablerefs{\citet{ivison02,ivison05,chapman03b,chapman05,capak04,clements04,egami04,greve04,smail04,fomalont06,kovacs06,laurent06,tacconi06,pope06,huynh07,hainline08phd}}
\end{tabular}
\end{table*}

\begin{table*}
\caption{Photometry upper limits of SMGs \label{tab:lim}	}
\centering
\begin{tabular}{lrrrrl}
\hline\hline
 & & $\lambda_{\rm obs}$ & flux & error & \\
		SMG & $z$ & ($\mu$m) & ($\mu$Jy) & ($\mu$Jy) & Reference \\
\hline
SMMJ030227.73+000653.5 & 1.408 & 70.000 & 13600.000 & 0.000
 & \citet{hainline08phd} \\ 
SMMJ030231.81+001031.3 & 1.316 & 0.365 & 0.059 & 0.000 & \citet{clements04} \\ 
SMMJ030231.81+001031.3 & 1.316 & 0.428 & 0.100 & 0.000 & \citet{chapman05} \\ 
SMMJ030231.81+001031.3 & 1.316 & 0.440 & 0.102 & 0.000 & \citet{clements04} \\ 
SMMJ030231.81+001031.3 & 1.316 & 0.656 & 0.331 & 0.000 & \citet{chapman05} \\ 
\hline
\tablecomments{When the error is equal to zero, the flux column denotes $3\sigma$ upper limit. Otherwise --- formal flux at the position of an SMG. This table is available in its entirety in a machine-readable form in the electronic edition of the Journal. A portion is shown here for guidance regarding its form and content.}
\tablerefs{\citet{ivison02,ivison05,chapman03b,chapman05,capak04,clements04,egami04,greve04,smail04,fomalont06,kovacs06,laurent06,tacconi06,pope06,huynh07,hainline08phd}}
\end{tabular}
\end{table*}

\longtabL{3}{
\begin{landscape}
\begin{scriptsize}
\begin{longtable}{cc rrrrr rrcc crc c l}
\caption{Properties of SMGs derived from the SED modeling \label{tab:grasilres}}
\\
\hline\hline
 &  & \multicolumn{4}{c}{SFR ($M_\odot$ yr$^{-1}$)} & SSFR & $\log M_*$ & $M_{\rm burst}/M_*$ & $M_*/L_K$ & $\log M_{d}$ & $\log L_{\rm IR}$ & $T_d$ & $A_V$ & & \\
\cline{3-6}
SMG & $z$ & SED & UV & IR & radio & (Gyr$^{-1}$)  & ($M_\odot$) & (\%) & ($M_\odot/L_\odot$) & ($M_\odot$) & ($L_\odot$) & (K) & (mag) & $q$ & AGN?	\\
(1)  & (2)              &  (3) & (4) &  (5) & (6) & (7) & (8) & (9) & (10) & (11) & (12) & (13) & (14) & (15) & (16)\\
\hline
\endfirsthead
\caption{continued.}\\
\hline\hline
 &  & \multicolumn{4}{c}{SFR ($M_\odot$ yr$^{-1}$)} & SSFR & $\log M_*$ & $M_{\rm burst}/M_*$ & $M_*/L_K$ & $\log M_{d}$ & $\log L_{\rm IR}$ & $T_d$ & $A_V$ & & \\
\cline{3-6}
SMG & $z$ & SED & UV & IR & radio & (Gyr$^{-1}$)  & ($M_\odot$) & (\%) & ($M_\odot/L_\odot$) & ($M_\odot$) & ($L_\odot$) & (K) & (mag) & $q$ & AGN?	\\
(1)  & (2)              &  (3) & (4) &  (5) & (6) & (7) & (8) & (9) & (10) & (11) & (12) & (13) & (14) & (15) & (16)\\
\hline
\endhead
\hline
\endfoot
\object{SMMJ030227.73+000653.5} & 1.408 &          825 & 7.91 &          713 & 
        1190 & 4.81 & 11.17 & 25.5 & 0.42 & 8.54 & 12.62 & 52.6 & 1.81 & 2.29 & 
rad\\
\object{SMMJ030231.81+001031.3} & 1.316 &          433 & 0.68 &          224 & 
         211 & 7.81 & 10.46 & 71.1 & 1.36 & 9.10 & 12.11 & 25.6 & 4.38 & 2.54 & 
\dots\\
\object{SMMJ030236.15+000817.1} & 2.435 &           91 & 4.18 &         1152 & 
         810 & 3.33 & 11.54 & 1.1 & 0.55 & 8.62 & 12.83 & 41.3 & 0.27 & 2.67 & 
mIR,rad\\
\object{SMMJ105151.69+572636.0} & 1.147 &          134 & 0.30 &          248 & 
         457 & 0.92 & 11.43 & 1.5 & 0.65 & 8.97 & 12.16 & 34.4 & 4.00 & 2.25 & 
rad\\
\object{SMMJ105155.47+572312.7} & 2.686 &          837 & 12.95 &          589 & 
        1110 & 1.91 & 11.49 & 12.5 & 0.66 & 8.90 & 12.54 & 38.7 & 2.29 & 2.24 & 
mIR\\
\object{SMMJ105158.02+571800.2} & 2.239 &          314 & 16.77 &          373 & 
        1561 & 0.44 & 11.93 & 0.6 & 0.60 & 9.15 & 12.34 & 33.2 & 1.12 & 1.89 & 
rad\\
\object{SMMJ105200.22+572420.2} & 0.689 &          163 & 0.26 &          118 & 
          59 & 0.74 & 11.20 & 4.5 & 1.56 & 8.72 & 11.84 & 33.2 & 2.58 & 2.82 & 
\dots\\
\object{SMMJ105201.25+572445.7} & 2.148 &          935 & 2.58 &          798 & 
        1044 & 3.40 & 11.37 & 18.1 & 0.95 & 9.01 & 12.67 & 45.1 & 2.91 & 2.40 & 
mIR\\
\object{SMMJ105207.49+571904.0} & 2.689 &         2195 & 3.81 &         1559 & 
        6679 & 1.93 & 11.91 & 12.5 & 1.18 & 8.80 & 12.96 & 45.1 & 2.85 & 1.88 & 
rad\\
\object{SMMJ105225.79+571906.4} & 2.372 &          217 & 18.89 &          329 & 
        2310 & 0.75 & 11.64 & 0.0 & 0.49 & 8.96 & 12.28 & 33.2 & 1.16 & 1.67 & 
rad\\
\object{SMMJ105227.58+572512.4} & 2.142 &          410 & 4.73 &          376 & 
         564 & 0.57 & 11.82 & 1.9 & 0.78 & 8.79 & 12.34 & 38.7 & 1.71 & 2.34 & 
\dots\\
\object{SMMJ105227.77+572218.2} & 1.956 &          515 & 1.55 &          447 & 
         473 & 0.89 & 11.70 & 4.5 & 1.16 & 9.01 & 12.42 & 33.2 & 2.17 & 2.49 & 
\dots\\
\object{SMMJ105230.73+572209.5} & 2.611 &         1255 & 45.92 &          975 & 
        1942 & 0.72 & 12.13 & 3.8 & 0.68 & 9.18 & 12.75 & 38.7 & 1.51 & 2.22 & 
\dots\\
\object{SMMJ105238.19+571651.1} & 1.852 &          919 & 21.87 &          659 & 
         734 & 4.54 & 11.16 & 28.8 & 0.37 & 8.85 & 12.58 & 38.7 & 2.13 & 2.47 & 
\dots\\
\object{SMMJ105238.30+572435.8} & 3.036 &         1558 & 7.40 &         1169 & 
        1921 & 0.69 & 12.23 & 3.9 & 0.85 & 9.04 & 12.83 & 45.1 & 1.86 & 2.30 & 
spec\\
\object{SMMJ123549.44+621536.8} & 2.203 &          237 & 29.23 &          335 & 
        1144 & 0.70 & 11.68 & 0.0 & 0.51 & 9.18 & 12.29 & 33.2 & 1.11 & 1.98 & 
X,rad\\
\object{SMMJ123553.26+621337.7} & 2.098 &         1460 & 6.05 &         1223 & 
         802 & 5.73 & 11.33 & 31.8 & 0.42 & 8.85 & 12.85 & 52.6 & 2.99 & 2.70 & 
SB\\
\object{SMMJ123555.14+620901.7} & 1.875 &          332 & 11.94 &          358 & 
        2252 & 0.61 & 11.76 & 1.0 & 0.61 & 8.98 & 12.32 & 33.2 & 1.41 & 1.72 & 
X,mIR,rad\\
\object{SMMJ123600.10+620253.5} & 2.710 &         3733 & 6.49 &         3066 & 
        6407 & 8.92 & 11.54 & 51.6 & 0.29 & 8.52 & 13.25 & 71.5 & 3.64 & 2.20 & 
rad\\
\object{SMMJ123600.15+621047.2} & 1.994 &         1449 & 3.04 &         1102 & 
        1602 & 2.53 & 11.64 & 15.2 & 0.89 & 9.03 & 12.81 & 38.7 & 2.80 & 2.35 & 
SB,rad\\
\object{SMMJ123606.72+621550.7} & 2.416 &          448 & 30.23 &          354 & 
         454 & 1.27 & 11.45 & 6.9 & 0.65 & 8.92 & 12.32 & 33.2 & 1.15 & 2.41 & 
X,spec,mIR\\
\object{SMMJ123606.85+621021.4} & 2.509 &         1291 & 12.59 &         1153 & 
        1531 & 1.17 & 11.99 & 5.6 & 0.79 & 9.34 & 12.83 & 33.2 & 1.58 & 2.39 & 
X,rad\\
\object{SMMJ123616.15+621513.7} & 2.578 &          968 & 2.91 &          754 & 
        1179 & 0.80 & 11.97 & 4.5 & 1.12 & 9.04 & 12.64 & 33.2 & 2.17 & 2.32 & X
\\
\object{SMMJ123618.33+621550.5} & 1.865 &          339 & 2.20 &          325 & 
        1585 & 0.66 & 11.70 & 2.4 & 1.10 & 9.25 & 12.28 & 28.5 & 2.33 & 1.83 & 
SB,rad\\
\object{SMMJ123621.27+621708.4} & 1.988 &          330 & 5.65 &          266 & 
        1797 & 0.32 & 11.91 & 1.5 & 1.30 & 9.43 & 12.19 & 24.4 & 1.92 & 1.69 & 
SB,rad\\
\object{SMMJ123622.65+621629.7} & 2.466 &         1981 & 6.78 &         1438 & 
        1403 & 1.72 & 11.92 & 10.8 & 0.90 & 8.90 & 12.92 & 45.1 & 2.40 & 2.53 & 
X\\
\object{SMMJ123629.13+621045.8} & 1.013 &          202 & 0.69 &          176 & 
         207 & 0.35 & 11.70 & 1.5 & 1.53 & 9.07 & 12.01 & 24.4 & 2.13 & 2.45 & X
\\
\object{SMMJ123632.61+620800.1} & 1.993 &         1067 & 17.30 &          973 & 
        1107 & 4.52 & 11.33 & 22.7 & 0.45 & 8.45 & 12.75 & 71.5 & 1.70 & 2.46 & 
X,spec,mIR,rad\\
\object{SMMJ123634.51+621241.0} & 1.219 &          309 & 4.49 &          289 & 
         901 & 0.64 & 11.66 & 2.4 & 1.06 & 8.93 & 12.23 & 28.5 & 2.16 & 2.02 & 
SB,rad\\
\object{SMMJ123635.59+621424.1} & 2.005 &         1921 & 14.81 &         1740 & 
        1087 & 7.23 & 11.38 & 37.1 & 0.25 & 8.45 & 13.01 & 71.5 & 2.31 & 2.72 & 
X,spec,mIR\\
\object{SMMJ123636.75+621156.1} & 0.557 &           11 & 0.94 &           21 & 
          24 & 0.69 & 10.47 & 0.8 & 0.67 & 9.36 & 11.08 & 15.2 & 1.04 & 2.44 & 
X,spec\\
\object{SMMJ123707.21+621408.1} & 2.484 &          361 & 4.01 &          338 & 
         905 & 0.38 & 11.95 & 1.3 & 0.92 & 8.95 & 12.29 & 33.2 & 1.65 & 2.09 & 
X,rad\\
\object{SMMJ123711.98+621325.7} & 1.992 &          371 & 2.50 &          337 & 
         658 & 1.72 & 11.29 & 8.0 & 0.61 & 8.64 & 12.29 & 45.1 & 2.30 & 2.23 & 
X,rad\\
\object{SMMJ123712.05+621212.3} & 2.914 &          150 & 4.96 &          247 & 
         604 & 0.20 & 12.10 & 0.0 & 1.32 & 9.53 & 12.16 & 24.4 & 1.91 & 2.13 & 
X,spec,rad\\
\object{SMMJ123716.01+620323.3} & 2.037 &          879 & 567.07 &         1091
 &         1399 & 1.28 & 11.93 & 0.0 & 0.31 & 8.74 & 12.80 & 45.1 & 0.18 & 2.41
 & X,spec\\
\object{SMMJ123721.87+621035.3} & 0.979 &           76 & 2.44 &           91 & 
          96 & 0.33 & 11.44 & 0.6 & 0.80 & 9.77 & 11.72 & 16.9 & 1.12 & 2.49 & 
X,spec\\
\object{SMMJ131201.17+424208.1} & 3.405 &         4375 & 37.82 &         3748 & 
        1992 & 16.26 & 11.36 & 88.3 & 0.18 & 8.14 & 13.34 & 113.3 & 3.08 & 2.79
 & spec\\
\object{SMMJ131208.82+424129.1} & 1.544 &          787 & 3.75 &          551 & 
         560 & 1.75 & 11.50 & 10.9 & 0.74 & 8.70 & 12.51 & 45.1 & 1.99 & 2.51 & 
spec\\
\object{SMMJ131212.69+424422.5} & 2.805 &         2095 & 2.18 &         1470 & 
        2710 & 2.22 & 11.82 & 14.4 & 1.29 & 8.89 & 12.93 & 38.7 & 3.44 & 2.25 & 
spec,rad\\
\object{SMMJ131215.27+423900.9} & 2.565 &         2361 & 3387.93 &           80
 &         1498 & 0.04 & 12.32 & 0.0 & 0.32 & 9.90 & 11.67 & 15.4 & 0.00 & 1.24
 & spec,mIR\\
\object{SMMJ131222.35+423814.1} & 2.565 &          600 & 512.57 &          510
 &          569 & 0.88 & 11.76 & 0.0 & 0.31 & 8.75 & 12.47 & 33.2 & 0.10 & 2.47
 & spec,mIR\\
\object{SMMJ131225.20+424344.5} & 1.038 &          238 & 8.94 &          252 & 
         205 & 2.26 & 11.05 & 7.2 & 0.36 & 8.42 & 12.17 & 38.7 & 1.46 & 2.60 & 
\dots\\
\object{SMMJ131225.73+423941.4} & 1.554 &          742 & 5.47 &          661 & 
        5188 & 3.94 & 11.22 & 19.1 & 0.59 & 8.62 & 12.59 & 45.1 & 2.14 & 1.62 & 
rad\\
\object{SMMJ131228.30+424454.8} & 2.931 &         1572 & 16.32 &         1131 & 
        1482 & 2.43 & 11.67 & 15.3 & 0.61 & 8.30 & 12.82 & 61.3 & 1.75 & 2.40 & 
\dots\\
\object{SMMJ131231.07+424609.0} & 2.713 &         1212 & 1.42 &          910 & 
         966 & 5.13 & 11.25 & 31.8 & 0.66 & 8.70 & 12.72 & 45.1 & 3.95 & 2.49 & 
\dots\\
\object{SMMJ131232.31+423949.5} & 2.320 &         2457 & 8.41 &         1660 & 
        1635 & 1.60 & 12.02 & 10.8 & 0.87 & 8.47 & 12.99 & 61.3 & 2.40 & 2.52 & 
mIR\\
\object{SMMJ131239.14+424155.7} & 2.242 &         1271 & 3.75 &         1068 & 
         795 & 3.32 & 11.51 & 17.9 & 0.80 & 9.01 & 12.79 & 38.7 & 2.87 & 2.64 & 
\dots\\
\object{SMMJ141741.81+522823.0} & 1.150 &          948 & 13.50 &          770 & 
         274 & 0.75 & 12.01 & 3.8 & 0.72 & 8.48 & 12.65 & 45.1 & 1.28 & 2.97 & 
mIR\\
\object{SMMJ141742.04+523025.7} & 0.661 &          403 & 11.08 &          267 & 
         216 & 2.02 & 11.12 & 13.8 & 0.40 & 8.23 & 12.19 & 45.1 & 1.74 & 2.61 & 
\dots\\
\object{SMMJ141750.50+523101.0} & 2.128 &          343 & 5.31 &          241 & 
         808 & 1.91 & 11.10 & 12.5 & 0.66 & 8.58 & 12.15 & 38.7 & 2.29 & 1.99 & 
\dots\\
\object{SMMJ141800.40+512820.3} & 1.913 &         1999 & 17.45 &         1095 & 
        1424 & 0.39 & 12.44 & 3.3 & 0.80 & 8.61 & 12.80 & 52.6 & 0.89 & 2.40 & 
\dots\\
\object{SMMJ141802.87+523011.1} & 2.127 &         1325 & 10.95 &         1127 & 
         552 & 11.82 & 10.98 & 63.5 & 0.74 & 8.34 & 12.82 & 61.3 & 2.87 & 2.83
 & \dots\\
\object{SMMJ141809.00+522803.8} & 2.712 &         1213 & 6.24 &          551 & 
        1641 & 0.33 & 12.23 & 3.3 & 0.83 & 8.64 & 12.51 & 45.1 & 1.09 & 2.04 & 
\dots\\
\object{SMMJ141813.54+522923.4} & 3.484 &          243 & 6.65 &          240 & 
        3967 & 1.16 & 11.31 & 4.2 & 0.60 & 8.86 & 12.15 & 33.2 & 1.35 & 1.30 & 
spec,rad\\
\object{SMMJ163627.94+405811.2} & 3.180 &         1332 & 18.32 &         1095 & 
        3210 & 6.32 & 11.24 & 34.0 & 0.65 & 8.82 & 12.80 & 45.1 & 2.41 & 2.05 & 
spec\\
\object{SMMJ163631.47+405546.9} & 2.283 &         1325 & 10.28 &          865 & 
        1646 & 1.77 & 11.69 & 12.5 & 0.96 & 8.94 & 12.70 & 38.7 & 2.62 & 2.24 & 
spec\\
\object{SMMJ163639.01+405635.9} & 1.495 &          250 & 7.51 &         1101 & 
        1002 & 6.49 & 11.23 & 4.8 & 0.53 & 8.94 & 12.81 & 32.7 & 1.46 & 2.56 & 
mIR,rad\\
\object{SMMJ163650.43+405734.5} & 2.378 &         1147 & 73.43 &         1191 & 
        4030 & 3.24 & 11.57 & 10.1 & 0.33 & 8.93 & 12.84 & 45.1 & 1.12 & 1.99 & 
spec,mIR,rad\\
\object{SMMJ163658.19+410523.8} & 2.454 &         1769 & 4.97 &         1485 & 
        1801 & 2.55 & 11.77 & 13.8 & 0.80 & 9.04 & 12.94 & 45.1 & 2.45 & 2.43 & 
\dots\\
\object{SMMJ163658.78+405728.1} & 1.190 &          129 & 8.41 &          171 & 
         274 & 0.56 & 11.49 & 0.9 & 0.59 & 9.13 & 11.10 & 24.4 & 1.14 & 2.31 & 
\dots\\
\object{SMMJ163704.34+410530.3} & 0.840 &          205 & 2.50 &          157 & 
          74 & 2.22 & 10.85 & 13.1 & 0.92 & 9.13 & 11.96 & 33.2 & 2.43 & 2.84 & 
\dots\\
\object{SMMJ163706.51+405313.8} & 2.374 &         2020 & 9.63 &         1478 & 
        1344 & 1.83 & 11.91 & 10.9 & 0.76 & 9.06 & 12.94 & 45.1 & 1.99 & 2.56 & 
spec\\
\object{SMMJ221724.69+001242.1} & 0.510 &          154 & 14.71 &           43 & 
          62 & 0.21 & 11.31 & 3.4 & 0.63 & 9.36 & 11.40 & 15.9 & 0.69 & 2.36 & 
\dots\\
\object{SMMJ221725.97+001238.9} & 3.094 &         2499 & 2.29 &         1935 & 
        1353 & 2.10 & 11.96 & 12.5 & 1.58 & 9.38 & 13.05 & 38.7 & 3.27 & 2.67 & 
\dots\\
\object{SMMJ221733.02+000906.0} & 0.926 &          447 & 0.72 &          381 & 
         333 & 0.88 & 11.64 & 4.5 & 1.56 & 9.16 & 12.35 & 33.2 & 2.58 & 2.58 & 
\dots\\
\object{SMMJ221733.12+001120.2} & 0.652 &           31 & 2.77 &           59 & 
          62 & 0.28 & 11.33 & 4.7 & 1.79 & 9.13 & 11.54 & 21.8 & 1.27 & 2.49 & 
rad\\
\object{SMMJ221733.91+001352.1} & 2.555 &          875 & 17.73 &          731 & 
         954 & 0.78 & 11.97 & 3.8 & 0.79 & 9.01 & 12.63 & 38.7 & 1.69 & 2.40 & 
\dots\\
\object{SMMJ221735.15+001537.2} & 3.098 &          594 & 8.51 &          536 & 
        1627 & 2.28 & 11.37 & 10.0 & 0.68 & 8.94 & 12.49 & 38.7 & 1.93 & 2.03 & 
\dots\\
\object{SMMJ221735.84+001558.9} & 3.089 &         1969 & 7.43 &         1668 & 
        1450 & 6.49 & 11.41 & 35.0 & 0.30 & 8.35 & 12.99 & 71.5 & 2.98 & 2.58 & 
\dots\\
\object{SMMJ221737.39+001025.1} & 2.614 &         2991 & 13.32 &         2641 & 
        2484 & 7.05 & 11.57 & 37.1 & 0.29 & 8.47 & 13.19 & 71.5 & 2.70 & 2.54 & 
\dots\\
\object{SMMJ221804.42+002154.4} & 2.517 &         1474 & 15.50 &         1239 & 
         908 & 6.30 & 11.29 & 33.5 & 0.55 & 8.85 & 12.86 & 52.6 & 2.11 & 2.65 & 
mIR\\
\object{SMMJ221806.77+001245.7} & 3.623 &         8825 & 29.59 &         7774 & 
       11225 & 20.81 & 11.57 & 109.9 & 0.23 & 8.25 & 13.66 & 113.3 & 3.76 & 2.36
 & rad\\
\hline
 \object{SMMJ030226.17+000624.5} & 0.080 &            3 & 0.06 &            2 & 
           4 & 0.07 & 10.53 & 0.4 & 1.08 & 8.28 & 10.11 & 11.4 & 0.33 & 2.24 & 
rad\\
\object{SMMJ030238.62+001106.3} & 0.276 &           14 & 0.04 &           36 & 
          44 & 59.74 & 8.78 & 113.3 & 0.36 & 8.32 & 11.32 & 25.5 & 4.22 & 2.43
 & spec,rad\\
\object{SMMJ030244.82+000632.3} & 0.176 &            6 & 0.59 &           11 & 
           7 & 0.37 & 10.46 & 0.0 & 0.76 & 8.22 & 10.80 & 20.1 & 1.44 & 2.69 & 
\dots\\
\object{SMMJ123651.76+621221.3} & 0.298 &            6 & 0.12 &           14 & 
           7 & 1.02 & 10.13 & 0.9 & 0.98 & 8.91 & 10.90 & 13.3 & 1.93 & 2.79 & 
\dots\\
\hline
\multicolumn{1}{r}{mean} & 2.002 &         1065 & 68.35 &          873 & 
        1429 & 3.51 & 11.71 & 15.6 & 0.75 & 9.02 & 12.71 & 40.1 & 2.03 & 2.34
 & \dots  \\
\multicolumn{1}{r}{median} & 2.148 &          825 & 6.78 &          659 & 
        1087 & 1.72 & 11.54 & 7.2 & 0.68 & 8.91 & 12.58 & 38.7 & 1.99 & 2.40
 & \dots  \\
\multicolumn{1}{r}{std dev} & 0.851 &         1271 & 395.38 &         1066 & 
        1731 & 7.44 & 0.55 & 23.2 & 0.36 & 0.36 & 0.60 & 18.3 & 0.95 & 0.34
 & \dots  \\
\multicolumn{1}{r}{min} & 0.080 &            3 & 0.04 &            2 & 
           4 & 0.04 & 8.78 & 0.0 & 0.18 & 8.14 & 10.11 & 11.4 & 0.00 & 1.24
 & \dots  \\
\multicolumn{1}{r}{max} & 3.623 &         8825 & 3387.93 &         7774 & 
       11225 & 59.74 & 12.44 & 113.3 & 1.79 & 9.90 & 13.66 & 113.3 & 4.38 & 2.97
 &  \dots \\
\hline
\end{longtable}
\end{scriptsize}
\begin{table}[b]
\tablecomments{Column~(1):~SMG name. Column~(2):~redshift \citep{chapman05}. Column~(3):~total star formation rate (SFR) for $0.15-120\,M_\odot$ stars averaged over the last 50 Myr derived from the SED model. Column~(4):~SFR from UV emission interpolated from the SED template \citep[using][]{kennicutt}. Column~(5):~SFR from IR emission (Column 12) used in all analysis throughout the paper \citep[using][]{kennicutt}. Column~(6):~SFR from radio emission derived directly from the radio data \citep[using][]{bell03}. Column~(7):~specific SFR$\mbox{}\equiv\mbox{SFR}_{\rm IR}/M_*$. Column~(8):~stellar mass. Column~(9):~Ratio of the mass of gas converted to star during the recent starburst episode to the total stellar mass.  There are values greater than 100\%, because the starburst episode is ongoing; 0\% means that non-starburst template was adopted. Column~(10):~stellar mass to light ratio (luminosity at rest-frame K was intepolated using the best SED model). Column~(11):~dust mass. Column~(12):~total $8$--$1000\,\mu$m infrared luminosity. Column~(13):~dust temperature. Column~(14):~Average extinction $A_V=2.5\log$($V$-band starlight unextinguished / $V$-band starlight observed). Column~(15):~FIR-radio correlation parameter (Section~\ref{sec:firr}). Column~(16):~AGN flag ---  X: X-ray identified AGN; SB: X-ray identified starburst \citep{alexander05}; spec: spectroscopically identified AGN or QSO \citep{chapman05}; mIR: mid-IR identified AGN (Section~\ref{sec:agn}); rad: radio datapoint is more than $3\sigma$ above the starburst model (Section~\ref{sec:agn}). The horizontal line divides the $z>0.5$ and $z<0.5$ samples. This table is available in a machine-readable form in the electronic edition of the Journal.}
\end{table}
\end{landscape}
}
\clearpage

\begin{figure*}
\begin{center}
\plotone{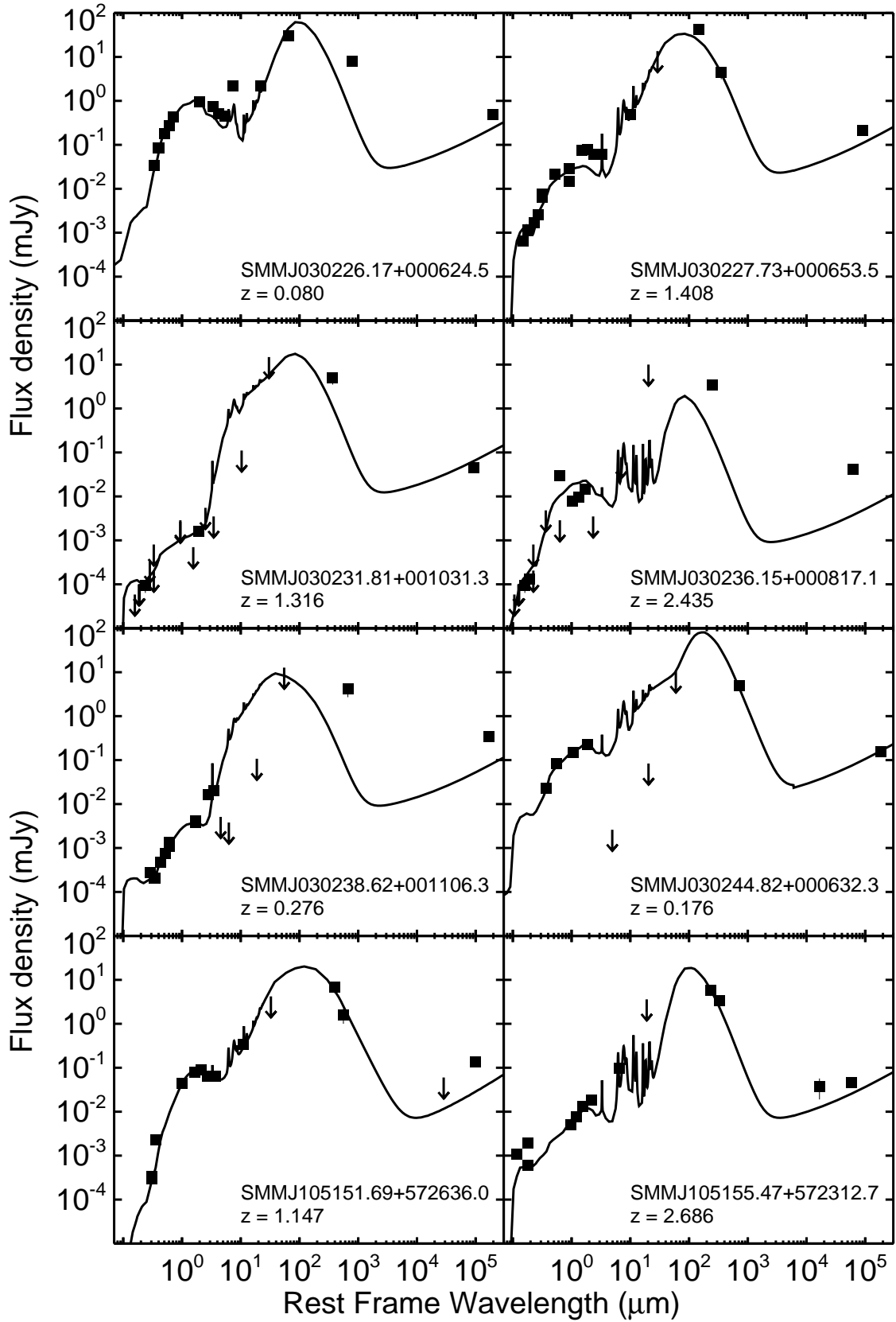}
\end{center}
\caption{Spectral energy distributions (SEDs) of SMGs. {\it Solid lines}: the best GRASIL fits. {\it Dashed lines}: SEDs of GRB hosts \citep{michalowski08} shown for comparison.
{\it Squares}: detections with errors, in most cases, smaller than the size of the symbols. {\it Arrows}: $3\sigma$ upper limit (values marked at the base). In the cases where our fits strongly underpredict the observed data at $850\,\mu$m, we adopted $L_{\rm IR}$ and $T_d$ of \citet{chapman05}.
}
\label{fig:sed}
\end{figure*}


\addtocounter{figure}{-1}
\begin{figure*}
\begin{center}
\plotone{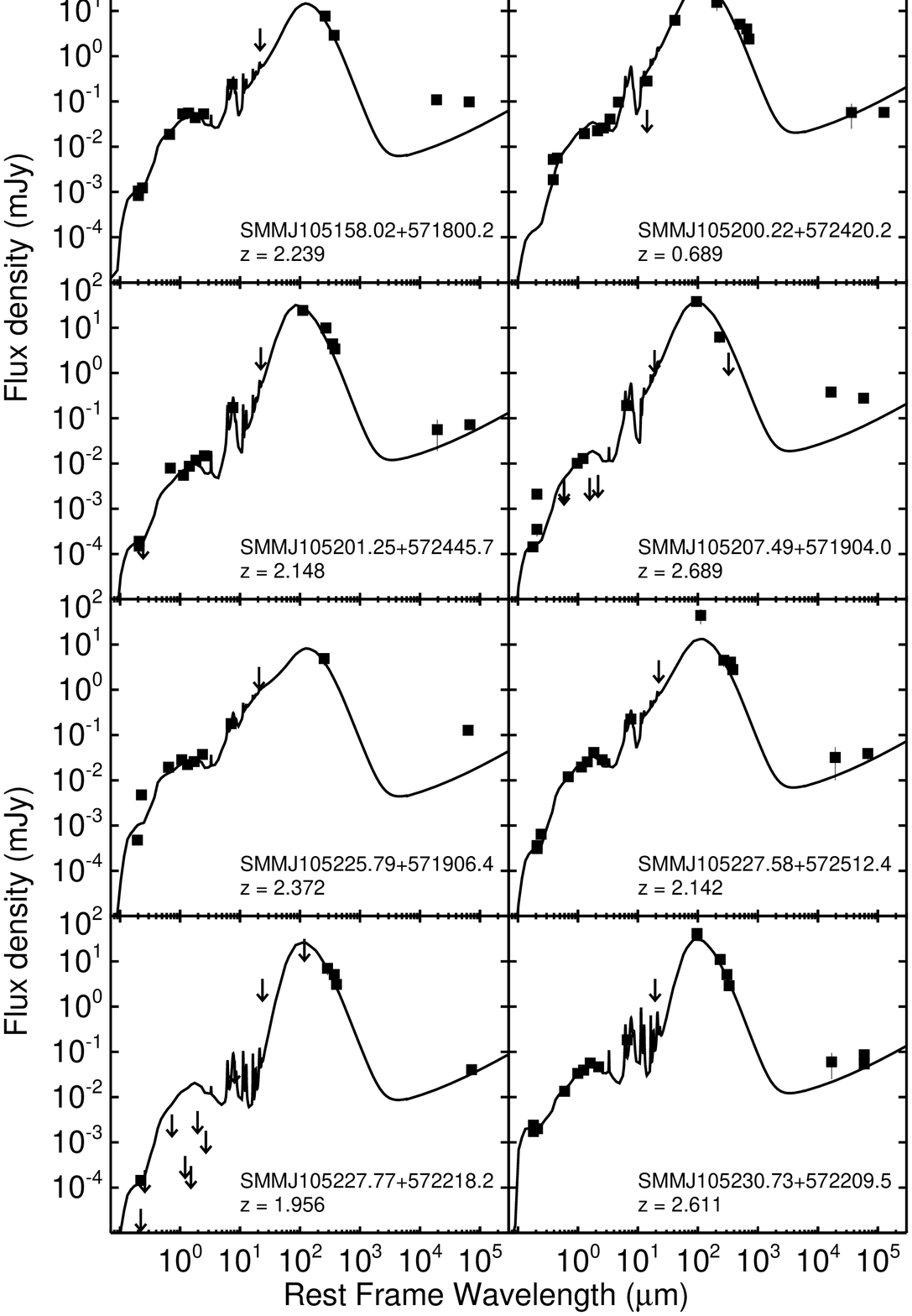}
\end{center}
\caption{(continued).}
\end{figure*}

\addtocounter{figure}{-1}
\begin{figure*}
\begin{center}
\plotone{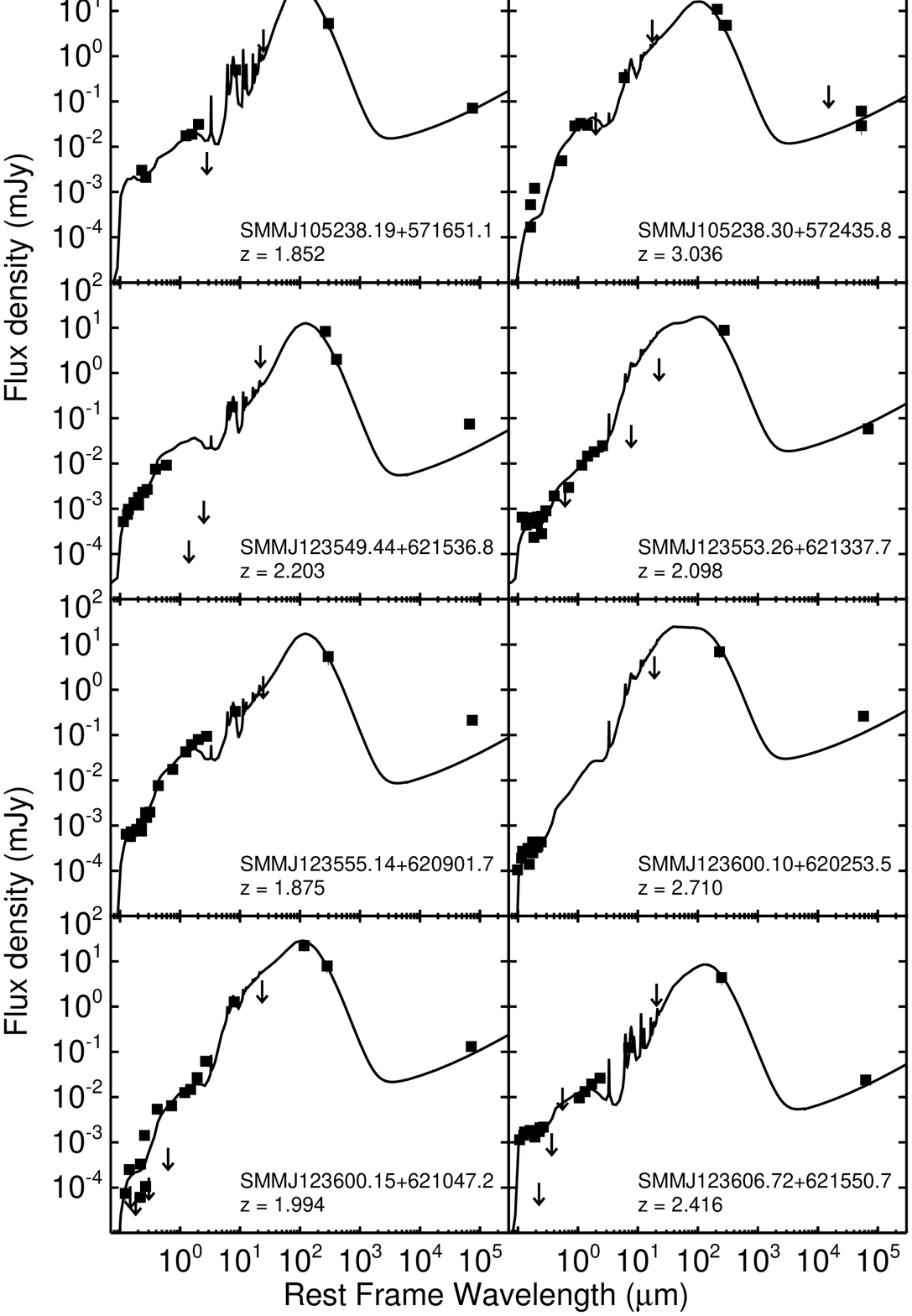}
\end{center}
\caption{(continued).}
\end{figure*}

\addtocounter{figure}{-1}
\begin{figure*}
\begin{center}
\plotone{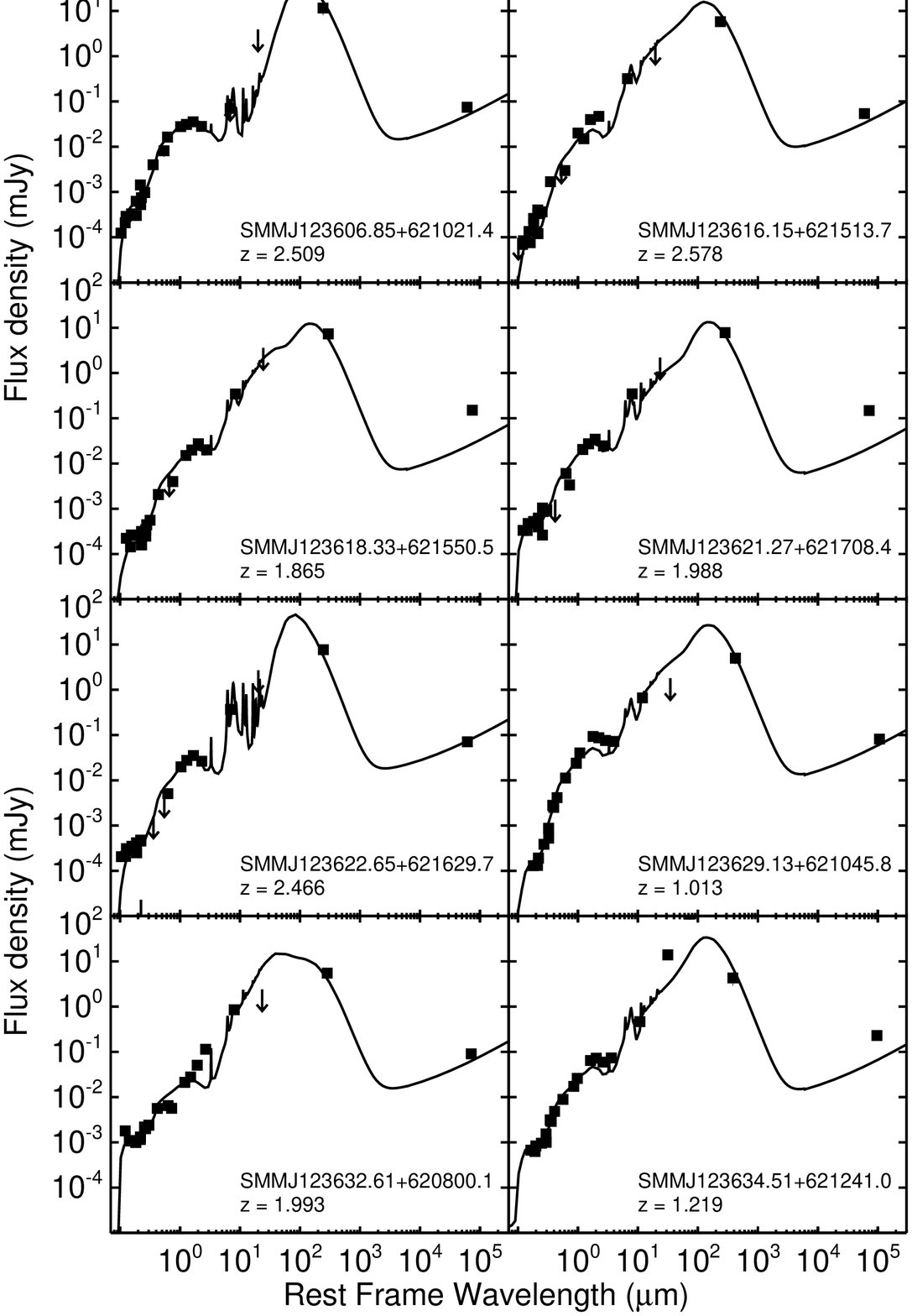}
\end{center}
\caption{(continued).}
\end{figure*}

\addtocounter{figure}{-1}
\begin{figure*}
\begin{center}
\plotone{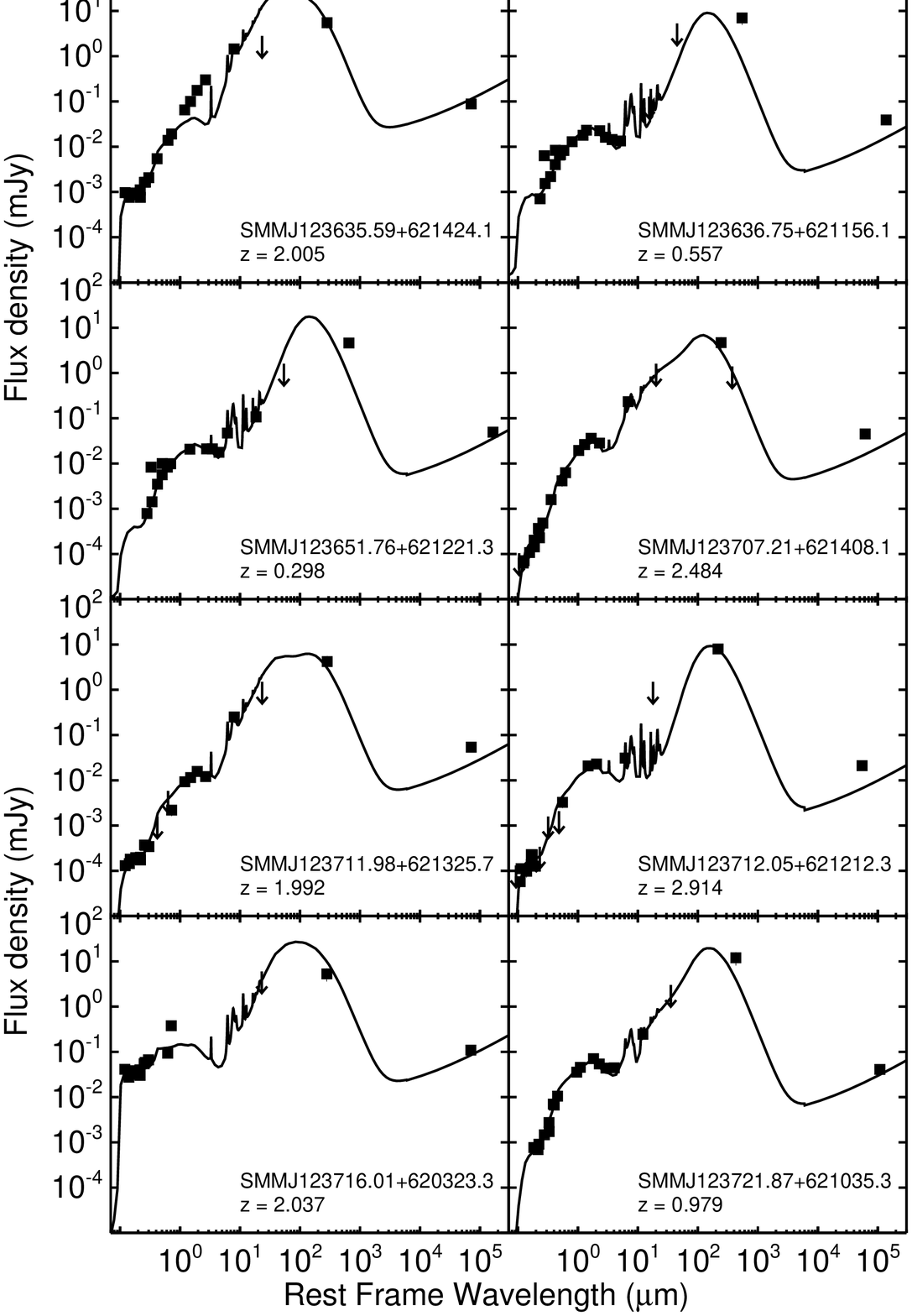}
\end{center}
\caption{(continued).}
\end{figure*}

\addtocounter{figure}{-1}
\begin{figure*}
\begin{center}
\plotone{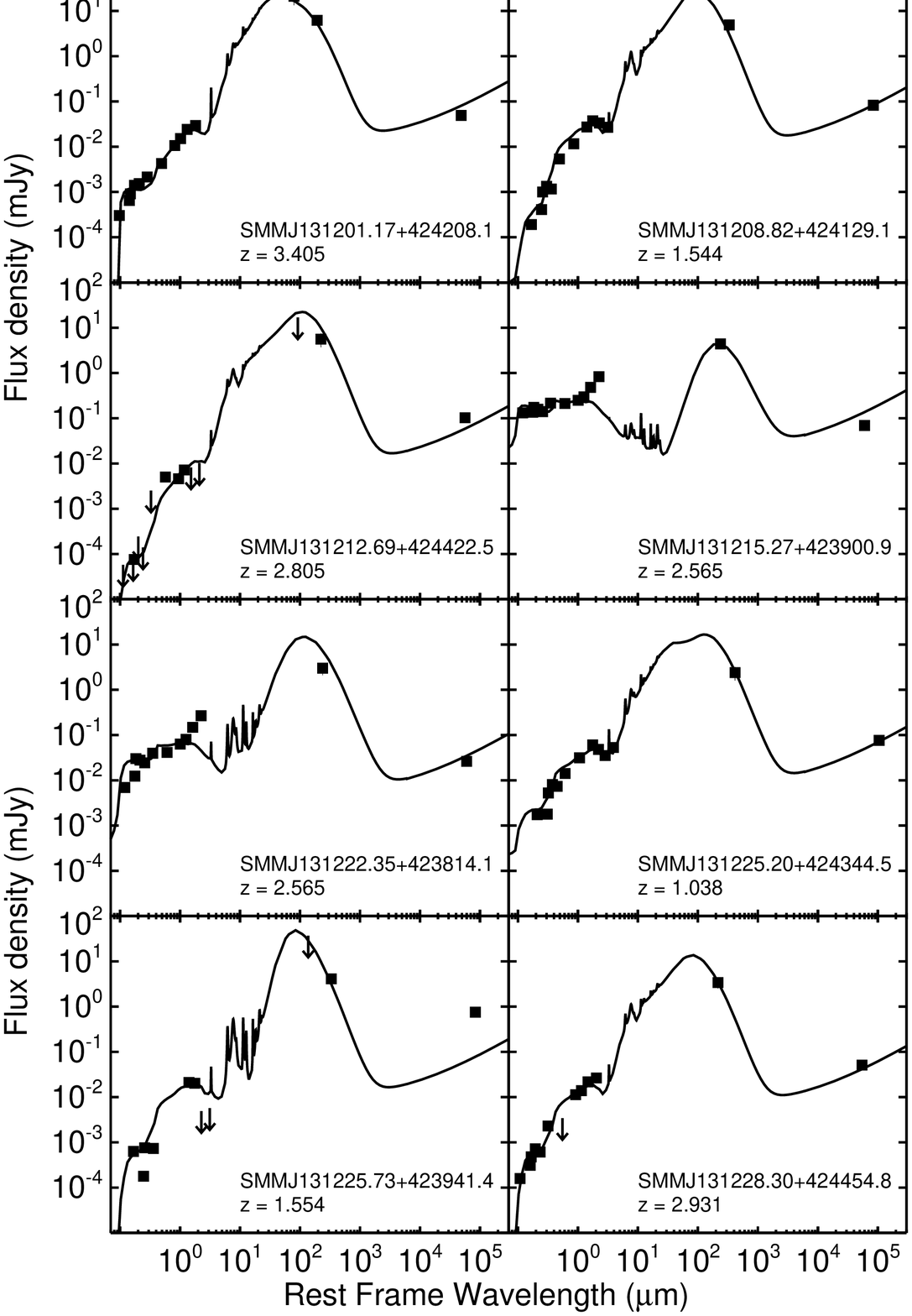}
\end{center}
\caption{(continued).}
\end{figure*}

\addtocounter{figure}{-1}
\begin{figure*}
\begin{center}
\plotone{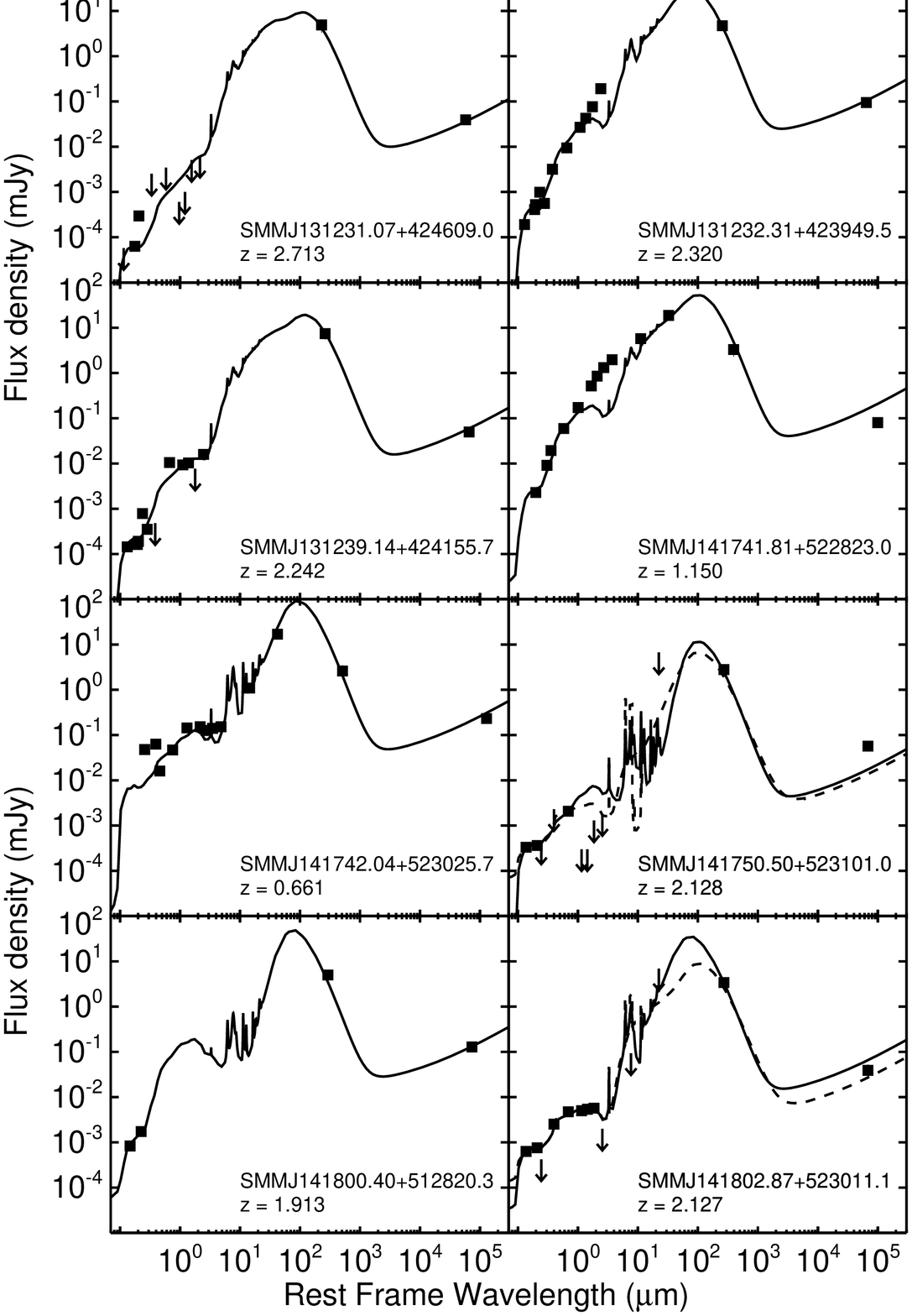}
\end{center}
\caption{(continued).}
\end{figure*}

\addtocounter{figure}{-1}
\begin{figure*}
\begin{center}
\plotone{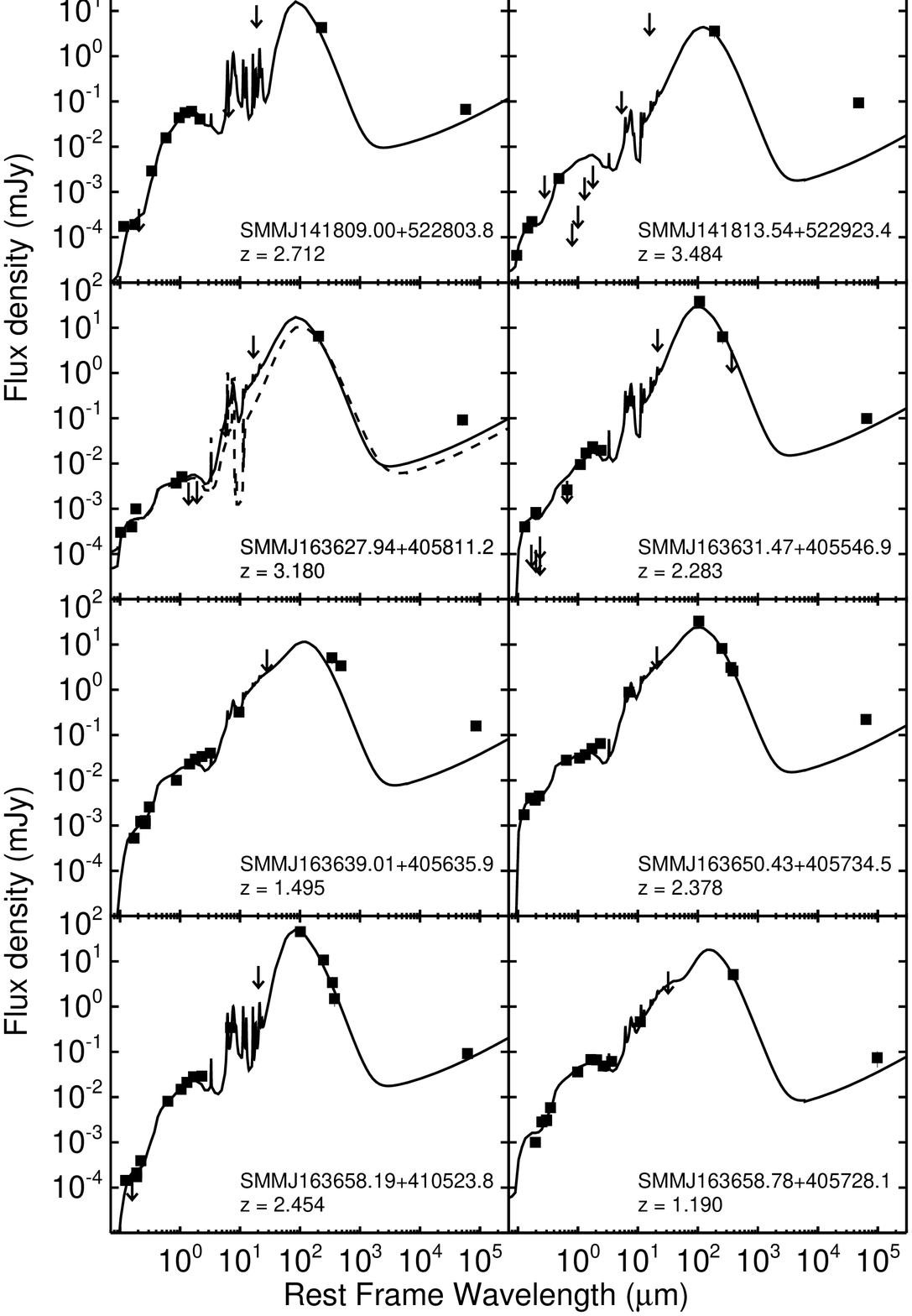}
\end{center}
\caption{(continued).}
\end{figure*}

\addtocounter{figure}{-1}
\begin{figure*}
\begin{center}
\plotone{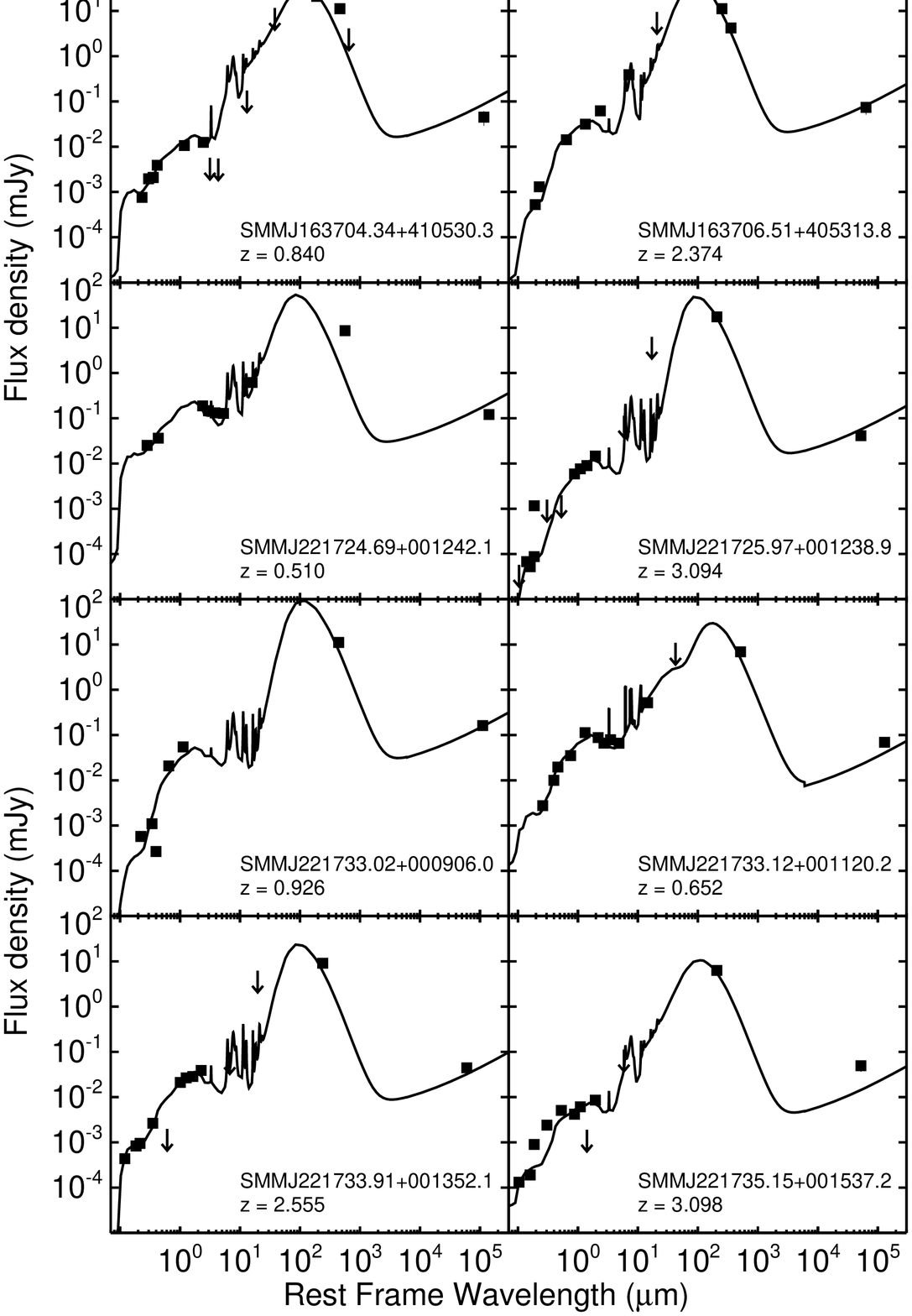}
\end{center}
\caption{(continued).}
\end{figure*}

\addtocounter{figure}{-1}
\begin{figure*}
\begin{center}
\plotone{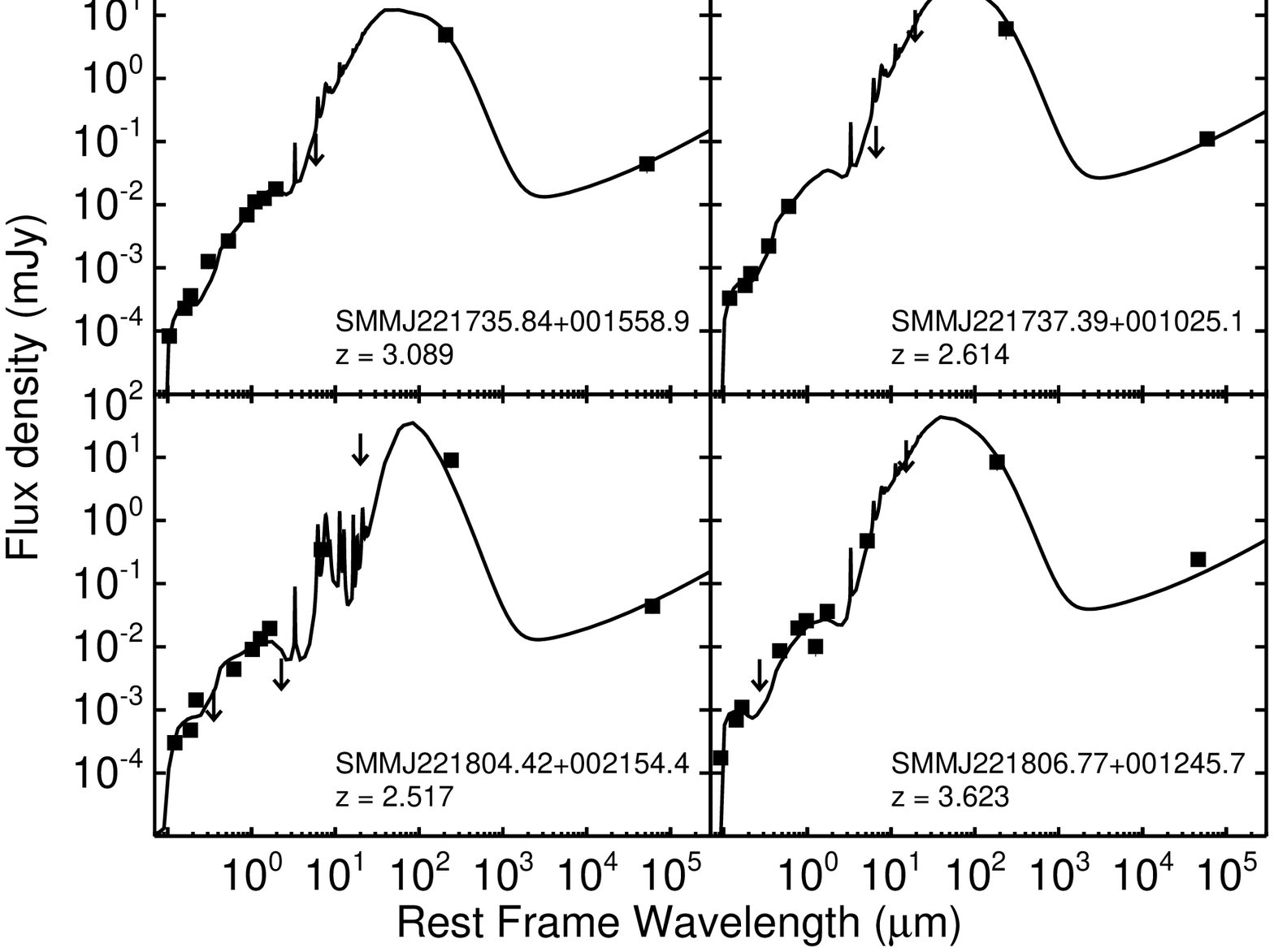}
\end{center}
\caption{(continued).}
\end{figure*}


\longtab{4}{
\begin{longtable}{cccccll}
\caption{Compilation of star formation rate density determinations in $M_\odot\,\mbox{yr}^{-1}\,\mbox{Mpc}^{-3}$ \label{tab:sfrd}}
\\
\hline\hline
 $z$ & $\Delta z$ & $\rho_{\rm SFR}$ & $-\Delta\rho_{\rm SFR}$ & $+\Delta\rho_{\rm SFR}$ & Estimator & Reference \\
\hline
\endfirsthead
\caption{continued.}\\
\hline\hline
 $z$ & $\Delta z$ & $\rho_{\rm SFR}$ & $-\Delta\rho_{\rm SFR}$ & $+\Delta\rho_{\rm SFR}$ & Estimator & Reference \\
\hline
\endhead
\hline
\endfoot
3.780&0.340&0.1690&0.0218&0.0250& UV & \citet{giavalisco04,hopkins04} \\
4.920&0.330&0.1089&0.0300&0.0414& UV & \citet{giavalisco04,hopkins04} \\
5.740&0.360&0.1194&0.0423&0.0655& UV & \citet{giavalisco04,hopkins04} \\
0.350&0.150&0.0356&0.0058&0.0070& UV & \citet{wilson02,hopkins04} \\
0.800&0.200&0.0656&0.0110&0.0133& UV & \citet{wilson02,hopkins04} \\
1.350&0.250&0.0925&0.0259&0.0361& UV & \citet{wilson02,hopkins04} \\
1.500&0.500&0.1954&0.0721&0.1143& UV & \citet{massarotti01,hopkins04} \\
2.750&0.750&0.3076&0.1135&0.1799& UV & \citet{massarotti01,hopkins04} \\
4.000&0.500&0.1300&0.0569&0.1012& UV & \citet{massarotti01,hopkins04} \\
0.150&0.150&0.0395&0.0043&0.0048& UV & \citet{sullivan00,hopkins04} \\
3.040&0.250&0.1603&0.0174&0.0196& UV & \citet{steidel99,hopkins04} \\
4.130&0.300&0.1245&0.0256&0.0322& UV & \citet{steidel99,hopkins04} \\
0.700&0.200&0.0481&0.0102&0.0130& UV & \citet{cowie99,hopkins04} \\
1.250&0.250&0.0652&0.0154&0.0201& UV & \citet{cowie99,hopkins04} \\
0.150&0.150&0.0428&0.0125&0.0176& UV & \citet{treyer98,hopkins04} \\
0.750&0.250&0.1019&0.0297&0.0420& UV & \citet{connolly97,hopkins04} \\
1.250&0.250&0.1368&0.0399&0.0564& UV & \citet{connolly97,hopkins04} \\
1.750&0.250&0.1062&0.0310&0.0438& UV & \citet{connolly97,hopkins04} \\
0.350&0.150&0.0289&0.0043&0.0051& UV & \citet{lilly96,hopkins04} \\
0.625&0.125&0.0542&0.0091&0.0110& UV & \citet{lilly96,hopkins04} \\
0.875&0.125&0.1050&0.0307&0.0433& UV & \citet{lilly96,hopkins04} \\
4.850&0.450&0.0350&0.0150&0.0150& UV & \citet{iwata03,vanbreukelen05} \\
0.700&0.300&0.0462&0.0060&0.0068& UV & \citet{cowie96,somerville01} \\
1.250&0.250&0.0668&0.0195&0.0276& UV & \citet{cowie96,somerville01} \\
0.350&0.150&0.0495&0.0258&0.0272& UV & \citet{sawicki97,somerville01} \\
0.750&0.250&0.0733&0.0109&0.0089& UV & \citet{sawicki97,somerville01} \\
1.500&0.500&0.0988&0.0087&0.0095& UV & \citet{sawicki97,somerville01} \\
2.500&0.500&0.2113&0.0273&0.0313& UV & \citet{sawicki97,somerville01} \\
3.500&0.500&0.0922&0.0190&0.0187& UV & \citet{sawicki97,somerville01} \\
0.250&0.250&0.0569&0.0226&0.0592& UV & \citet{pascarelle98,somerville01} \\
0.750&0.250&0.0638&0.0176&0.0446& UV & \citet{pascarelle98,somerville01} \\
1.250&0.250&0.0901&0.0218&0.0665& UV & \citet{pascarelle98,somerville01} \\
1.750&0.250&0.0922&0.0223&0.0681& UV & \citet{pascarelle98,somerville01} \\
2.500&0.500&0.0556&0.0213&0.0503& UV & \citet{pascarelle98,somerville01} \\
3.500&0.500&0.0519&0.0240&0.0616& UV & \citet{pascarelle98,somerville01} \\
4.500&0.500&0.0653&0.0374&0.1145& UV & \citet{pascarelle98,somerville01} \\
5.500&0.500&0.0285&0.0166&0.0824& UV & \citet{pascarelle98,somerville01} \\
0.315&0.115&0.0373&0.0005&0.0005& UV & \citet{mobasher09} \\
0.540&0.110&0.0533&0.0005&0.0006& UV & \citet{mobasher09} \\
0.765&0.115&0.0957&0.0006&0.0006& UV & \citet{mobasher09} \\
0.990&0.110&0.1082&0.0006&0.0006& UV & \citet{mobasher09} \\
3.800&0.350&0.0891&0.0097&0.0109& UV & \citet{bouwens07} \\
5.000&0.350&0.0331&0.0043&0.0049& UV & \citet{bouwens07} \\
5.900&0.300&0.0224&0.0038&0.0045& UV & \citet{bouwens07} \\
4.000&0.500&0.0362&0.0050&0.0050& UV & \citet{ouchi04} \\
4.700&0.500&0.0300&0.0175&0.0175& UV & \citet{ouchi04} \\
4.900&0.300&0.0138&0.0069&0.0069& UV & \citet{ouchi04} \\
4.000&0.500&0.0300&0.0025&0.0025& UV & \citet{ouchi04} \\
4.700&0.500&0.0200&0.0088&0.0088& UV & \citet{ouchi04} \\
4.900&0.300&0.0088&0.0044&0.0044& UV & \citet{ouchi04} \\
5.850&0.250&0.0034&0.0014&0.0014& UV & \citet{stanway03} \\
1.000&0.500&0.2080&0.0990&0.0990& UV & \citet{thompson06} \\
2.000&0.500&0.3980&0.1800&0.1800& UV & \citet{thompson06} \\
3.000&0.500&0.3220&0.1600&0.1600& UV & \citet{thompson06} \\
4.000&0.500&0.0940&0.0390&0.0390& UV & \citet{thompson06} \\
5.000&0.500&0.0410&0.0160&0.0160& UV & \citet{thompson06} \\
6.000&0.500&0.1260&0.0740&0.0740& UV & \citet{thompson06} \\
2.280&0.330&0.1778&0.1407&0.6733& UV & \citet{ly09} \\
4.000&0.300&0.1301&0.0194&0.0227& UV & \citet{yoshida06,ly09} \\
4.700&0.300&0.0745&0.0316&0.0550& UV & \citet{yoshida06,ly09} \\
2.300&0.400&0.1500&0.0223&0.0262& UV & \citet{reddy08,ly09} \\
0.050&0.050&0.0126&0.0026&0.0033& UV & \citet{wyder05} \\
3.200&0.140&0.1600&0.0798&0.1592& UV & \citet{shim07} \\
0.330&0.040&0.0353&0.0059&0.0071& UV & \citet{dahlen07} \\
0.545&0.085&0.0996&0.0148&0.0174& UV & \citet{dahlen07} \\
1.125&0.205&0.1283&0.0240&0.0295& UV & \citet{dahlen07} \\
1.750&0.130&0.1898&0.0390&0.0491& UV & \citet{dahlen07} \\
2.225&0.145&0.1407&0.0364&0.0491& UV & \citet{dahlen07} \\
2.750&0.750&0.1800&0.0337&0.0414& UV & \citet{wadadekar06} \\
0.900&0.500&0.0989&0.0221&0.0285&[\ion{O}{2}]& \citet{teplitz03,hopkins04} \\
0.025&0.025&0.0122&0.0036&0.0050&[\ion{O}{2}]& \citet{gallego02,hopkins04} \\
0.200&0.100&0.0136&0.0032&0.0062&[\ion{O}{2}]& \citet{hogg98,hopkins04} \\
0.300&0.100&0.0119&0.0023&0.0031&[\ion{O}{2}]& \citet{hogg98,hopkins04} \\
0.400&0.100&0.0536&0.0095&0.0147&[\ion{O}{2}]& \citet{hogg98,hopkins04} \\
0.500&0.100&0.0955&0.0137&0.0180&[\ion{O}{2}]& \citet{hogg98,hopkins04} \\
0.600&0.100&0.0649&0.0086&0.0117&[\ion{O}{2}]& \citet{hogg98,hopkins04} \\
0.700&0.100&0.0535&0.0083&0.0120&[\ion{O}{2}]& \citet{hogg98,hopkins04} \\
0.800&0.100&0.0566&0.0085&0.0116&[\ion{O}{2}]& \citet{hogg98,hopkins04} \\
0.900&0.100&0.0714&0.0112&0.0165&[\ion{O}{2}]& \citet{hogg98,hopkins04} \\
1.000&0.100&0.1146&0.0208&0.0320&[\ion{O}{2}]& \citet{hogg98,hopkins04} \\
1.100&0.100&0.0899&0.0242&0.0523&[\ion{O}{2}]& \citet{hogg98,hopkins04} \\
1.200&0.100&0.0859&0.0286&0.0859&[\ion{O}{2}]& \citet{hogg98,hopkins04} \\
0.375&0.125&0.0197&0.0033&0.0034&[\ion{O}{2}]& \citet{hammer97,hopkins04} \\
0.625&0.125&0.0594&0.0174&0.0171&[\ion{O}{2}]& \citet{hammer97,hopkins04} \\
0.875&0.125&0.1396&0.0814&0.0817&[\ion{O}{2}]& \citet{hammer97,hopkins04} \\
0.401&0.011&0.0240&0.0080&0.0080&[\ion{O}{3}]& \citet{hippelein03} \\
0.636&0.010&0.0720&0.0160&0.0160&[\ion{O}{3}]& \citet{hippelein03} \\
0.881&0.014&0.1070&0.0350&0.0350&[\ion{O}{2}]& \citet{hippelein03} \\
1.193&0.018&0.2280&0.0550&0.0550&[\ion{O}{2}]& \citet{hippelein03} \\
2.750&0.750&0.2773&0.0810&0.1144&H$\beta$& \citet{pettini98,hopkins04} \\
0.025&0.025&0.0249&0.0056&0.0072&H$\alpha$& \citet{perezgonzalez03,hopkins04} \\
0.800&0.300&0.1172&0.0262&0.0338&H$\alpha$& \citet{tresse02,hopkins04} \\
2.200&0.050&0.2655&0.0641&0.0845&H$\alpha$& \citet{moorwood00,hopkins04} \\
1.250&0.550&0.2350&0.0137&0.0145&H$\alpha$& \citet{hopkins00,hopkins04} \\
0.150&0.150&0.0151&0.0020&0.0022&H$\alpha$& \citet{sullivan00,hopkins04} \\
0.900&0.100&0.1067&0.0294&0.0440&H$\alpha$& \citet{glazebrook99,hopkins04} \\
1.300&0.600&0.2799&0.0676&0.0891&H$\alpha$& \citet{yan99,hopkins04} \\
0.200&0.100&0.0324&0.0042&0.0048&H$\alpha$& \citet{tresse98,hopkins04} \\
0.022&0.022&0.0126&0.0046&0.0074&H$\alpha$& \citet{gallego95,hopkins04} \\
0.043&0.043&0.0240&0.0026&0.0029&H$\alpha$& \citet{gronwall99,somerville01} \\
0.243&0.009&0.0360&0.0120&0.0060&H$\alpha$& \citet{fujita03b} \\
0.245&0.007&0.0240&0.0060&0.0060&H$\alpha$& \citet{hippelein03} \\
0.100&0.100&0.0192&0.0042&0.0014&H$\alpha$& \citet{brinchmann04} \\
0.010&0.010&0.0158&0.0033&0.0071&H$\alpha$& \citet{hanish06} \\
0.840&0.030&0.1500&0.0200&0.0200&H$\alpha$& \citet{sobral09} \\
2.230&0.150&0.1700&0.0900&0.1600&H$\alpha$& \citet{geach08} \\
0.242&0.009&0.0180&0.0040&0.0070&H$\alpha$& \citet{shioya08} \\
0.840&0.030&0.1700&0.0300&0.0300&H$\alpha$& \citet{villar08} \\
2.300&0.400&0.3484&0.0869&0.0869&H$\alpha$& \citet{reddy08} \\
3.050&0.350&0.2141&0.0450&0.0450&H$\alpha$& \citet{reddy08} \\
0.350&0.150&0.0365&0.0169&0.0314& mid-IR & \citet{flores99,hopkins04} \\
0.625&0.125&0.0678&0.0297&0.0527& mid-IR & \citet{flores99,hopkins04} \\
0.875&0.125&0.1337&0.0602&0.1096& mid-IR & \citet{flores99,hopkins04} \\
0.215&0.215&0.0300&0.0100&0.0200& mid-IR & \citet{mann02} \\
0.515&0.085&0.0700&0.0100&0.0200& mid-IR & \citet{mann02} \\
0.100&0.100&0.0175&0.0049&0.0049& mid-IR & \citet{pozzi04} \\
0.300&0.100&0.0301&0.0140&0.0140& mid-IR & \citet{pozzi04} \\
0.300&0.100&0.0197&0.0067&0.0066& mid-IR & \citet{zheng07} \\
0.500&0.100&0.0349&0.0088&0.0088& mid-IR & \citet{zheng07} \\
0.700&0.100&0.0570&0.0096&0.0101& mid-IR & \citet{zheng07} \\
0.900&0.100&0.0616&0.0120&0.0121& mid-IR & \citet{zheng07} \\
2.300&0.400&0.2091&0.0357&0.0357& mid-IR & \citet{reddy08} \\
3.050&0.350&0.1124&0.0211&0.0211& mid-IR & \citet{reddy08} \\
1.000&0.100&0.2000&0.0300&0.0300& mid-IR & \citet{caputi07} \\
2.000&0.300&0.1100&0.0200&0.0200& mid-IR & \citet{caputi07} \\
0.450&0.150&0.2750&0.0220&0.0220& mid-IR & \citet{santini09} \\
0.800&0.200&0.4870&0.0170&0.0170& mid-IR & \citet{santini09} \\
1.250&0.250&0.7550&0.0290&0.0290& mid-IR & \citet{santini09} \\
2.000&0.500&1.6590&0.0580&0.0580& mid-IR & \citet{santini09} \\
0.100&0.100&0.0180&0.0025&0.0029& mid-IR & \citet{perezgonzalez05} \\
0.100&0.100&0.0163&0.0023&0.0027& mid-IR & \citet{perezgonzalez05} \\
0.100&0.100&0.0171&0.0024&0.0028& mid-IR & \citet{perezgonzalez05} \\
0.300&0.100&0.0384&0.0019&0.0020& mid-IR & \citet{perezgonzalez05} \\
0.300&0.100&0.0356&0.0017&0.0018& mid-IR & \citet{perezgonzalez05} \\
0.300&0.100&0.0330&0.0016&0.0017& mid-IR & \citet{perezgonzalez05} \\
0.500&0.100&0.0818&0.0060&0.0064& mid-IR & \citet{perezgonzalez05} \\
0.500&0.100&0.0927&0.0089&0.0098& mid-IR & \citet{perezgonzalez05} \\
0.500&0.100&0.0589&0.0043&0.0046& mid-IR & \citet{perezgonzalez05} \\
0.700&0.100&0.1052&0.0052&0.0054& mid-IR & \citet{perezgonzalez05} \\
0.700&0.100&0.1459&0.0140&0.0155& mid-IR & \citet{perezgonzalez05} \\
0.700&0.100&0.0951&0.0047&0.0049& mid-IR & \citet{perezgonzalez05} \\
0.900&0.100&0.1319&0.0580&0.1036& mid-IR & \citet{perezgonzalez05} \\
0.900&0.100&0.1877&0.0852&0.1559& mid-IR & \citet{perezgonzalez05} \\
0.900&0.100&0.1255&0.0552&0.0985& mid-IR & \citet{perezgonzalez05} \\
1.200&0.200&0.1741&0.0127&0.0137& mid-IR & \citet{perezgonzalez05} \\
1.200&0.200&0.2477&0.0400&0.0478& mid-IR & \citet{perezgonzalez05} \\
1.200&0.200&0.1423&0.0136&0.0151& mid-IR & \citet{perezgonzalez05} \\
1.600&0.200&0.1574&0.0151&0.0167& mid-IR & \citet{perezgonzalez05} \\
1.600&0.200&0.2076&0.0379&0.0464& mid-IR & \citet{perezgonzalez05} \\
1.600&0.200&0.1353&0.0130&0.0143& mid-IR & \citet{perezgonzalez05} \\
2.000&0.200&0.1319&0.0344&0.0466& mid-IR & \citet{perezgonzalez05} \\
2.000&0.200&0.2239&0.0705&0.1028& mid-IR & \citet{perezgonzalez05} \\
2.000&0.200&0.1388&0.0362&0.0490& mid-IR & \citet{perezgonzalez05} \\
2.400&0.200&0.1785&0.0432&0.0570& mid-IR & \citet{perezgonzalez05} \\
2.400&0.200&0.3524&0.1109&0.1618& mid-IR & \citet{perezgonzalez05} \\
2.400&0.200&0.2296&0.0556&0.0733& mid-IR & \citet{perezgonzalez05} \\
0.698 & 0.618 & 0.0102 & 0.0014 & 0.0014& submm & This work \\
1.775 & 0.367 & 0.0228 & 0.0027 & 0.0027& submm & This work \\
2.357 & 0.209 & 0.0486 & 0.0054 & 0.0054& submm & This work \\
3.101 & 0.522 & 0.0341 & 0.0040 & 0.0040& submm & This work \\
2.000&1.000&0.1476&0.0607&0.0973& submm & \citet{barger00,hopkins04} \\
4.500&1.500&0.1901&0.1195&0.2454& submm & \citet{barger00,hopkins04} \\
0.057&0.041&0.0206&0.0016&0.0024& submm & \citet{pascale09} \\
0.138&0.040&0.0292&0.0022&0.0040& submm & \citet{pascale09} \\
0.250&0.073&0.0192&0.0026&0.0036& submm & \citet{pascale09} \\
0.454&0.132&0.0511&0.0022&0.0067& submm & \citet{pascale09} \\
0.824&0.239&0.0785&0.0073&0.0086& submm & \citet{pascale09} \\
2.281&1.219&0.1104&0.0092&0.0140& submm & \citet{pascale09} \\
0.005&0.005&0.0109&0.0007&0.0008& radio & \citet{condon02,hopkins04} \\
0.080&0.080&0.0187&0.0035&0.0038& radio & \citet{sadler02,hopkins04} \\
0.010&0.010&0.0177&0.0036&0.0036& radio & \citet{serjeant02,hopkins04} \\
0.070&0.070&0.0120&0.0025&0.0031& radio & \citet{machalski00,hopkins04} \\
0.206&0.196&0.0408&0.0157&0.0155& radio & \citet{haarsma00,hopkins04} \\
0.464&0.054&0.0667&0.0246&0.0254& radio & \citet{haarsma00,hopkins04} \\
0.623&0.075&0.0764&0.0344&0.0340& radio & \citet{haarsma00,hopkins04} \\
0.804&0.080&0.1315&0.0446&0.0459& radio & \citet{haarsma00,hopkins04} \\
1.600&0.640&0.1641&0.0557&0.0522& radio & \citet{haarsma00,hopkins04} \\
0.005&0.005&0.0209&0.0000&0.0000& radio & \citet{condon89,hopkins04} \\
0.152&0.149&0.0220&0.0010&0.0010& radio & \citet{mauch07} \\
0.310&0.210&0.0331&0.0074&0.0076& radio & \citet{seymour08} \\
0.810&0.290&0.0851&0.0262&0.0271& radio & \citet{seymour08} \\
1.500&0.400&0.1479&0.0666&0.0812& radio & \citet{seymour08} \\
2.450&0.550&0.1202&0.0756&0.1036& radio & \citet{seymour08} \\
0.100&0.100&0.0087&0.0063&0.0062& radio & \citet{dunne09} \\
0.300&0.100&0.0292&0.0107&0.0106& radio & \citet{dunne09} \\
0.500&0.100&0.0385&0.0118&0.0106& radio & \citet{dunne09} \\
0.700&0.100&0.0700&0.0175&0.0164& radio & \citet{dunne09} \\
0.900&0.100&0.0781&0.0202&0.0172& radio & \citet{dunne09} \\
1.100&0.100&0.1124&0.0262&0.0249& radio & \citet{dunne09} \\
1.300&0.100&0.1327&0.0287&0.0257& radio & \citet{dunne09} \\
1.500&0.100&0.2043&0.0369&0.0343& radio & \citet{dunne09} \\
1.700&0.100&0.1788&0.0339&0.0277& radio & \citet{dunne09} \\
1.900&0.100&0.1582&0.0256&0.0245& radio & \citet{dunne09} \\
2.250&0.250&0.1061&0.0141&0.0125& radio & \citet{dunne09} \\
2.750&0.250&0.1049&0.0150&0.0123& radio & \citet{dunne09} \\
3.250&0.250&0.0583&0.0078&0.0069& radio & \citet{dunne09} \\
4.250&0.750&0.0184&0.0032&0.0026& radio & \citet{dunne09} \\
0.150&0.150&0.0383&0.0131&0.0251& X-ray & \citet{georgakakis03,hopkins04} \\
\hline
2.750 & 0.750 & $>$0.0607 & & & UV & \citet{madau96,hopkins04}\\
4.000 & 0.500 & $>$0.0189 & & & UV & \citet{madau96,hopkins04}\\
2.750&0.750&$>$0.0290& & &UV& \citet{madau98,vanbreukelen05} \\
4.000&0.500&$>$0.0110& & &UV& \citet{madau98,vanbreukelen05} \\
3.200&0.140&$>$0.0033& & &UV& \citet{shim07} \\
2.200&0.350&$>$0.0372& & &UV& \citet{sawicki06,sawicki06b,ly09} \\
2.960&0.260&$>$0.0370& & &UV& \citet{sawicki06,sawicki06b,ly09} \\
4.130&0.260&$>$0.0161& & &UV& \citet{sawicki06,sawicki06b,ly09} \\
3.500&0.500&$>$0.0442& & &UV& \citet{paltani07,ly09} \\
2.700&0.700&$>$0.0282& & &UV& \citet{bouwens03} \\
3.850&0.450&$>$0.0166& & &UV& \citet{bouwens03} \\
4.700&0.200&$>$0.0147& & &UV& \citet{bouwens03} \\
6.000&0.200&$>$0.0360& & &UV& \citet{bouwens03b} \\
5.900&0.200&$>$0.0070& & &UV& \citet{bouwens04} \\
5.900&0.200&$>$0.0073& & &UV& \citet{bouwens06} \\
5.900&0.200&$>$0.0221& & &UV& \citet{bouwens06} \\
6.000&0.400&$>$0.0050& & &UV& \citet{bunker04} \\
3.050&0.350&$>$0.0321& & &UV& \citet{reddy08} \\
5.000&0.500&$>$0.0137& & &UV& \citet{iwata07} \\
5.900&0.300&$>$0.0003& & &UV& \citet{shimasaku05} \\
5.850&0.250&$>$0.0034& & &UV& \citet{stanway03} \\
2.259&0.053&$>$0.0054& & &Ly$\alpha$& \citet{nilsson09} \\
2.379&0.023&$>$0.0024& & &Ly$\alpha$& \citet{palunas04} \\
3.110&0.020&$>$0.0120& & &Ly$\alpha$& \citet{gronwall07} \\
3.156&0.025&$>$0.0130& & &Ly$\alpha$& \citet{nilsson07} \\
3.135&0.045&$>$0.0043& & &Ly$\alpha$& \citet{ouchi08} \\
3.140&0.040&$>$0.0300& & &Ly$\alpha$& \citet{kudritzki00} \\
3.400&0.030&$>$0.0060& & &Ly$\alpha$& \citet{hu98} \\
3.438&0.033&$>$0.0100& & &Ly$\alpha$& \citet{cowiehu98} \\
3.463&0.982&$>$0.0220& & &Ly$\alpha$& \citet{vanbreukelen05} \\
3.690&0.060&$>$0.0021& & &Ly$\alpha$& \citet{ouchi08} \\
3.700&0.220&$>$0.0004& & &Ly$\alpha$& \citet{fujita03,vanbreukelen05} \\
4.500&0.064&$>$0.0100& & &Ly$\alpha$& \citet{hu98} \\
4.860&0.030&$>$0.0063& & &Ly$\alpha$& \citet{ouchi03,vanbreukelen05} \\
5.690&0.090&$>$0.0032& & &Ly$\alpha$& \citet{ouchi08} \\
5.700&0.100&$>$0.0012& & &Ly$\alpha$& \citet{ajiki03} \\
5.700&0.050&$>$0.0018& & &Ly$\alpha$& \citet{malhotra04} \\
5.700&0.050&$>$0.0023& & &Ly$\alpha$& \citet{shimasaku06} \\
5.700&0.050&$>$0.0007& & &Ly$\alpha$& \citet{murayama07} \\
5.735&0.062&$>$0.0005& & &Ly$\alpha$& \citet{rhoads03} \\
6.500&0.050&$>$0.0036& & &Ly$\alpha$& \citet{malhotra04} \\
6.550&0.050&$>$0.0006& & &Ly$\alpha$& \citet{taniguchi05} \\
6.578&0.002&$>$0.0005& & &Ly$\alpha$& \citet{kodaira03} \\
3.000 & 1.000 & $>$0.0818 & & & submm & \citet{hughes98,hopkins04}\\
\hline
\end{longtable}
\begin{table}[b]
\tablecomments{Lower limits indicate value not corrected for extinction. The data with double reference were taken directly form the compilation given in the second reference. This table is available in a machine-readable form in the electronic edition of the Journal.}
\end{table}
}

\longtab{5}{
\begin{longtable}{cccccll}
\caption{Compilation of stellar mass density determinations in $\log M_\odot\,\mbox{Mpc}^{-3}$ \label{tab:mstard}   }
\\
\hline\hline
 $z$ & $\Delta z$ & $\rho_{*}$ & $-\Delta\rho_{*}$ & $+\Delta\rho_{*}$  & Reference \\
\hline
\endfirsthead
\caption{continued.}\\
\hline\hline
 $z$ & $\Delta z$ & $\rho_{*}$ & $-\Delta\rho_{*}$ & $+\Delta\rho_{*}$  & Reference \\
\hline
\endhead
\hline
\endfoot
0.698 & 0.618 & 7.12 &0.08 &0.13& This work \\
1.775 & 0.367 & 7.18 &0.11 &0.16& This work \\
2.357 & 0.209 & 7.61 &0.08 &0.12& This work \\
3.101 & 0.522 & 7.28 &0.07 &0.12& This work \\
0.100&0.100&8.48&0.10&0.10& \citet{borch06} \\
0.300&0.100&8.34&0.15&0.15& \citet{borch06} \\
0.500&0.100&8.32&0.11&0.11& \citet{borch06} \\
0.700&0.100&8.33&0.10&0.10& \citet{borch06} \\
0.900&0.100&8.17&0.18&0.18& \citet{borch06} \\
0.100&0.100&8.49&0.05&0.04& \citet{rudnick03} \\
1.120&0.480&8.14&0.10&0.11& \citet{rudnick03} \\
2.010&0.400&7.48&0.16&0.12& \citet{rudnick03} \\
2.800&0.400&7.49&0.14&0.12& \citet{rudnick03} \\
0.950&0.450&8.46&0.07&0.07& \citet{dickinson03} \\
1.700&0.300&8.06&0.13&0.17& \citet{dickinson03} \\
2.250&0.250&7.58&0.07&0.11& \citet{dickinson03} \\
2.750&0.250&7.52&0.14&0.23& \citet{dickinson03} \\
0.950&0.450&8.61&0.07&0.07& \citet{dickinson03} \\
1.700&0.300&8.22&0.12&0.16& \citet{dickinson03} \\
2.250&0.250&8.01&0.08&0.09& \citet{dickinson03} \\
2.750&0.250&7.89&0.15&0.20& \citet{dickinson03} \\
0.950&0.450&8.52&0.08&0.07& \citet{dickinson03} \\
1.700&0.300&7.97&0.17&0.17& \citet{dickinson03} \\
2.250&0.250&7.36&0.08&0.11& \citet{dickinson03} \\
2.750&0.250&7.27&0.18&0.27& \citet{dickinson03} \\
0.375&0.125&8.65&0.17&0.12& \citet{cohen02} \\
0.650&0.150&8.65&0.01&0.08& \citet{cohen02} \\
0.925&0.125&8.62&0.09&0.08& \citet{cohen02} \\
0.500&0.100&8.83&0.04&0.04& \citet{drory04,drory05} \\
0.700&0.100&8.76&0.04&0.04& \citet{drory04,drory05} \\
0.900&0.100&8.60&0.04&0.04& \citet{drory04,drory05} \\
1.100&0.100&8.55&0.04&0.04& \citet{drory04,drory05} \\
0.500&0.250&8.50&0.27&0.27& \citet{drory04,drory05} \\
0.500&0.250&8.51&0.17&0.17& \citet{drory04,drory05} \\
1.000&0.250&8.42&0.12&0.12& \citet{drory04,drory05} \\
1.000&0.250&8.29&0.19&0.19& \citet{drory04,drory05} \\
1.500&0.250&8.38&0.16&0.16& \citet{drory04,drory05} \\
1.500&0.250&8.03&0.16&0.16& \citet{drory04,drory05} \\
2.000&0.250&8.09&0.19&0.19& \citet{drory04,drory05} \\
2.000&0.250&8.04&0.20&0.20& \citet{drory04,drory05} \\
2.625&0.375&8.15&0.19&0.19& \citet{drory04,drory05} \\
2.625&0.375&7.78&0.20&0.20& \citet{drory04,drory05} \\
3.500&0.500&7.92&0.17&0.17& \citet{drory04,drory05} \\
3.500&0.500&7.68&0.20&0.20& \citet{drory04,drory05} \\
4.500&0.500&7.37&0.26&0.26& \citet{drory04,drory05} \\
4.500&0.500&7.43&0.20&0.20& \citet{drory04,drory05} \\
0.900&0.100&8.18&0.13&0.10& \citet{glazebrook04} \\
1.200&0.100&7.82&0.13&0.01& \citet{glazebrook04} \\
1.450&0.150&8.08&0.09&0.08& \citet{glazebrook04} \\
1.800&0.200&7.69&0.13&0.11& \citet{glazebrook04} \\
0.500&0.100&8.32&0.03&0.03& \citet{fontana03,fontana04,fontana06} \\
0.700&0.100&8.53&0.02&0.02& \citet{fontana03,fontana04,fontana06} \\
0.900&0.100&8.16&0.03&0.03& \citet{fontana03,fontana04,fontana06} \\
1.150&0.150&8.26&0.02&0.02& \citet{fontana03,fontana04,fontana06} \\
1.450&0.150&7.96&0.03&0.03& \citet{fontana03,fontana04,fontana06} \\
1.800&0.200&7.90&0.04&0.04& \citet{fontana03,fontana04,fontana06} \\
2.500&0.500&7.60&0.04&0.04& \citet{fontana03,fontana04,fontana06} \\
3.500&0.500&7.23&0.12&0.12& \citet{fontana03,fontana04,fontana06} \\
4.500&0.500&7.73&0.12&0.12& \citet{fontana03,fontana04,fontana06} \\
5.500&0.500&7.84&0.12&0.12& \citet{fontana03,fontana04,fontana06} \\
0.450&0.250&8.51&0.04&0.24& \citet{fontana03,fontana04,fontana06} \\
0.850&0.150&8.44&0.05&0.20& \citet{fontana03,fontana04,fontana06} \\
1.250&0.250&8.19&0.11&0.13& \citet{fontana03,fontana04,fontana06} \\
1.750&0.250&7.86&0.24&0.24& \citet{fontana03,fontana04,fontana06} \\
2.250&0.250&7.65&0.24&0.24& \citet{fontana03,fontana04,fontana06} \\
0.500&0.250&8.64&0.17&0.24& \citet{fontana03,fontana04,fontana06} \\
1.000&0.250&8.29&0.31&0.35& \citet{fontana03,fontana04,fontana06} \\
1.625&0.375&7.87&0.28&0.35& \citet{fontana03,fontana04,fontana06} \\
2.250&0.250&7.92&0.42&0.26& \citet{fontana03,fontana04,fontana06} \\
2.850&0.350&7.90&0.20&0.38& \citet{fontana03,fontana04,fontana06} \\
0.100&0.100&8.75&0.07&0.06& \citet{cole01} \\
0.350&0.150&8.53&0.13&0.09& \citet{brinchmannellis00} \\
0.625&0.125&8.56&0.05&0.06& \citet{brinchmannellis00} \\
0.875&0.125&8.48&0.04&0.06& \citet{brinchmannellis00} \\
0.100&0.100&8.75&0.12&0.12& \citet{perezgonzalez08} \\
0.300&0.100&8.61&0.06&0.06& \citet{perezgonzalez08} \\
0.500&0.100&8.57&0.04&0.04& \citet{perezgonzalez08} \\
0.700&0.100&8.52&0.05&0.05& \citet{perezgonzalez08} \\
0.900&0.100&8.44&0.05&0.05& \citet{perezgonzalez08} \\
1.150&0.150&8.35&0.05&0.05& \citet{perezgonzalez08} \\
1.450&0.150&8.18&0.07&0.07& \citet{perezgonzalez08} \\
1.800&0.200&8.02&0.07&0.07& \citet{perezgonzalez08} \\
2.250&0.250&7.87&0.09&0.09& \citet{perezgonzalez08} \\
2.750&0.250&7.76&0.18&0.18& \citet{perezgonzalez08} \\
3.250&0.250&7.63&0.14&0.14& \citet{perezgonzalez08} \\
3.750&0.250&7.49&0.13&0.13& \citet{perezgonzalez08} \\
0.150&0.150&8.76&0.13&0.11& \citet{salucci99} \\
0.097&0.084&8.73&0.07&0.06& \citet{driver07} \\
0.150&0.150&8.72&0.01&0.01& \citet{bell03b} \\
0.950&0.450&8.45&0.07&0.06& \citet{conselice05b} \\
1.700&0.300&7.74&0.22&0.15& \citet{conselice05b} \\
2.250&0.250&7.36&0.40&0.21& \citet{conselice05b} \\
2.750&0.250&7.30&0.92&0.27& \citet{conselice05b} \\
0.500&0.250&8.57&0.03&0.03& \citet{elsner08} \\
1.000&0.250&8.37&0.02&0.02& \citet{elsner08} \\
1.500&0.250&8.22&0.03&0.03& \citet{elsner08} \\
2.000&0.250&8.10&0.04&0.04& \citet{elsner08} \\
2.500&0.250&7.93&0.04&0.04& \citet{elsner08} \\
3.500&0.500&7.59&0.05&0.05& \citet{elsner08} \\
4.500&0.500&6.90&0.08&0.08& \citet{elsner08} \\
3.960&0.290&7.01&0.06&0.05& \citet{stark09} \\
4.790&0.250&6.63&0.07&0.06& \citet{stark09} \\
6.010&0.250&6.29&0.09&0.07& \citet{stark09} \\
5.000&0.600&6.78&0.08&0.22& \citet{stark07} \\
6.000&0.300&6.40&0.00&0.51& \citet{eyles07} \\
6.000&0.500&6.59&0.55&0.24& \citet{yan06} \\
0.100&0.100&8.59&0.04&0.04& \citet{rudnick06} \\
0.500&0.500&8.05&0.03&0.07& \citet{rudnick06} \\
1.300&0.300&7.87&0.04&0.07& \citet{rudnick06} \\
2.000&0.400&7.76&0.06&0.06& \citet{rudnick06} \\
2.800&0.400&7.59&0.11&0.06& \citet{rudnick06} \\
0.100&0.100&8.51&0.07&0.07& \citet{marchesini09} \\
1.650&0.350&7.91&0.15&0.02& \citet{marchesini09} \\
2.500&0.500&7.55&0.18&0.12& \citet{marchesini09} \\
3.500&0.500&7.27&0.39&0.93& \citet{marchesini09} \\
0.550&0.150&8.31&0.07&0.07& \citet{bundy06} \\
0.875&0.125&8.30&0.10&0.10& \citet{bundy06} \\
1.200&0.200&8.15&0.10&0.10& \citet{bundy06} \\
0.300&0.100&8.46&0.03&0.03& \citet{ilbert09} \\
0.500&0.100&8.22&0.02&0.02& \citet{ilbert09} \\
0.700&0.100&8.25&0.02&0.02& \citet{ilbert09} \\
0.900&0.100&8.32&0.01&0.01& \citet{ilbert09} \\
1.100&0.100&8.09&0.02&0.02& \citet{ilbert09} \\
1.350&0.150&7.93&0.01&0.01& \citet{ilbert09} \\
1.750&0.250&7.72&0.09&0.14& \citet{ilbert09} \\
0.300&0.100&8.79&0.17&0.15& \citet{arnouts07} \\
0.500&0.100&8.63&0.11&0.12& \citet{arnouts07} \\
0.700&0.100&8.63&0.10&0.10& \citet{arnouts07} \\
0.900&0.100&8.73&0.13&0.13& \citet{arnouts07} \\
1.100&0.100&8.54&0.11&0.11& \citet{arnouts07} \\
1.350&0.150&8.43&0.12&0.12& \citet{arnouts07} \\
1.750&0.250&8.17&0.12&0.12& \citet{arnouts07} \\
0.325&0.225&8.66&0.10&0.10& \citet{franceschini06} \\
0.725&0.175&8.61&0.10&0.10& \citet{franceschini06} \\
1.150&0.250&8.45&0.10&0.10& \citet{franceschini06} \\
0.225&0.175&8.45&0.01&0.01& \citet{pozzetti07} \\
0.550&0.150&8.34&0.02&0.02& \citet{pozzetti07} \\
0.800&0.100&8.22&0.01&0.01& \citet{pozzetti07} \\
1.050&0.150&8.14&0.01&0.01& \citet{pozzetti07} \\
1.400&0.200&8.04&0.02&0.02& \citet{pozzetti07} \\
2.050&0.450&8.05&0.01&0.01& \citet{pozzetti07} \\
0.025&0.025&8.81&0.05&0.05& \citet{kochanek01} \\
0.100&0.100&8.72&0.03&0.03& \citet{driver06} \\
1.100&0.400&8.47&0.11&0.11& \citet{gwyn05} \\
1.750&0.250&8.38&0.21&0.21& \citet{gwyn05} \\
2.500&0.500&8.21&0.14&0.14& \citet{gwyn05} \\
4.500&1.500&7.93&0.11&0.11& \citet{gwyn05} \\
1.250&0.250&8.37&0.07&0.06& \citet{caputi06} \\
1.750&0.250&8.12&0.07&0.06& \citet{caputi06} \\
\hline
\end{longtable}
\begin{table}[b]
\tablecomments{This table is available in a machine-readable form in the electronic edition of the Journal.}
\end{table}
}


\begin{thebibliography}{270}
\expandafter\ifx\csname natexlab\endcsname\relax\def\natexlab#1{#1}\fi
\expandafter\ifx\csname url\endcsname\relax
  \def\url#1{{\tt #1}}\fi
\expandafter\ifx\csname urlprefix\endcsname\relax\def\urlprefix{URL }\fi

\bibitem[{{Ajiki} et~al.(2003){Ajiki}, {Taniguchi}, {Fujita} et~al.}]{ajiki03}
{Ajiki} M., et~al., 2003, \aj, 126, 2091

\bibitem[{{Alexander} et~al.(2005){Alexander}, {Bauer}, {Chapman}
  et~al.}]{alexander05}
{Alexander} D.M., {Bauer} F.E., {Chapman} S.C., {Smail} I., {Blain} A.W.,
  {Brandt} W.N., {Ivison} R.J., 2005, \apj, 632, 736

\bibitem[{{Alexander} et~al.(2008){Alexander}, {Brandt}, {Smail}
  et~al.}]{alexander08}
{Alexander} D.M., et~al., 2008, \aj, 135, 1968

\bibitem[{{Almaini} et~al.(2003){Almaini}, {Scott}, {Dunlop}
  et~al.}]{almaini03}
{Almaini} O., et~al., 2003, \mnras, 338, 303

\bibitem[{{Appleton} et~al.(2004){Appleton}, {Fadda}, {Marleau}
  et~al.}]{appleton04}
{Appleton} P.N., et~al., 2004, \apjs, 154, 147

\bibitem[{{Aravena} et~al.(2010){Aravena}, {Bertoldi}, {Carilli}
  et~al.}]{aravena10}
{Aravena} M., et~al., 2010, \apjl, 708, L36

\bibitem[{{Arnouts} et~al.(2007){Arnouts}, {Walcher}, {Le F{\`e}vre}
  et~al.}]{arnouts07}
{Arnouts} S., et~al., 2007, \aap, 476, 137

\bibitem[{{Ashby} et~al.(2006){Ashby}, {Dye}, {Huang} et~al.}]{ashby06}
{Ashby} M.L.N., et~al., 2006, \apj, 644, 778

\bibitem[{{Austermann} et~al.(2009){Austermann}, {Aretxaga}, {Hughes}
  et~al.}]{austermann09}
{Austermann} J.E., et~al., 2009, \mnras, 393, 1573

\bibitem[{{Barger} et~al.(1999){Barger}, {Cowie}, \& {Sanders}}]{barger99}
{Barger} A.J., {Cowie} L.L., {Sanders} D.B., 1999, \apjl, 518, L5

\bibitem[{{Barger} et~al.(2000){Barger}, {Cowie}, \& {Richards}}]{barger00}
{Barger} A.J., {Cowie} L.L., {Richards} E.A., 2000, \aj, 119, 2092

\bibitem[{{Bell}(2003)}]{bell03}
{Bell} E.F., 2003, \apj, 586, 794

\bibitem[{{Bell} et~al.(2003){Bell}, {McIntosh}, {Katz}, \&
  {Weinberg}}]{bell03b}
{Bell} E.F., {McIntosh} D.H., {Katz} N., {Weinberg} M.D., 2003, \apjs, 149, 289

\bibitem[{{Bell} et~al.(2007){Bell}, {Zheng}, {Papovich} et~al.}]{bell07}
{Bell} E.F., {Zheng} X.Z., {Papovich} C., {Borch} A., {Wolf} C., {Meisenheimer}
  K., 2007, \apj, 663, 834

\bibitem[{{Berciano Alba} et~al.(2010){Berciano Alba}, {Koopmans}, {Garrett},
  {Wucknitz}, \& {Limousin}}]{bercianoalba10}
{Berciano Alba} A., {Koopmans} L.V.E., {Garrett} M.A., {Wucknitz} O.,
  {Limousin} M., 2010, \aap, 509, 54

\bibitem[{{Beswick} et~al.(2008){Beswick}, {Muxlow}, {Thrall}, {Richards}, \&
  {Garrington}}]{beswick08}
{Beswick} R.J., {Muxlow} T.W.B., {Thrall} H., {Richards} A.M.S., {Garrington}
  S.T., 2008, \mnras, 385, 1143

\bibitem[{{Blain}(1997)}]{blain97}
{Blain} A.W., 1997, \mnras, 290, 553

\bibitem[{{Blain} \& {Longair}(1996)}]{blain96}
{Blain} A.W., {Longair} M.S., 1996, \mnras, 279, 847

\bibitem[{{Blain} et~al.(1999){Blain}, {Kneib}, {Ivison}, \& {Smail}}]{blain99}
{Blain} A.W., {Kneib} J., {Ivison} R.J., {Smail} I., 1999, \apjl, 512, L87

\bibitem[{{Blain} et~al.(2002){Blain}, {Smail}, {Ivison}, {Kneib}, \&
  {Frayer}}]{blain02}
{Blain} A.W., {Smail} I., {Ivison} R.J., {Kneib} J.P., {Frayer} D.T., 2002,
  \physrep, 369, 111

\bibitem[{{Blain} et~al.(2004){Blain}, {Chapman}, {Smail}, \&
  {Ivison}}]{blainT}
{Blain} A.W., {Chapman} S.C., {Smail} I., {Ivison} R., 2004, \apj, 611, 52

\bibitem[{{Borch} et~al.(2006){Borch}, {Meisenheimer}, {Bell} et~al.}]{borch06}
{Borch} A., et~al., 2006, \aap, 453, 869

\bibitem[{{Borys} et~al.(2005){Borys}, {Smail}, {Chapman} et~al.}]{borys05}
{Borys} C., {Smail} I., {Chapman} S.C., {Blain} A.W., {Alexander} D.M.,
  {Ivison} R.J., 2005, \apj, 635, 853

\bibitem[{{Bouwens} et~al.(2003{\natexlab{a}}){Bouwens}, {Broadhurst}, \&
  {Illingworth}}]{bouwens03}
{Bouwens} R., {Broadhurst} T., {Illingworth} G., 2003{\natexlab{a}}, \apj, 593,
  640

\bibitem[{{Bouwens} et~al.(2003{\natexlab{b}}){Bouwens}, {Illingworth},
  {Rosati} et~al.}]{bouwens03b}
{Bouwens} R.J., et~al., 2003{\natexlab{b}}, \apj, 595, 589

\bibitem[{{Bouwens} et~al.(2004){Bouwens}, {Illingworth}, {Thompson}
  et~al.}]{bouwens04}
{Bouwens} R.J., et~al., 2004, \apjl, 606, L25

\bibitem[{{Bouwens} et~al.(2006){Bouwens}, {Illingworth}, {Blakeslee}, \&
  {Franx}}]{bouwens06}
{Bouwens} R.J., {Illingworth} G.D., {Blakeslee} J.P., {Franx} M., 2006, \apj,
  653, 53

\bibitem[{{Bouwens} et~al.(2007){Bouwens}, {Illingworth}, {Franx}, \&
  {Ford}}]{bouwens07}
{Bouwens} R.J., {Illingworth} G.D., {Franx} M., {Ford} H., 2007, \apj, 670, 928

\bibitem[{{Boyle} et~al.(2007){Boyle}, {Cornwell}, {Middelberg}
  et~al.}]{boyle07}
{Boyle} B.J., {Cornwell} T.J., {Middelberg} E., {Norris} R.P., {Appleton} P.N.,
  {Smail} I., 2007, \mnras, 376, 1182

\bibitem[{{Brinchmann} \& {Ellis}(2000)}]{brinchmannellis00}
{Brinchmann} J., {Ellis} R.S., 2000, \apjl, 536, L77

\bibitem[{{Brinchmann} et~al.(2004){Brinchmann}, {Charlot}, {White}
  et~al.}]{brinchmann04}
{Brinchmann} J., {Charlot} S., {White} S.D.M., {Tremonti} C., {Kauffmann} G.,
  {Heckman} T., {Brinkmann} J., 2004, \mnras, 351, 1151

\bibitem[{{Bundy} et~al.(2006){Bundy}, {Ellis}, {Conselice} et~al.}]{bundy06}
{Bundy} K., et~al., 2006, \apj, 651, 120

\bibitem[{{Bunker} et~al.(2004){Bunker}, {Stanway}, {Ellis}, \&
  {McMahon}}]{bunker04}
{Bunker} A.J., {Stanway} E.R., {Ellis} R.S., {McMahon} R.G., 2004, \mnras, 355,
  374

\bibitem[{{Capak} et~al.(2004){Capak}, {Cowie}, {Hu} et~al.}]{capak04}
{Capak} P., et~al., 2004, \aj, 127, 180

\bibitem[{{Capak} et~al.(2008){Capak}, {Carilli}, {Lee} et~al.}]{capak08}
{Capak} P., et~al., 2008, \apjl, 681, L53

\bibitem[{{Caputi} et~al.(2006){Caputi}, {McLure}, {Dunlop}, {Cirasuolo}, \&
  {Schael}}]{caputi06}
{Caputi} K.I., {McLure} R.J., {Dunlop} J.S., {Cirasuolo} M., {Schael} A.M.,
  2006, \mnras, 366, 609

\bibitem[{{Caputi} et~al.(2007){Caputi}, {Lagache}, {Yan} et~al.}]{caputi07}
{Caputi} K.I., et~al., 2007, \apj, 660, 97

\bibitem[{{Castro Cer{\'o}n} et~al.(2006){Castro Cer{\'o}n}, Micha{\l}owski,
  Hjorth et~al.}]{castroceron06}
{Castro Cer{\'o}n} J.M., Micha{\l}owski M., Hjorth J., Watson D.J., Fynbo
  J.P.U., Gorosabel J., 2006, \apj L, 653, L85

\bibitem[{{Castro Cer{\'o}n} et~al.(2009){Castro Cer{\'o}n}, {Micha{\l}owski},
  {Hjorth} et~al.}]{castroceron09}
{Castro Cer{\'o}n} J.M., {Micha{\l}owski} M.J., {Hjorth} J., {Malesani} D.,
  {Gorosabel} J., {Watson} D., {Fynbo} J.P.U., 2009, \apj, submitted, {\tt
  arXiv:0803.2235v1 [astro-ph]}

\bibitem[{{Chapman} et~al.(2001){Chapman}, {Richards}, {Lewis}, {Wilson}, \&
  {Barger}}]{chapman01}
{Chapman} S.C., {Richards} E.A., {Lewis} G.F., {Wilson} G., {Barger} A.J.,
  2001, \apjl, 548, L147

\bibitem[{{Chapman} et~al.(2003{\natexlab{a}}){Chapman}, {Barger}, {Cowie}
  et~al.}]{chapman03c}
{Chapman} S.C., et~al., 2003{\natexlab{a}}, \apj, 585, 57

\bibitem[{{Chapman} et~al.(2003{\natexlab{b}}){Chapman}, {Windhorst},
  {Odewahn}, {Yan}, \& {Conselice}}]{chapman03b}
{Chapman} S.C., {Windhorst} R., {Odewahn} S., {Yan} H., {Conselice} C.,
  2003{\natexlab{b}}, \apj, 599, 92

\bibitem[{{Chapman} et~al.(2004){Chapman}, {Smail}, {Windhorst}, {Muxlow}, \&
  {Ivison}}]{chapman04b}
{Chapman} S.C., {Smail} I., {Windhorst} R., {Muxlow} T., {Ivison} R.J., 2004,
  \apj, 611, 732

\bibitem[{{Chapman} et~al.(2005){Chapman}, {Blain}, {Smail}, \&
  {Ivison}}]{chapman05}
{Chapman} S.C., {Blain} A.W., {Smail} I., {Ivison} R.J., 2005, \apj, 622, 772

\bibitem[{{Clements} et~al.(2004){Clements}, {Eales}, {Wojciechowski}
  et~al.}]{clements04}
{Clements} D., et~al., 2004, \mnras, 351, 447

\bibitem[{{Clements} et~al.(2008){Clements}, {Vaccari}, {Babbedge}
  et~al.}]{clements08}
{Clements} D.L., et~al., 2008, \mnras, 387, 247

\bibitem[{{Cohen}(2002)}]{cohen02}
{Cohen} J.G., 2002, \apj, 567, 672

\bibitem[{{Cole} et~al.(2001){Cole}, {Norberg}, {Baugh} et~al.}]{cole01}
{Cole} S., et~al., 2001, \mnras, 326, 255

\bibitem[{{Condon}(1989)}]{condon89}
{Condon} J.J., 1989, \apj, 338, 13

\bibitem[{{Condon}(1992)}]{condon}
{Condon} J.J., 1992, \araa, 30, 575

\bibitem[{{Condon} et~al.(2002){Condon}, {Cotton}, \& {Broderick}}]{condon02}
{Condon} J.J., {Cotton} W.D., {Broderick} J.J., 2002, \aj, 124, 675

\bibitem[{{Connolly} et~al.(1997){Connolly}, {Szalay}, {Dickinson}, {Subbarao},
  \& {Brunner}}]{connolly97}
{Connolly} A.J., {Szalay} A.S., {Dickinson} M., {Subbarao} M.U., {Brunner}
  R.J., 1997, \apjl, 486, L11

\bibitem[{{Conselice} et~al.(2005){Conselice}, {Blackburne}, \&
  {Papovich}}]{conselice05b}
{Conselice} C.J., {Blackburne} J.A., {Papovich} C., 2005, \apj, 620, 564

\bibitem[{{Coppin} et~al.(2006){Coppin}, {Chapin}, {Mortier} et~al.}]{coppin06}
{Coppin} K., et~al., 2006, \mnras, 372, 1621

\bibitem[{{Coppin} et~al.(2008){Coppin}, {Halpern}, {Scott} et~al.}]{coppin08}
{Coppin} K., et~al., 2008, \mnras, 384, 1597

\bibitem[{{Coppin} et~al.(2009){Coppin}, {Smail}, {Alexander}
  et~al.}]{coppin09}
{Coppin} K.E.K., et~al., 2009, \mnras, 395, 1905

\bibitem[{{Courty} et~al.(2007){Courty}, {Bj{\"o}rnsson}, \&
  {Gudmundsson}}]{courty07}
{Courty} S., {Bj{\"o}rnsson} G., {Gudmundsson} E.H., 2007, \mnras, 376, 1375

\bibitem[{{Cowie} \& {Hu}(1998)}]{cowiehu98}
{Cowie} L.L., {Hu} E.M., 1998, \aj, 115, 1319

\bibitem[{{Cowie} et~al.(1996){Cowie}, {Songaila}, {Hu}, \& {Cohen}}]{cowie96}
{Cowie} L.L., {Songaila} A., {Hu} E.M., {Cohen} J.G., 1996, \aj, 112, 839

\bibitem[{{Cowie} et~al.(1999){Cowie}, {Songaila}, \& {Barger}}]{cowie99}
{Cowie} L.L., {Songaila} A., {Barger} A.J., 1999, \aj, 118, 603

\bibitem[{{Cowie} et~al.(2002){Cowie}, {Barger}, \& {Kneib}}]{cowie02}
{Cowie} L.L., {Barger} A.J., {Kneib} J., 2002, \aj, 123, 2197

\bibitem[{{Daddi} et~al.(2000){Daddi}, {Cimatti}, \& {Renzini}}]{daddi00}
{Daddi} E., {Cimatti} A., {Renzini} A., 2000, \aap, 362, L45

\bibitem[{{Daddi} et~al.(2009{\natexlab{a}}){Daddi}, {Dannerbauer}, {Krips}
  et~al.}]{daddi09b}
{Daddi} E., {Dannerbauer} H., {Krips} M., {Walter} F., {Dickinson} M., {Elbaz}
  D., {Morrison} G.E., 2009{\natexlab{a}}, \apjl, 695, L176

\bibitem[{{Daddi} et~al.(2009{\natexlab{b}}){Daddi}, {Dannerbauer}, {Stern}
  et~al.}]{daddi09}
{Daddi} E., et~al., 2009{\natexlab{b}}, \apj, 694, 1517

\bibitem[{{Dahlen} et~al.(2007){Dahlen}, {Mobasher}, {Dickinson}
  et~al.}]{dahlen07}
{Dahlen} T., {Mobasher} B., {Dickinson} M., {Ferguson} H.C., {Giavalisco} M.,
  {Kretchmer} C., {Ravindranath} S., 2007, \apj, 654, 172

\bibitem[{{Dav{\'e}} et~al.(2009){Dav{\'e}}, {Finlator}, {Oppenheimer}
  et~al.}]{dave09}
{Dav{\'e}} R., {Finlator} K., {Oppenheimer} B.D., {Fardal} M., {Katz} N.,
  {Kere{\v s}} D., {Weinberg} D.H., 2009, \mnras, submitted, {\tt
  arXiv:0909.4078 [astro-ph.CO]}

\bibitem[{{de Ravel} et~al.(2009){de Ravel}, {Le F{\`e}vre}, {Tresse}
  et~al.}]{deravel09}
{de Ravel} L., et~al., 2009, \aap, 498, 379

\bibitem[{{de Vries} et~al.(2007){de Vries}, {Hodge}, {Becker}, {White}, \&
  {Helfand}}]{devries07}
{de Vries} W.H., {Hodge} J.A., {Becker} R.H., {White} R.L., {Helfand} D.J.,
  2007, \aj, 134, 457

\bibitem[{{Devlin} et~al.(2009){Devlin}, {Ade}, {Aretxaga} et~al.}]{devlin09}
{Devlin} M.J., et~al., 2009, \nat, 458, 737

\bibitem[{{Dickinson} et~al.(2003){Dickinson}, {Papovich}, {Ferguson}, \&
  {Budav{\'a}ri}}]{dickinson03}
{Dickinson} M., {Papovich} C., {Ferguson} H.C., {Budav{\'a}ri} T., 2003, \apj,
  587, 25

\bibitem[{{Driver} et~al.(2006){Driver}, {Allen}, {Graham} et~al.}]{driver06}
{Driver} S.P., et~al., 2006, \mnras, 368, 414

\bibitem[{{Driver} et~al.(2007){Driver}, {Popescu}, {Tuffs} et~al.}]{driver07}
{Driver} S.P., {Popescu} C.C., {Tuffs} R.J., {Liske} J., {Graham} A.W., {Allen}
  P.D., {de Propris} R., 2007, \mnras, 379, 1022

\bibitem[{{Drory} et~al.(2004){Drory}, {Bender}, {Feulner} et~al.}]{drory04}
{Drory} N., {Bender} R., {Feulner} G., {Hopp} U., {Maraston} C., {Snigula} J.,
  {Hill} G.J., 2004, \apj, 608, 742

\bibitem[{{Drory} et~al.(2005){Drory}, {Salvato}, {Gabasch} et~al.}]{drory05}
{Drory} N., {Salvato} M., {Gabasch} A., {Bender} R., {Hopp} U., {Feulner} G.,
  {Pannella} M., 2005, \apjl, 619, L131

\bibitem[{{Dunne} et~al.(2009){Dunne}, {Ivison}, {Maddox} et~al.}]{dunne09}
{Dunne} L., et~al., 2009, \mnras, 394, 3

\bibitem[{{Dye} et~al.(2008){Dye}, {Eales}, {Aretxaga} et~al.}]{dye08}
{Dye} S., et~al., 2008, \mnras, 386, 1107

\bibitem[{{Dye} et~al.(2009){Dye}, {Ade}, {Bock} et~al.}]{dye09}
{Dye} S., et~al., 2009, \apj, 703, 285

\bibitem[{{Eales} et~al.(1999){Eales}, {Lilly}, {Gear} et~al.}]{eales99}
{Eales} S., {Lilly} S., {Gear} W., {Dunne} L., {Bond} J.R., {Hammer} F., {Le
  F{\`e}vre} O., {Crampton} D., 1999, \apj, 515, 518

\bibitem[{{Eales} et~al.(2009){Eales}, {Chapin}, {Devlin} et~al.}]{eales09}
{Eales} S., et~al., 2009, \apj, 707, 1779

\bibitem[{{Egami} et~al.(2004){Egami}, {Dole}, {Huang} et~al.}]{egami04}
{Egami} E., et~al., 2004, \apjs, 154, 130

\bibitem[{{Elsner} et~al.(2008){Elsner}, {Feulner}, \& {Hopp}}]{elsner08}
{Elsner} F., {Feulner} G., {Hopp} U., 2008, \aap, 477, 503

\bibitem[{{Erb} et~al.(2006){Erb}, {Steidel}, {Shapley} et~al.}]{erb06}
{Erb} D.K., {Steidel} C.C., {Shapley} A.E., {Pettini} M., {Reddy} N.A.,
  {Adelberger} K.L., 2006, \apj, 646, 107

\bibitem[{{Eyles} et~al.(2007){Eyles}, {Bunker}, {Ellis} et~al.}]{eyles07}
{Eyles} L.P., {Bunker} A.J., {Ellis} R.S., {Lacy} M., {Stanway} E.R., {Stark}
  D.P., {Chiu} K., 2007, \mnras, 374, 910

\bibitem[{{Flores} et~al.(1999){Flores}, {Hammer}, {Thuan} et~al.}]{flores99}
{Flores} H., et~al., 1999, \apj, 517, 148

\bibitem[{{Fomalont} et~al.(2006){Fomalont}, {Kellermann}, {Cowie}
  et~al.}]{fomalont06}
{Fomalont} E.B., {Kellermann} K.I., {Cowie} L.L., {Capak} P., {Barger} A.J.,
  {Partridge} R.B., {Windhorst} R.A., {Richards} E.A., 2006, \apjs, 167, 103

\bibitem[{{Fontana} et~al.(2003){Fontana}, {Donnarumma}, {Vanzella}
  et~al.}]{fontana03}
{Fontana} A., et~al., 2003, \apjl, 594, L9

\bibitem[{{Fontana} et~al.(2004){Fontana}, {Pozzetti}, {Donnarumma}
  et~al.}]{fontana04}
{Fontana} A., et~al., 2004, \aap, 424, 23

\bibitem[{{Fontana} et~al.(2006){Fontana}, {Salimbeni}, {Grazian}
  et~al.}]{fontana06}
{Fontana} A., et~al., 2006, \aap, 459, 745

\bibitem[{{Franceschini} et~al.(2006){Franceschini}, {Rodighiero}, {Cassata}
  et~al.}]{franceschini06}
{Franceschini} A., et~al., 2006, \aap, 453, 397

\bibitem[{{Frayer} et~al.(2004){Frayer}, {Chapman}, {Yan} et~al.}]{frayer04}
{Frayer} D.T., et~al., 2004, \apjs, 154, 137

\bibitem[{{Fujita} et~al.(2003{\natexlab{a}}){Fujita}, {Ajiki}, {Shioya}
  et~al.}]{fujita03}
{Fujita} S.S., et~al., 2003{\natexlab{a}}, \aj, 125, 13

\bibitem[{{Fujita} et~al.(2003{\natexlab{b}}){Fujita}, {Ajiki}, {Shioya}
  et~al.}]{fujita03b}
{Fujita} S.S., et~al., 2003{\natexlab{b}}, \apjl, 586, L115

\bibitem[{{Gallego} et~al.(1995){Gallego}, {Zamorano}, {Aragon-Salamanca}, \&
  {Rego}}]{gallego95}
{Gallego} J., {Zamorano} J., {Aragon-Salamanca} A., {Rego} M., 1995, \apjl,
  455, L1

\bibitem[{{Gallego} et~al.(2002){Gallego}, {Garc{\'{\i}}a-Dab{\'o}},
  {Zamorano}, {Arag{\'o}n-Salamanca}, \& {Rego}}]{gallego02}
{Gallego} J., {Garc{\'{\i}}a-Dab{\'o}} C.E., {Zamorano} J.,
  {Arag{\'o}n-Salamanca} A., {Rego} M., 2002, \apjl, 570, L1

\bibitem[{{Garn} et~al.(2009){Garn}, {Green}, {Riley}, \& {Alexander}}]{garn09}
{Garn} T., {Green} D.A., {Riley} J.M., {Alexander} P., 2009, \mnras, 397, 1101

\bibitem[{{Garrett}(2002)}]{garrett02}
{Garrett} M.A., 2002, \aap, 384, L19

\bibitem[{{Geach} et~al.(2008){Geach}, {Smail}, {Best} et~al.}]{geach08}
{Geach} J.E., {Smail} I., {Best} P.N., {Kurk} J., {Casali} M., {Ivison} R.J.,
  {Coppin} K., 2008, \mnras, 388, 1473

\bibitem[{{Genzel} et~al.(2008){Genzel}, {Burkert}, {Bouch{\'e}}
  et~al.}]{genzel08}
{Genzel} R., et~al., 2008, \apj, 687, 59

\bibitem[{{Georgakakis} et~al.(2003){Georgakakis}, {Hopkins}, {Sullivan}
  et~al.}]{georgakakis03}
{Georgakakis} A., {Hopkins} A.M., {Sullivan} M., {Afonso} J., {Georgantopoulos}
  I., {Mobasher} B., {Cram} L.E., 2003, \mnras, 345, 939

\bibitem[{{Giavalisco} et~al.(2004){Giavalisco}, {Dickinson}, {Ferguson}
  et~al.}]{giavalisco04}
{Giavalisco} M., et~al., 2004, \apjl, 600, L103

\bibitem[{{Glazebrook} et~al.(1999){Glazebrook}, {Blake}, {Economou}, {Lilly},
  \& {Colless}}]{glazebrook99}
{Glazebrook} K., {Blake} C., {Economou} F., {Lilly} S., {Colless} M., 1999,
  \mnras, 306, 843

\bibitem[{{Glazebrook} et~al.(2004){Glazebrook}, {Abraham}, {McCarthy}
  et~al.}]{glazebrook04}
{Glazebrook} K., et~al., 2004, \nat, 430, 181

\bibitem[{{Greve} et~al.(2004){Greve}, {Ivison}, {Bertoldi} et~al.}]{greve04}
{Greve} T.R., {Ivison} R.J., {Bertoldi} F., {Stevens} J.A., {Dunlop} J.S.,
  {Lutz} D., {Carilli} C.L., 2004, \mnras, 354, 779

\bibitem[{{Greve} et~al.(2005){Greve}, {Bertoldi}, {Smail} et~al.}]{greve05}
{Greve} T.R., et~al., 2005, \mnras, 359, 1165

\bibitem[{{Gronwall}(1999)}]{gronwall99}
{Gronwall} C., 1999, AIPC, 470, 335

\bibitem[{{Gronwall} et~al.(2007){Gronwall}, {Ciardullo}, {Hickey}
  et~al.}]{gronwall07}
{Gronwall} C., et~al., 2007, \apj, 667, 79

\bibitem[{{Gruppioni} et~al.(2003){Gruppioni}, {Pozzi}, {Zamorani}
  et~al.}]{gruppioni03}
{Gruppioni} C., {Pozzi} F., {Zamorani} G., {Ciliegi} P., {Lari} C., {Calabrese}
  E., {La Franca} F., {Matute} I., 2003, \mnras, 341, L1

\bibitem[{{Gwyn} \& {Hartwick}(2005)}]{gwyn05}
{Gwyn} S.D.J., {Hartwick} F.D.A., 2005, \aj, 130, 1337

\bibitem[{{Haarsma} et~al.(2000){Haarsma}, {Partridge}, {Windhorst}, \&
  {Richards}}]{haarsma00}
{Haarsma} D.B., {Partridge} R.B., {Windhorst} R.A., {Richards} E.A., 2000,
  \apj, 544, 641

\bibitem[{{Hainline}(2008)}]{hainline08phd}
{Hainline} L.J., 2008, Multi-Wavelength Properties of Submillimeter-Selected
  Galaxies, Ph.D. thesis, California Institute of Technology

\bibitem[{{Hainline} et~al.(2006){Hainline}, {Blain}, {Greve}
  et~al.}]{hainline06}
{Hainline} L.J., {Blain} A.W., {Greve} T.R., {Chapman} S.C., {Smail} I.,
  {Ivison} R.J., 2006, \apj, 650, 614

\bibitem[{{Hainline} et~al.(2009){Hainline}, {Blain}, {Smail}
  et~al.}]{hainline09}
{Hainline} L.J., {Blain} A.W., {Smail} I., {Frayer} D.T., {Chapman} S.C.,
  {Ivison} R.J., {Alexander} D.M., 2009, \apj, 699, 1610

\bibitem[{{Hammer} et~al.(1997){Hammer}, {Flores}, {Lilly} et~al.}]{hammer97}
{Hammer} F., et~al., 1997, \apj, 481, 49

\bibitem[{{Hanish} et~al.(2006){Hanish}, {Meurer}, {Ferguson}
  et~al.}]{hanish06}
{Hanish} D.J., et~al., 2006, \apj, 649, 150

\bibitem[{{Helou} et~al.(1985){Helou}, {Soifer}, \& {Rowan-Robinson}}]{helou85}
{Helou} G., {Soifer} B.T., {Rowan-Robinson} M., 1985, \apjl, 298, L7

\bibitem[{{Hippelein} et~al.(2003){Hippelein}, {Maier}, {Meisenheimer}
  et~al.}]{hippelein03}
{Hippelein} H., et~al., 2003, \aap, 402, 65

\bibitem[{{Hogg} et~al.(1998){Hogg}, {Cohen}, {Blandford}, \& {Pahre}}]{hogg98}
{Hogg} D.W., {Cohen} J.G., {Blandford} R., {Pahre} M.A., 1998, \apj, 504, 622

\bibitem[{{Holland} et~al.(1999){Holland}, {Robson}, {Gear}
  et~al.}]{hollandscuba}
{Holland} W.S., et~al., 1999, \mnras, 303, 659

\bibitem[{{Hopkins}(2004)}]{hopkins04}
{Hopkins} A.M., 2004, \apj, 615, 209

\bibitem[{{Hopkins} \& {Beacom}(2006)}]{hopkins06}
{Hopkins} A.M., {Beacom} J.F., 2006, \apj, 651, 142

\bibitem[{{Hopkins} et~al.(2000){Hopkins}, {Connolly}, \& {Szalay}}]{hopkins00}
{Hopkins} A.M., {Connolly} A.J., {Szalay} A.S., 2000, \aj, 120, 2843

\bibitem[{{Hu} et~al.(1998){Hu}, {Cowie}, \& {McMahon}}]{hu98}
{Hu} E.M., {Cowie} L.L., {McMahon} R.G., 1998, \apjl, 502, L99

\bibitem[{{Hughes} et~al.(1998){Hughes}, {Serjeant}, {Dunlop}
  et~al.}]{hughes98}
{Hughes} D.H., et~al., 1998, \nat, 394, 241

\bibitem[{{Huynh} et~al.(2007){Huynh}, {Pope}, {Frayer}, \& {Scott}}]{huynh07}
{Huynh} M.T., {Pope} A., {Frayer} D.T., {Scott} D., 2007, \apj, 659, 305

\bibitem[{{Ibar} et~al.(2008){Ibar}, {Cirasuolo}, {Ivison} et~al.}]{ibar08}
{Ibar} E., et~al., 2008, \mnras, 386, 953

\bibitem[{{Iglesias-P{\'a}ramo} et~al.(2007){Iglesias-P{\'a}ramo}, {Buat},
  {Hern{\'a}ndez-Fern{\'a}ndez} et~al.}]{iglesias07}
{Iglesias-P{\'a}ramo} J., et~al., 2007, \apj, 670, 279

\bibitem[{{Ilbert} et~al.(2010){Ilbert}, {Salvato}, {Le Floc'h}
  et~al.}]{ilbert09}
{Ilbert} O., et~al., 2010, \apj, 709, 644

\bibitem[{{Ivison} et~al.(2000){Ivison}, {Dunlop}, {Smail} et~al.}]{ivison00}
{Ivison} R.J., {Dunlop} J.S., {Smail} I., {Dey} A., {Liu} M.C., {Graham} J.R.,
  2000, \apj, 542, 27

\bibitem[{{Ivison} et~al.(2002){Ivison}, {Greve}, {Smail} et~al.}]{ivison02}
{Ivison} R.J., et~al., 2002, \mnras, 337, 1

\bibitem[{{Ivison} et~al.(2004){Ivison}, {Greve}, {Serjeant} et~al.}]{ivison04}
{Ivison} R.J., et~al., 2004, \apjs, 154, 124

\bibitem[{{Ivison} et~al.(2005){Ivison}, {Smail}, {Dunlop} et~al.}]{ivison05}
{Ivison} R.J., et~al., 2005, \mnras, 364, 1025

\bibitem[{{Ivison} et~al.(2009){Ivison}, {Alexander}, {Biggs}
  et~al.}]{ivison09}
{Ivison} R.J., et~al., 2009, \mnras, 1794--+

\bibitem[{{Iwata} et~al.(2003){Iwata}, {Ohta}, {Tamura} et~al.}]{iwata03}
{Iwata} I., {Ohta} K., {Tamura} N., {Ando} M., {Wada} S., {Watanabe} C.,
  {Akiyama} M., {Aoki} K., 2003, \pasj, 55, 415

\bibitem[{{Iwata} et~al.(2007){Iwata}, {Ohta}, {Tamura} et~al.}]{iwata07}
{Iwata} I., {Ohta} K., {Tamura} N., {Akiyama} M., {Aoki} K., {Ando} M.,
  {Kiuchi} G., {Sawicki} M., 2007, \mnras, 376, 1557

\bibitem[{{Kennicutt}(1998)}]{kennicutt}
{Kennicutt} R.C., 1998, \araa, 36, 189

\bibitem[{{Knudsen} et~al.(2008{\natexlab{a}}){Knudsen}, {Kneib}, \&
  {Egami}}]{knudsen08b}
{Knudsen} K.K., {Kneib} J.P., {Egami} E., 2008{\natexlab{a}}, In: {Chary} R.R.,
  {Teplitz} H.I., {Sheth} K. (eds.) Infrared Diagnostics of Galaxy Evolution,
  vol. 381 of Astronomical Society of the Pacific Conference Series, 372

\bibitem[{{Knudsen} et~al.(2008{\natexlab{b}}){Knudsen}, {van der Werf}, \&
  {Kneib}}]{knudsen08}
{Knudsen} K.K., {van der Werf} P.P., {Kneib} J.P., 2008{\natexlab{b}}, \mnras,
  384, 1611

\bibitem[{{Knudsen} et~al.(2010){Knudsen}, {Kneib}, {Richard}, {Petitpas}, \&
  {Egami}}]{knudsen09}
{Knudsen} K.K., {Kneib} J., {Richard} J., {Petitpas} G., {Egami} E., 2010,
  \apj, 709, 210

\bibitem[{{Kochanek} et~al.(2001){Kochanek}, {Pahre}, {Falco}
  et~al.}]{kochanek01}
{Kochanek} C.S., et~al., 2001, \apj, 560, 566

\bibitem[{{Kodaira} et~al.(2003){Kodaira}, {Taniguchi}, {Kashikawa}
  et~al.}]{kodaira03}
{Kodaira} K., et~al., 2003, \pasj, 55, L17

\bibitem[{{Kov{\'a}cs} et~al.(2006){Kov{\'a}cs}, {Chapman}, {Dowell}
  et~al.}]{kovacs06}
{Kov{\'a}cs} A., {Chapman} S.C., {Dowell} C.D., {Blain} A.W., {Ivison} R.J.,
  {Smail} I., {Phillips} T.G., 2006, \apj, 650, 592

\bibitem[{{Kudritzki} et~al.(2000){Kudritzki}, {M{\'e}ndez}, {Feldmeier}
  et~al.}]{kudritzki00}
{Kudritzki} R.P., et~al., 2000, \apj, 536, 19

\bibitem[{{Lacki} \& {Thompson}(2009)}]{lacki09b}
{Lacki} B.C., {Thompson} T.A., 2009, \apj, submitted, {\tt arXiv:0910.0478
  [astro-ph.CO]}

\bibitem[{{Lacki} et~al.(2009){Lacki}, {Thompson}, \& {Quataert}}]{lacki09}
{Lacki} B.C., {Thompson} T.A., {Quataert} E., 2009, \apj, submitted, {\tt
  arXiv:0907.4161 [astro-ph.CO]}

\bibitem[{{Laurent} et~al.(2006){Laurent}, {Glenn}, {Egami} et~al.}]{laurent06}
{Laurent} G.T., et~al., 2006, \apj, 643, 38

\bibitem[{{Law} et~al.(2009){Law}, {Steidel}, {Erb} et~al.}]{law09}
{Law} D.R., {Steidel} C.C., {Erb} D.K., {Larkin} J.E., {Pettini} M., {Shapley}
  A.E., {Wright} S.A., 2009, \apj, 697, 2057

\bibitem[{{Lilly} et~al.(1996){Lilly}, {Le Fevre}, {Hammer}, \&
  {Crampton}}]{lilly96}
{Lilly} S.J., {Le Fevre} O., {Hammer} F., {Crampton} D., 1996, \apjl, 460, L1

\bibitem[{{Lilly} et~al.(1999){Lilly}, {Eales}, {Gear} et~al.}]{lilly99}
{Lilly} S.J., {Eales} S.A., {Gear} W.K.P., {Hammer} F., {Le F{\`e}vre} O.,
  {Crampton} D., {Bond} J.R., {Dunne} L., 1999, \apj, 518, 641

\bibitem[{{Ly} et~al.(2009){Ly}, {Malkan}, {Treu} et~al.}]{ly09}
{Ly} C., et~al., 2009, \apj, 697, 1410

\bibitem[{{Machalski} \& {Godlowski}(2000)}]{machalski00}
{Machalski} J., {Godlowski} W., 2000, \aap, 360, 463

\bibitem[{{Madau} et~al.(1996){Madau}, {Ferguson}, {Dickinson}
  et~al.}]{madau96}
{Madau} P., {Ferguson} H.C., {Dickinson} M.E., {Giavalisco} M., {Steidel} C.C.,
  {Fruchter} A., 1996, \mnras, 283, 1388

\bibitem[{{Madau} et~al.(1998){Madau}, {Pozzetti}, \& {Dickinson}}]{madau98}
{Madau} P., {Pozzetti} L., {Dickinson} M., 1998, \apj, 498, 106

\bibitem[{{Malhotra} \& {Rhoads}(2004)}]{malhotra04}
{Malhotra} S., {Rhoads} J.E., 2004, \apjl, 617, L5

\bibitem[{{Mann} et~al.(2002){Mann}, {Oliver}, {Carballo} et~al.}]{mann02}
{Mann} R.G., et~al., 2002, \mnras, 332, 549

\bibitem[{{Marchesini} et~al.(2009){Marchesini}, {van Dokkum}, {F{\"o}rster
  Schreiber} et~al.}]{marchesini09}
{Marchesini} D., {van Dokkum} P.G., {F{\"o}rster Schreiber} N.M., {Franx} M.,
  {Labb{\'e}} I., {Wuyts} S., 2009, \apj, 701, 1765

\bibitem[{{Marleau} et~al.(2007){Marleau}, {Fadda}, {Appleton}
  et~al.}]{marleau07}
{Marleau} F.R., {Fadda} D., {Appleton} P.N., {Noriega-Crespo} A., {Im} M.,
  {Clancy} D., 2007, \apj, 663, 218

\bibitem[{{Massarotti} et~al.(2001){Massarotti}, {Iovino}, \&
  {Buzzoni}}]{massarotti01}
{Massarotti} M., {Iovino} A., {Buzzoni} A., 2001, \apjl, 559, L105

\bibitem[{{Mauch} \& {Sadler}(2007)}]{mauch07}
{Mauch} T., {Sadler} E.M., 2007, \mnras, 375, 931

\bibitem[{{Men{\'e}ndez-Delmestre} et~al.(2007){Men{\'e}ndez-Delmestre},
  {Blain}, {Alexander} et~al.}]{menendezdelmestre07}
{Men{\'e}ndez-Delmestre} K., et~al., 2007, \apjl, 655, L65

\bibitem[{{Men{\'e}ndez-Delmestre} et~al.(2009){Men{\'e}ndez-Delmestre},
  {Blain}, {Smail} et~al.}]{menendezdelmestre09}
{Men{\'e}ndez-Delmestre} K., et~al., 2009, \apj, 699, 667

\bibitem[{{Mentuch} et~al.(2009){Mentuch}, {Abraham}, {Glazebrook}
  et~al.}]{mentuch09}
{Mentuch} E., et~al., 2009, \apj, 706, 1020

\bibitem[{{Micha{\l}owski} \& {Hjorth}(2007)}]{michalowski06}
{Micha{\l}owski} M.J., {Hjorth} J., 2007, {in AIP Conf. Ser. 924, The
  Multicolored Landscape of Compact Objects and Their Explosive Origins, ed. L.
  A. Antonelli et al. (Melville, NY: AIP), 143}

\bibitem[{{Micha{\l}owski} et~al.(2008){Micha{\l}owski}, {Hjorth}, {Castro
  Cer{\'o}n}, \& {Watson}}]{michalowski08}
{Micha{\l}owski} M.J., {Hjorth} J., {Castro Cer{\'o}n} J.M., {Watson} D., 2008,
  \apj, 672, 817

\bibitem[{{Micha{\l}owski} et~al.(2009){Micha{\l}owski}, {Hjorth}, {Malesani}
  et~al.}]{michalowski09}
{Micha{\l}owski} M.J., et~al., 2009, \apj, 693, 347

\bibitem[{{Micha{\l}owski} et~al.(2010){Micha{\l}owski}, {Watson}, \&
  {Hjorth}}]{michalowski10smg4}
{Micha{\l}owski} M.J., {Watson} D., {Hjorth} J., 2010, \apj, {accepted {\tt
  arXiv:1002.2636}}

\bibitem[{{Miller} \& {Owen}(2001)}]{miller01}
{Miller} N.A., {Owen} F.N., 2001, \aj, 121, 1903

\bibitem[{{Mobasher} et~al.(2009){Mobasher}, {Dahlen}, {Hopkins}
  et~al.}]{mobasher09}
{Mobasher} B., et~al., 2009, \apj, 690, 1074

\bibitem[{{Moorwood} et~al.(2000){Moorwood}, {van der Werf}, {Cuby}, \&
  {Oliva}}]{moorwood00}
{Moorwood} A.F.M., {van der Werf} P.P., {Cuby} J.G., {Oliva} E., 2000, \aap,
  362, 9

\bibitem[{{Murayama} et~al.(2007){Murayama}, {Taniguchi}, {Scoville}
  et~al.}]{murayama07}
{Murayama} T., et~al., 2007, \apjs, 172, 523

\bibitem[{{Murphy}(2009)}]{murphy09b}
{Murphy} E.J., 2009, \apj, 706, 482

\bibitem[{{Murphy} et~al.(2009){Murphy}, {Chary}, {Alexander}
  et~al.}]{murphy09}
{Murphy} E.J., {Chary} R.R., {Alexander} D.M., {Dickinson} M., {Magnelli} B.,
  {Morrison} G., {Pope} A., {Teplitz} H.I., 2009, \apj, 698, 1380

\bibitem[{{Narayanan} et~al.(2009){Narayanan}, {Cox}, {Hayward}, {Younger}, \&
  {Hernquist}}]{narayanan09}
{Narayanan} D., {Cox} T.J., {Hayward} C.C., {Younger} J.D., {Hernquist} L.,
  2009, \mnras, 400, 1919

\bibitem[{{Narayanan} et~al.(2010){Narayanan}, {Hayward}, {Cox}
  et~al.}]{narayanan10}
{Narayanan} D., {Hayward} C.C., {Cox} T.J., {Hernquist} L., {Jonsson} P.,
  {Younger} J.D., {Groves} B., 2010, \mnras, 401, 1613

\bibitem[{{Nilsson} et~al.(2007){Nilsson}, {M{\o}ller}, {M{\"o}ller}
  et~al.}]{nilsson07}
{Nilsson} K.K., et~al., 2007, \aap, 471, 71

\bibitem[{{Nilsson} et~al.(2009){Nilsson}, {Tapken}, {M{\o}ller}
  et~al.}]{nilsson09}
{Nilsson} K.K., {Tapken} C., {M{\o}ller} P., {Freudling} W., {Fynbo} J.P.U.,
  {Meisenheimer} K., {Laursen} P., {{\"O}stlin} G., 2009, \aap, 498, 13

\bibitem[{{Ouchi} et~al.(2003){Ouchi}, {Shimasaku}, {Furusawa}
  et~al.}]{ouchi03}
{Ouchi} M., et~al., 2003, \apj, 582, 60

\bibitem[{{Ouchi} et~al.(2004){Ouchi}, {Shimasaku}, {Okamura} et~al.}]{ouchi04}
{Ouchi} M., et~al., 2004, \apj, 611, 660

\bibitem[{{Ouchi} et~al.(2008){Ouchi}, {Shimasaku}, {Akiyama} et~al.}]{ouchi08}
{Ouchi} M., et~al., 2008, \apjs, 176, 301

\bibitem[{{Paltani} et~al.(2007){Paltani}, {Le F{\`e}vre}, {Ilbert}
  et~al.}]{paltani07}
{Paltani} S., et~al., 2007, \aap, 463, 873

\bibitem[{{Palunas} et~al.(2004){Palunas}, {Teplitz}, {Francis}, {Williger}, \&
  {Woodgate}}]{palunas04}
{Palunas} P., {Teplitz} H.I., {Francis} P.J., {Williger} G.M., {Woodgate} B.E.,
  2004, \apj, 602, 545

\bibitem[{{Pascale} et~al.(2009){Pascale}, {Ade}, {Bock} et~al.}]{pascale09}
{Pascale} E., et~al., 2009, \apj, 707, 1740

\bibitem[{{Pascarelle} et~al.(1998){Pascarelle}, {Lanzetta}, \&
  {Fern{\'a}ndez-Soto}}]{pascarelle98}
{Pascarelle} S.M., {Lanzetta} K.M., {Fern{\'a}ndez-Soto} A., 1998, \apjl, 508,
  L1

\bibitem[{{Perera} et~al.(2008){Perera}, {Chapin}, {Austermann}
  et~al.}]{perera08}
{Perera} T.A., et~al., 2008, \mnras, 391, 1227

\bibitem[{{P{\'e}rez-Gonz{\'a}lez} et~al.(2003){P{\'e}rez-Gonz{\'a}lez},
  {Zamorano}, {Gallego}, {Arag{\'o}n-Salamanca}, \& {Gil de
  Paz}}]{perezgonzalez03}
{P{\'e}rez-Gonz{\'a}lez} P.G., {Zamorano} J., {Gallego} J.,
  {Arag{\'o}n-Salamanca} A., {Gil de Paz} A., 2003, \apj, 591, 827

\bibitem[{{P{\'e}rez-Gonz{\'a}lez} et~al.(2005){P{\'e}rez-Gonz{\'a}lez},
  {Rieke}, {Egami} et~al.}]{perezgonzalez05}
{P{\'e}rez-Gonz{\'a}lez} P.G., et~al., 2005, \apj, 630, 82

\bibitem[{{P{\'e}rez-Gonz{\'a}lez} et~al.(2008){P{\'e}rez-Gonz{\'a}lez},
  {Rieke}, {Villar} et~al.}]{perezgonzalez08}
{P{\'e}rez-Gonz{\'a}lez} P.G., et~al., 2008, \apj, 675, 234

\bibitem[{{Pettini} et~al.(1998){Pettini}, {Kellogg}, {Steidel}
  et~al.}]{pettini98}
{Pettini} M., {Kellogg} M., {Steidel} C.C., {Dickinson} M., {Adelberger} K.L.,
  {Giavalisco} M., 1998, \apj, 508, 539

\bibitem[{{Pope} et~al.(2006){Pope}, {Scott}, {Dickinson} et~al.}]{pope06}
{Pope} A., et~al., 2006, \mnras, 370, 1185

\bibitem[{{Pope} et~al.(2008){Pope}, {Chary}, {Alexander} et~al.}]{pope08}
{Pope} A., et~al., 2008, \apj, 675, 1171

\bibitem[{{Pozzetti} et~al.(2007){Pozzetti}, {Bolzonella}, {Lamareille}
  et~al.}]{pozzetti07}
{Pozzetti} L., et~al., 2007, \aap, 474, 443

\bibitem[{{Pozzi} et~al.(2004){Pozzi}, {Gruppioni}, {Oliver} et~al.}]{pozzi04}
{Pozzi} F., et~al., 2004, \apj, 609, 122

\bibitem[{{Rawat} et~al.(2008){Rawat}, {Hammer}, {Kembhavi}, \&
  {Flores}}]{rawat08}
{Rawat} A., {Hammer} F., {Kembhavi} A.K., {Flores} H., 2008, \apj, 681, 1089

\bibitem[{{Reddy} et~al.(2006){Reddy}, {Steidel}, {Fadda} et~al.}]{reddy06}
{Reddy} N.A., {Steidel} C.C., {Fadda} D., {Yan} L., {Pettini} M., {Shapley}
  A.E., {Erb} D.K., {Adelberger} K.L., 2006, \apj, 644, 792

\bibitem[{{Reddy} et~al.(2008){Reddy}, {Steidel}, {Pettini} et~al.}]{reddy08}
{Reddy} N.A., {Steidel} C.C., {Pettini} M., {Adelberger} K.L., {Shapley} A.E.,
  {Erb} D.K., {Dickinson} M., 2008, \apjs, 175, 48

\bibitem[{{Rhoads} et~al.(2003){Rhoads}, {Dey}, {Malhotra} et~al.}]{rhoads03}
{Rhoads} J.E., et~al., 2003, \aj, 125, 1006

\bibitem[{{Rieke} et~al.(2009){Rieke}, {Alonso-Herrero}, {Weiner}
  et~al.}]{rieke09}
{Rieke} G.H., {Alonso-Herrero} A., {Weiner} B.J., {P{\'e}rez-Gonz{\'a}lez}
  P.G., {Blaylock} M., {Donley} J.L., {Marcillac} D., 2009, \apj, 692, 556

\bibitem[{{Rudnick} et~al.(2003){Rudnick}, {Rix}, {Franx} et~al.}]{rudnick03}
{Rudnick} G., et~al., 2003, \apj, 599, 847

\bibitem[{{Rudnick} et~al.(2006){Rudnick}, {Labb{\'e}}, {F{\"o}rster Schreiber}
  et~al.}]{rudnick06}
{Rudnick} G., et~al., 2006, \apj, 650, 624

\bibitem[{{Sadler} et~al.(2002){Sadler}, {Jackson}, {Cannon} et~al.}]{sadler02}
{Sadler} E.M., et~al., 2002, \mnras, 329, 227

\bibitem[{{Sajina} et~al.(2008){Sajina}, {Yan}, {Lutz} et~al.}]{sajina08}
{Sajina} A., et~al., 2008, \apj, 683, 659

\bibitem[{{Salpeter}(1955)}]{salpeter}
{Salpeter} E.E., 1955, \apj, 121, 161

\bibitem[{{Salucci} \& {Persic}(1999)}]{salucci99}
{Salucci} P., {Persic} M., 1999, \mnras, 309, 923

\bibitem[{{Santini} et~al.(2009){Santini}, {Fontana}, {Grazian}
  et~al.}]{santini09}
{Santini} P., et~al., 2009, \aap, 504, 751

\bibitem[{{Sawicki} \& {Thompson}(2006{\natexlab{a}})}]{sawicki06}
{Sawicki} M., {Thompson} D., 2006{\natexlab{a}}, \apj, 642, 653

\bibitem[{{Sawicki} \& {Thompson}(2006{\natexlab{b}})}]{sawicki06b}
{Sawicki} M., {Thompson} D., 2006{\natexlab{b}}, \apj, 648, 299

\bibitem[{{Sawicki} et~al.(1997){Sawicki}, {Lin}, \& {Yee}}]{sawicki97}
{Sawicki} M.J., {Lin} H., {Yee} H.K.C., 1997, \aj, 113, 1

\bibitem[{{Schinnerer} et~al.(2008){Schinnerer}, {Carilli}, {Capak}
  et~al.}]{schinnerer08}
{Schinnerer} E., et~al., 2008, \apjl, 689, L5

\bibitem[{{Scott} et~al.(2008){Scott}, {Austermann}, {Perera} et~al.}]{scott08}
{Scott} K.S., et~al., 2008, \mnras, 385, 2225

\bibitem[{{Scott} et~al.(2002){Scott}, {Fox}, {Dunlop} et~al.}]{scott}
{Scott} S.E., et~al., 2002, \mnras, 331, 817

\bibitem[{{Serjeant} et~al.(2002){Serjeant}, {Gruppioni}, \&
  {Oliver}}]{serjeant02}
{Serjeant} S., {Gruppioni} C., {Oliver} S., 2002, \mnras, 330, 621

\bibitem[{{Serjeant} et~al.(2008){Serjeant}, {Dye}, {Mortier}
  et~al.}]{serjeant08}
{Serjeant} S., et~al., 2008, \mnras, 386, 1907

\bibitem[{{Seymour} et~al.(2008){Seymour}, {Dwelly}, {Moss} et~al.}]{seymour08}
{Seymour} N., et~al., 2008, \mnras, 386, 1695

\bibitem[{{Seymour} et~al.(2009){Seymour}, {Huynh}, {Dwelly}
  et~al.}]{seymour09}
{Seymour} N., {Huynh} M., {Dwelly} T., {Symeonidis} M., {Hopkins} A., {McHardy}
  I.M., {Page} M.J., {Rieke} G., 2009, \mnras, 398, 1573

\bibitem[{{Shim} et~al.(2007){Shim}, {Im}, {Choi}, {Yan}, \&
  {Storrie-Lombardi}}]{shim07}
{Shim} H., {Im} M., {Choi} P., {Yan} L., {Storrie-Lombardi} L., 2007, \apj,
  669, 749

\bibitem[{{Shimasaku} et~al.(2005){Shimasaku}, {Ouchi}, {Furusawa}
  et~al.}]{shimasaku05}
{Shimasaku} K., {Ouchi} M., {Furusawa} H., {Yoshida} M., {Kashikawa} N.,
  {Okamura} S., 2005, \pasj, 57, 447

\bibitem[{{Shimasaku} et~al.(2006){Shimasaku}, {Kashikawa}, {Doi}
  et~al.}]{shimasaku06}
{Shimasaku} K., et~al., 2006, \pasj, 58, 313

\bibitem[{{Shioya} et~al.(2008){Shioya}, {Taniguchi}, {Sasaki}
  et~al.}]{shioya08}
{Shioya} Y., et~al., 2008, \apjs, 175, 128

\bibitem[{{Silva} et~al.(1998){Silva}, {Granato}, {Bressan}, \&
  {Danese}}]{silva98}
{Silva} L., {Granato} G.L., {Bressan} A., {Danese} L., 1998, \apj, 509, 103

\bibitem[{{Smail} et~al.(2002){Smail}, {Ivison}, {Blain}, \& {Kneib}}]{smail02}
{Smail} I., {Ivison} R.J., {Blain} A.W., {Kneib} J.P., 2002, \mnras, 331, 495

\bibitem[{{Smail} et~al.(2004){Smail}, {Chapman}, {Blain}, \&
  {Ivison}}]{smail04}
{Smail} I., {Chapman} S.C., {Blain} A.W., {Ivison} R.J., 2004, \apj, 616, 71

\bibitem[{{Sobral} et~al.(2009){Sobral}, {Best}, {Geach} et~al.}]{sobral09}
{Sobral} D., et~al., 2009, \mnras, 398, 75

\bibitem[{{Somerville} et~al.(2001){Somerville}, {Primack}, \&
  {Faber}}]{somerville01}
{Somerville} R.S., {Primack} J.R., {Faber} S.M., 2001, \mnras, 320, 504

\bibitem[{{Stanway} et~al.(2003){Stanway}, {Bunker}, \& {McMahon}}]{stanway03}
{Stanway} E.R., {Bunker} A.J., {McMahon} R.G., 2003, \mnras, 342, 439

\bibitem[{{Stark} et~al.(2007){Stark}, {Bunker}, {Ellis}, {Eyles}, \&
  {Lacy}}]{stark07}
{Stark} D.P., {Bunker} A.J., {Ellis} R.S., {Eyles} L.P., {Lacy} M., 2007, \apj,
  659, 84

\bibitem[{{Stark} et~al.(2009){Stark}, {Ellis}, {Bunker} et~al.}]{stark09}
{Stark} D.P., {Ellis} R.S., {Bunker} A., {Bundy} K., {Targett} T., {Benson} A.,
  {Lacy} M., 2009, \apj, 697, 1493

\bibitem[{{Steidel} et~al.(1999){Steidel}, {Adelberger}, {Giavalisco},
  {Dickinson}, \& {Pettini}}]{steidel99}
{Steidel} C.C., {Adelberger} K.L., {Giavalisco} M., {Dickinson} M., {Pettini}
  M., 1999, \apj, 519, 1

\bibitem[{{Sullivan} et~al.(2000){Sullivan}, {Treyer}, {Ellis}
  et~al.}]{sullivan00}
{Sullivan} M., {Treyer} M.A., {Ellis} R.S., {Bridges} T.J., {Milliard} B.,
  {Donas} J., 2000, \mnras, 312, 442

\bibitem[{{Swinbank} et~al.(2004){Swinbank}, {Smail}, {Chapman}
  et~al.}]{swinbank04}
{Swinbank} A.M., {Smail} I., {Chapman} S.C., {Blain} A.W., {Ivison} R.J.,
  {Keel} W.C., 2004, \apj, 617, 64

\bibitem[{{Swinbank} et~al.(2006){Swinbank}, {Chapman}, {Smail}
  et~al.}]{swinbank06}
{Swinbank} A.M., {Chapman} S.C., {Smail} I., {Lindner} C., {Borys} C., {Blain}
  A.W., {Ivison} R.J., {Lewis} G.F., 2006, \mnras, 371, 465

\bibitem[{{Swinbank} et~al.(2008){Swinbank}, {Lacey}, {Smail}
  et~al.}]{swinbank08}
{Swinbank} A.M., et~al., 2008, \mnras, 391, 420

\bibitem[{{Tacconi} et~al.(2006){Tacconi}, {Neri}, {Chapman}
  et~al.}]{tacconi06}
{Tacconi} L.J., et~al., 2006, \apj, 640, 228

\bibitem[{{Tacconi} et~al.(2008){Tacconi}, {Genzel}, {Smail}
  et~al.}]{tacconi08}
{Tacconi} L.J., et~al., 2008, \apj, 680, 246

\bibitem[{{Takagi} et~al.(2004){Takagi}, {Hanami}, \& {Arimoto}}]{takagi04}
{Takagi} T., {Hanami} H., {Arimoto} N., 2004, \mnras, 355, 424

\bibitem[{{Takata} et~al.(2006){Takata}, {Sekiguchi}, {Smail}
  et~al.}]{takata06}
{Takata} T., {Sekiguchi} K., {Smail} I., {Chapman} S.C., {Geach} J.E.,
  {Swinbank} A.M., {Blain} A., {Ivison} R.J., 2006, \apj, 651, 713

\bibitem[{{Tamura} et~al.(2009){Tamura}, {Kohno}, {Nakanishi}
  et~al.}]{tamura09}
{Tamura} Y., et~al., 2009, \nat, 459, 61

\bibitem[{{Taniguchi} et~al.(2005){Taniguchi}, {Ajiki}, {Nagao}
  et~al.}]{taniguchi05}
{Taniguchi} Y., et~al., 2005, \pasj, 57, 165

\bibitem[{{Teplitz} et~al.(2003){Teplitz}, {Collins}, {Gardner}, {Hill}, \&
  {Rhodes}}]{teplitz03}
{Teplitz} H.I., {Collins} N.R., {Gardner} J.P., {Hill} R.S., {Rhodes} J., 2003,
  \apj, 589, 704

\bibitem[{{Thompson} et~al.(2006){Thompson}, {Eisenstein}, {Fan}
  et~al.}]{thompson06}
{Thompson} R.I., {Eisenstein} D., {Fan} X., {Dickinson} M., {Illingworth} G.,
  {Kennicutt} R.C. Jr., 2006, \apj, 647, 787

\bibitem[{{Trentham} et~al.(1999){Trentham}, {Blain}, \&
  {Goldader}}]{trentham99}
{Trentham} N., {Blain} A.W., {Goldader} J., 1999, \mnras, 305, 61

\bibitem[{{Tresse} \& {Maddox}(1998)}]{tresse98}
{Tresse} L., {Maddox} S.J., 1998, \apj, 495, 691

\bibitem[{{Tresse} et~al.(2002){Tresse}, {Maddox}, {Le F{\`e}vre}, \&
  {Cuby}}]{tresse02}
{Tresse} L., {Maddox} S.J., {Le F{\`e}vre} O., {Cuby} J.G., 2002, \mnras, 337,
  369

\bibitem[{{Treyer} et~al.(1998){Treyer}, {Ellis}, {Milliard}, {Donas}, \&
  {Bridges}}]{treyer98}
{Treyer} M.A., {Ellis} R.S., {Milliard} B., {Donas} J., {Bridges} T.J., 1998,
  \mnras, 300, 303

\bibitem[{{Valiante} et~al.(2007){Valiante}, {Lutz}, {Sturm}
  et~al.}]{valiante07}
{Valiante} E., {Lutz} D., {Sturm} E., {Genzel} R., {Tacconi} L.J., {Lehnert}
  M.D., {Baker} A.J., 2007, \apj, 660, 1060

\bibitem[{{van Breukelen} et~al.(2005){van Breukelen}, {Jarvis}, \&
  {Venemans}}]{vanbreukelen05}
{van Breukelen} C., {Jarvis} M.J., {Venemans} B.P., 2005, \mnras, 359, 895

\bibitem[{{van de Ven} et~al.(2003){van de Ven}, {van Dokkum}, \&
  {Franx}}]{ven03}
{van de Ven} G., {van Dokkum} P.G., {Franx} M., 2003, \mnras, 344, 924

\bibitem[{{van Dokkum} \& {Franx}(2001)}]{dokkum01}
{van Dokkum} P.G., {Franx} M., 2001, \apj, 553, 90

\bibitem[{{Villar} et~al.(2008){Villar}, {Gallego}, {P{\'e}rez-Gonz{\'a}lez}
  et~al.}]{villar08}
{Villar} V., {Gallego} J., {P{\'e}rez-Gonz{\'a}lez} P.G., {Pascual} S.,
  {Noeske} K., {Koo} D.C., {Barro} G., {Zamorano} J., 2008, \apj, 677, 169

\bibitem[{{Vlahakis} et~al.(2007){Vlahakis}, {Eales}, \& {Dunne}}]{vlahakis07}
{Vlahakis} C., {Eales} S., {Dunne} L., 2007, \mnras, 379, 1042

\bibitem[{{Wadadekar} et~al.(2006){Wadadekar}, {Casertano}, \& {de
  Mello}}]{wadadekar06}
{Wadadekar} Y., {Casertano} S., {de Mello} D., 2006, \aj, 132, 1023

\bibitem[{{Wall} et~al.(2008){Wall}, {Pope}, \& {Scott}}]{wall08}
{Wall} J.V., {Pope} A., {Scott} D., 2008, \mnras, 383, 435

\bibitem[{{Wang} et~al.(2006){Wang}, {Cowie}, \& {Barger}}]{wang06}
{Wang} W., {Cowie} L.L., {Barger} A.J., 2006, \apj, 647, 74

\bibitem[{{Wang} et~al.(2004){Wang}, {Cowie}, \& {Barger}}]{wang04}
{Wang} W.H., {Cowie} L.L., {Barger} A.J., 2004, \apj, 613, 655

\bibitem[{{Wang} et~al.(2009){Wang}, {Barger}, \& {Cowie}}]{wang09}
{Wang} W.H., {Barger} A.J., {Cowie} L.L., 2009, \apj, 690, 319

\bibitem[{{Watabe} et~al.(2009){Watabe}, {Risaliti}, {Salvati}
  et~al.}]{watabe09}
{Watabe} Y., {Risaliti} G., {Salvati} M., {Nardini} E., {Sani} E., {Marconi}
  A., 2009, \mnras, 396, L1

\bibitem[{{Webb} et~al.(2003{\natexlab{a}}){Webb}, {Eales}, {Lilly}
  et~al.}]{webb03b}
{Webb} T.M., et~al., 2003{\natexlab{a}}, \apj, 587, 41

\bibitem[{{Webb} et~al.(2003{\natexlab{b}}){Webb}, {Lilly}, {Clements}
  et~al.}]{webb03}
{Webb} T.M.A., {Lilly} S.J., {Clements} D.L., {Eales} S., {Yun} M., {Brodwin}
  M., {Dunne} L., {Gear} W.K., 2003{\natexlab{b}}, \apj, 597, 680

\bibitem[{{Wei{\ss}} et~al.(2009{\natexlab{a}}){Wei{\ss}}, {Ivison}, {Downes}
  et~al.}]{weiss09}
{Wei{\ss}} A., {Ivison} R.J., {Downes} D., {Walter} F., {Cirasuolo} M.,
  {Menten} K.M., 2009{\natexlab{a}}, \apjl, 705, L45

\bibitem[{{Wei{\ss}} et~al.(2009{\natexlab{b}}){Wei{\ss}}, {Kov{\'a}cs},
  {Coppin} et~al.}]{weiss09b}
{Wei{\ss}} A., et~al., 2009{\natexlab{b}}, \apj, 707, 1201

\bibitem[{{Wilson} et~al.(2002){Wilson}, {Cowie}, {Barger}, \&
  {Burke}}]{wilson02}
{Wilson} G., {Cowie} L.L., {Barger} A.J., {Burke} D.J., 2002, \aj, 124, 1258

\bibitem[{{Wyder} et~al.(2005){Wyder}, {Treyer}, {Milliard} et~al.}]{wyder05}
{Wyder} T.K., et~al., 2005, \apjl, 619, L15

\bibitem[{{Yan} et~al.(2006){Yan}, {Dickinson}, {Giavalisco} et~al.}]{yan06}
{Yan} H., {Dickinson} M., {Giavalisco} M., {Stern} D., {Eisenhardt} P.R.M.,
  {Ferguson} H.C., 2006, \apj, 651, 24

\bibitem[{{Yan} et~al.(1999){Yan}, {McCarthy}, {Freudling} et~al.}]{yan99}
{Yan} L., {McCarthy} P.J., {Freudling} W., {Teplitz} H.I., {Malumuth} E.M.,
  {Weymann} R.J., {Malkan} M.A., 1999, \apjl, 519, L47

\bibitem[{{Yang} et~al.(2007){Yang}, {Greve}, {Dowell}, \& {Borys}}]{yang07}
{Yang} M., {Greve} T.R., {Dowell} C.D., {Borys} C., 2007, \apj, 660, 1198

\bibitem[{{Yoshida} et~al.(2006){Yoshida}, {Shimasaku}, {Kashikawa}
  et~al.}]{yoshida06}
{Yoshida} M., et~al., 2006, \apj, 653, 988

\bibitem[{{Younger} et~al.(2007){Younger}, {Fazio}, {Huang} et~al.}]{younger07}
{Younger} J.D., et~al., 2007, \apj, 671, 1531

\bibitem[{{Younger} et~al.(2008){Younger}, {Fazio}, {Wilner}
  et~al.}]{younger08b}
{Younger} J.D., et~al., 2008, \apj, 688, 59

\bibitem[{{Younger} et~al.(2009{\natexlab{a}}){Younger}, {Fazio}, {Huang}
  et~al.}]{younger09b}
{Younger} J.D., et~al., 2009{\natexlab{a}}, \apj, 704, 803

\bibitem[{{Younger} et~al.(2009{\natexlab{b}}){Younger}, {Omont}, {Fiolet}
  et~al.}]{younger09}
{Younger} J.D., et~al., 2009{\natexlab{b}}, \mnras, 394, 1685

\bibitem[{{Yun} et~al.(2001){Yun}, {Reddy}, \& {Condon}}]{yun01}
{Yun} M.S., {Reddy} N.A., {Condon} J.J., 2001, \apj, 554, 803

\bibitem[{{Zheng} et~al.(2007){Zheng}, {Bell}, {Papovich} et~al.}]{zheng07}
{Zheng} X.Z., {Bell} E.F., {Papovich} C., {Wolf} C., {Meisenheimer} K., {Rix}
  H.W., {Rieke} G.H., {Somerville} R., 2007, \apjl, 661, L41

\end{thebibliography}
\end{document}